\documentclass[11pt]{article}
\usepackage[left=1in,top=1in,right=1in,bottom=1in,head=0in,nofoot]{geometry}

\usepackage{setspace,url,bm,amsmath} 
\usepackage{scalerel}

\usepackage{titlesec} 
\titlelabel{\thetitle.\quad} 
\titleformat*{\section}{\bf\Large\center}

\usepackage{float}
\usepackage{bbm}
\usepackage{latexsym}
\usepackage{caption}

\usepackage[utf8]{inputenc}
\usepackage{amsmath, amssymb, amsthm, bbm, mathrsfs, mathtools}
\usepackage[colorlinks,citecolor=blue,urlcolor=blue, hypertexnames=false]{hyperref}
\usepackage[linesnumbered,ruled,vlined]{algorithm2e}
\usepackage{verbatim}
\usepackage{multirow}
\usepackage{booktabs, array, threeparttable}

\usepackage{rotating}
\usepackage{graphicx}
\usepackage{subfigure}
\usepackage{float}
\usepackage{caption}
\usepackage{xcolor}
\usepackage{amsfonts}
\usepackage{xr}
\usepackage{tabularx, multicol}
\usepackage{enumitem}

\usepackage{etoolbox}
\usepackage{tabularx, booktabs, graphicx}

\makeatletter
\newcommand{\ostar}{\mathbin{\mathpalette\make@circled\star}}
\newcommand{\make@circled}[2]{%
  \ooalign{$\m@th#1\smallbigcirc{#1}$\cr\hidewidth$\m@th#1#2$\hidewidth\cr}%
}
\newcommand{\smallbigcirc}[1]{%
  \vcenter{\hbox{\scalebox{0.77778}{$\m@th#1\bigcirc$}}}%
}
\makeatother

\newcommand{\lt}{\left}
\newcommand{\rt}{\right}

\newcommand{\commenting}[1]{}

\renewcommand{\hat}{\widehat}
\renewcommand{\tilde}{\widetilde}

\newcommand{\Var}[1]{{\operatorname{Var}\left\{#1\right\}}}
\newcommand{\COV}[1]{{\operatorname{Cov}\left\{#1\right\}}}
\newcommand{\Cov}[2]{{\operatorname{Cov}\left\{#1,#2\right\}}}

\newcommand{\myvec}[1]{{\operatorname{vec}(#1)}}
\newcommand{\E}[1]{{\bbE\left\{#1\right\}}}
\newcommand{\Prob}[1]{{\bbP\left\{#1\right\}}}
\newcommand{\trace}[1]{{\operatorname{Tr}\left(#1\right)}}
\newcommand{\tr}[2]{\left\langle#1\, ,\, #2\right\rangle}
\newcommand{\ind}[1]{\boldsymbol{1}\left\{#1\right\}}

\newcommand{\innerprod}[2]{\left\langle#1\, ,\, #2\right\rangle}

\ifdef{\see}{\renewcommand{\see}[1]{\text{ (#1)}}}{\newcommand{\see}[1]{\text{ (#1)}}}

\newcommand{\diag}[1]{\mathrm{Diag}\left\{#1\right\}}

\def\boxit#1{\vbox{\hrule\hbox{\vrule\kern6pt\vbox{\kern6pt#1\kern6pt}\kern6pt\vrule}\hrule}}

\newcolumntype{P}[1]{>{\centering\arraybackslash}p{#1}}
\newcolumntype{M}[1]{>{\centering\arraybackslash}m{#1}}
\newcolumntype{L}[1]{>{\raggedright\arraybackslash}m{#1}}

\newcommand{\cA}{{\mathcal{A}}}

\newcommand{\cG}{{\mathcal{G}}}
\newcommand{\cL}{{\mathcal{L}}}

\newcommand{\cN}{{\mathcal{N}}}

\newcommand{\cQ}{{\mathcal{Q}}}

\newcommand{\bbP}{{\mathbb{P}}}
\newcommand{\bbT}{{\mathbb{T}}}
\newcommand{\bbS}{{\mathbb{S}}}
\newcommand{\bbR}{{\mathbb{R}}}
\newcommand{\bbB}{{\mathbb{B}}}
\newcommand{\bbE}{{\mathbb{E}}}

\newcommand{\bc}{{\boldsymbol{c}}}

\newcommand{\bF}{{\boldsymbol{F}}}

\newcommand{\bS}{{\boldsymbol{S}}}

\newcommand{\bV}{{\boldsymbol{V}}}

\newcommand{\bY}{{\boldsymbol{Y}}}

\newcommand{\hgamma}{{\hat{\gamma}}}

\newcommand{\hv}{{\hat{v}}}

\newcommand{\hS}{{\hat{S}}}
\newcommand{\hV}{{\hat{V}}}
\newcommand{\hY}{{\hat{Y}}}

\newcommand{\GG}[1]{}

\usepackage{amsthm}
\usepackage{amssymb}
\usepackage{amsmath}
\usepackage{color}

\usepackage{comment}
\theoremstyle{definition}

\newtheorem*{theorem*}{Theorem}
\newtheorem{theorem}{Theorem}
\newtheorem*{rmk*}{remark}
\newtheorem{proposition}{Proposition}
\newtheorem{lemma}{Lemma}
\newtheorem{example}{Example}

\newtheorem{condition}{Condition}

\newtheorem{definition}{Definition}
\newtheorem{remark}{Remark}
\newtheorem{corollary}{Corollary}
\newtheorem*{corollary*}{Corollary}

\usepackage{natbib} 
\bibpunct{(}{)}{;}{a}{}{,} 

\usepackage{etoolbox} 
\apptocmd{\sloppy}{\hbadness 10000\relax}{}{} 

\usepackage{multibib}
\newcites{sec}{References}

\usepackage{listings}

\usepackage{lscape}

\RequirePackage[normalem]{ulem}

\allowdisplaybreaks

\begin{document}
\onehalfspacing

\title{\bf  
 Berry--Esseen bounds for design-based causal inference with possibly diverging treatment levels and varying group sizes
} 
\author{Lei Shi and Peng Ding
\footnote{Lei Shi, Division of Biostatistics, University of California, Berkeley, CA 94720  (E-mail: leishi@berkeley.edu). Peng Ding, Department of Statistics, University of California, Berkeley, CA 94720 (E-mail: pengdingpku@berkeley.edu).
}
}
\date{}
 
\maketitle

\begin{abstract}
\citet{neyman1923application} introduced the randomization model, which contains the notation of potential outcomes to define causal effects and a framework for large-sample inference based on the design of the experiment. However, the existing theory for this framework is far from complete, especially when the number of treatment levels diverges and the treatment group sizes vary. We provide a unified discussion of statistical inference under the randomization model with general treatment group sizes. We formulate the estimator in terms of a linear permutation statistic and use results based on Stein's method to derive various Berry--Esseen bounds on the linear and quadratic functions of the estimator. These new Berry--Esseen bounds serve as the basis for design-based causal inference with possibly diverging treatment levels and a diverging number of causal parameters of interest. We also fill an important gap by proposing novel variance estimators for experiments with possibly many treatment levels without replications. Equipped with the newly developed results, design-based causal inference in general settings becomes more convenient with stronger theoretical guarantees.
\end{abstract}


\newpage
 
\section{Motivation: randomization-based causal inference}
\label{sec:motivation}

\subsection{Existing results}

In a seminal paper, \citet{neyman1923application} introduced the notation of potential outcomes to define causal effects.  More importantly, he also proposed a framework for statistical inference of causal effects based on the design of the experiment. In particular, he considered an experiment with $N$ units and $Q$ treatment arms, where the number of units under treatment $q$ equals $N_q$, with $\sum_{q=1}^Q N_q= N$. Corresponding to treatment level $q $, unit $i$ has the potential outcome $Y_i(q)$, where $i=1, \ldots, N$ and $q = 1, \ldots, Q$. Despite its simplicity, the following completely randomized experiment has been widely used in practice and has generated rich theoretical results. Definition \ref{def:cr} below characterizes the joint distribution of $Z =  (Z_1, \ldots, Z_N)$ under complete randomization, where $Z_i \in \{1, \ldots, Q\}$ is the treatment indicator for unit $i$. 


\begin{definition}[Complete randomization]\label{def:cr}
Fix treatment group sizes $N_1,\ldots, N_Q$ with $\sum_{q=1}^Q N_q= N$. The treatment vector $Z$ is uniform over all its possible values. 
\end{definition}

Mathematically, Definition \ref{def:cr} implies that $ \bbP( {Z} = {z} ) = N_1 ! \cdots N_Q! / N! $ for all possible values of ${z} = (z_1, \ldots, z_N ) $ such that $\sum_{i=1}^N \ind{z_i = q} = N_q$.  Computationally,  Definition \ref{def:cr} implies that ${Z}$ is from a random permutation of $N_1$ $1$'s, $\ldots $, $N_Q$ $Q$'s. \citet{neyman1923application} formulated complete randomization based on an urn model, which is equivalent to Definition \ref{def:cr}.    The experiment reveals one of the potential outcomes, which is the observed outcome $  Y_i = Y_i(Z_i) = \sum_{q=1}^Q Y_i(q) \ind{Z_i = q}$ for each unit $i$. 
%
%

In \citet{neyman1923application}'s framework, all potential outcomes are fixed and only the treatment indicators are random according to Definition \ref{def:cr}. \citet[][Chapter 9]{scheffe1959analysis} called it the {\it randomization model}. Under this model, it is conventional to call the resulting inference as {\it randomization inference} or {\it design-based inference}. It has become increasingly popular in both theory and practice \citep[e.g.,][]{kempthorne52, copas1973randomization, robins1988confidence, rosenbaum2002observational, hinkelmann07, freedman2008Aregression, freedman2008Bregression, lin2013agnostic, dasgupta2015causal, imbens15, ATHEY201773, fogarty2018regression, guo2021generalized}. We focus on \citet{neyman1923application}'s framework throughout the paper. 


A central goal in  \citet{neyman1923application} 's framework is to use the observed data $(Z_i, Y_i)_{i=1}^N$ to make inference of causal effects defined by the potential outcomes.  Define 
$$
\overline{Y}(q) = N^{-1}\sum_{i=1}^N Y_i(q) ,\quad
S(q,q') =  (N-1)^{-1} \sum_{i=1}^N(Y_i(q)-\overline{Y}(q))(Y_i(q')-\overline{Y}(q'))
$$
as the average value of the potential outcomes under treatment $q$ and the covariance of the potential outcomes under treatments $q$ and $q'$, respectively. 
Define the average potential outcome vector as $\overline{Y} = (\overline{Y}(1), \ldots,  \overline{Y}(Q))^\top  \in \bbR^Q$, and define the covariance matrix of the potential outcomes as $S = (S(q,q'))_{q,q'=1,\ldots, Q}$. The parameter of interest is a linear transformation of $\overline{Y}$:
\begin{align*}
    \gamma = F^\top \overline{Y}  
\end{align*}
for a pre-specified $F\in\bbR^{Q\times H}$. We call the matrix $F$ the coefficient matrix. 
In many problems, $F$ is a contrast matrix with columns orthogonal to $(1,\ldots,1)^\top$. Despite the simple form of $\gamma$, it can answer questions from a wide range of applications.
For instance, \citet{neyman1923application} considered pairwise differences in means, and \citet{dasgupta2015causal} and \citet{mukerjee2018using} considered linear combinations of the mean vector. Recently, \citet{li2017general} unified the literature by studying the properties of the linear 
 moment estimator for $\gamma$ under complete randomization. In particular, define the sample mean and variance of the observed  $Y_i(q)$'s as
\begin{align}\label{eqn:hS}
 \hY_q = N_q^{-1}\sum_{Z_i = q}Y_i,\quad \hS(q,q) =  (N_q-1)^{-1} \sum_{Z_i = q} (Y_i - {\hY_q} )^2,
\end{align}
respectively. Define 
\begin{align}\label{eqn:hY-hV}
\hY = (\hY_1,\ldots,\hY_Q)^\top\in\bbR^Q,\quad  \hV_\hY = \diag{N_q^{-1}\hS(q,q)}_{q\in[Q]} \in \bbR^{Q\times Q}
\end{align}
as the vector of sample averages and the diagonal matrix of the sample variances across all  arms, respectively. Under complete randomization, the random vector $ \hY  $ has mean and covariance 
\begin{align}\label{eqn:mean-var-Gamma}
\bbE\{\hY \} = \overline{Y}  ,\quad 
\COV{\hY} =  V_\hY = \diag{N_q^{-1}S(q,q)}_{q\in[Q]} - N^{-1}S,
\end{align} 
and moreover, $\hV_\hY$ is a conservative estimator for  $V_\hY$ in the sense that $ \bbE\{\hV_\hY\} - V_\hY = N^{-1}S$ is positive semidefinite (see \citet{li2017general} for a review). Therefore,  
    \begin{align}\label{eqn:estimates}
        \hgamma = F^\top \hY, \quad \hV_\hgamma = F^\top \hV_\hY F
    \end{align}
are an unbiased point estimator for $    \gamma$ and a conservative covariance estimator for $V_\hgamma = \COV{\hgamma} =  F^\top V_\hY F $, respectively. \citet{li2017general} also used the established combinatorial or rank central limit theorems (CLTs) \citep{hajek1960limiting, hoeffding1951combinatorial, fraser1956vector} to prove the asymptotic normality of $   \hgamma $ and the validity of the associated large-sample Wald-type inference, under certain regularity conditions.

\subsection{Open questions}
\label{sec::open-questions}

Despite the long history of \citet{neyman1923application}'s randomization model, the theory for randomization-based causal inference is far from complete. Technically, \citet{li2017general}'s review only covered the first regime \ref{regime:R1} below, and even there, finer results such as Berry--Esseen bounds (BEBs) have not been rigorously established for the most general setting. For other regimes below, many basic results are still missing in the literature.

\begin{enumerate}[label={({R\arabic*})}]
    \item\label{regime:R1} Small $Q$ and large $N_q$'s.  In this regime, the number of arms is small and the sample size in each arm is large. Asymptotically, as $N \to \infty$, we have that $Q$ is a fixed integer and $N_q/N\to e_q\in(0,1)$ for all $q=1,\ldots, Q$. 
    \cite{li2017general} showed that, under \ref{regime:R1} and some regularity conditions on the potential outcomes, we have 
\begin{equation}\label{eq::basic-asymptotic-inference}
        V_\hgamma^{-1/2}(\hgamma - \gamma) \rightsquigarrow \cN(0, I_H), \quad  N\hV_\hgamma - N\bbE\{\hV_\hgamma\} = o_\bbP(1),
\end{equation}
which ensures that the large sample Wald-type inference based on the Normal approximation is conservative.    \cite{li2017general}'s results are asymptotic. An important theoretical question is to quantify the finite-sample properties of $\hgamma$ by deriving non-asymptotic results.

    \item\label{regime:R2} Large $Q$ and large $N_q$'s. In this regime, each arm has adequate units for the variance estimation, but the number of arms is also large.  Asymptotically, as $N \to \infty$, we have $Q\to \infty$ and $N_q \to \infty$ for all $q=1,\ldots, Q$. Consequently, the limiting values of some $N_q/N$'s must be 0. The point estimates and variance estimators in \eqref{eqn:estimates} are still well-defined in this regime. We might expect that the asymptotic results in \eqref{eq::basic-asymptotic-inference} still hold because of large $N_q$'s. However, previous theoretical results do not cover this seemingly easy case due to the possibly diverging dimension of $F$.

    \item\label{regime:R3} Large $Q$ and small $N_q$'s. In this regime, the number of arms is large but the sample size within each arm is small. Asymptotically, as $N \to \infty$, we have $Q\to \infty$ and $2\le N_q\le \overline{n} $ for all $q$'s and some fixed $\overline{n} \ge 2$. This regime is well suited for many factorial experiments (see Example \ref{eg::factorial-design} below), in which the total number of factor combinations can be much larger than the number of replications in each combination \citep[e.g.,][]{mukerjee2006modern, wu2021experiments}. Although the point estimate and variance estimator in \eqref{eqn:estimates} are still well-defined, we do not expect a simple CLT based on the joint asymptotic normality of $\hY$ due to the small $N_q$'s. Nevertheless, $\widehat{\gamma} = F^\top \hY$, as a linear transformation of $\hY$, can still satisfy the CLT for some choice of $F$. This regime is a reminiscence of the so-called {\it proportional asymptotics} in regression analysis, and even there, statistical inference is still not satisfactory in general \citep[e.g.,][]{el2013robust, lei2018asymptotics, el2018can}. Technically, we need to analyze       $ F^\top \hY$ with the dimension of $\hY$ proportional to the sample size under the randomization model. This is a gap in the literature.

    \item\label{regime:R4} Large $Q$ and $N_q = 1 $ for all $q = 1, \ldots, Q$. This regime is much harder than \ref{regime:R3} because the variance estimator in \eqref{eqn:estimates} is not even well defined due to the lack of replications within each arm \citep[e.g.,][]{espinosa2016bayesian}. Therefore, we need to answer two fundamental questions. First, does $ F^\top \hY$ still satisfy the CLT for some $F$? Second, how do we estimate the variance of $ F^\top \hY$? These two questions are the basis for large-sample Wald-type inference in this regime. Neither has been covered by existing results.

    \item\label{regime:R5} Mixture of \ref{regime:R1}--\ref{regime:R4}. In the most general case, it is possible that the number of treatment levels diverges and the group sizes within different treatment arms vary a lot. Theoretically, we can partition the treatment levels into different types corresponding to the four regimes above. Understanding \ref{regime:R5} relies on understanding \ref{regime:R1}--\ref{regime:R4}.    Due to the difficulties in \ref{regime:R1}--\ref{regime:R4} mentioned above,  a rigorous analysis of \ref{regime:R5} requires a deeper understanding of the randomization model. This is another gap in the literature.

\end{enumerate}

\subsection{Classification of designs based on treatment group sizes}

For descriptive convenience, we define \ref{regime:R1}--\ref{regime:R4} as \textit{nearly uniform designs} and \ref{regime:R5} as \textit{general designs}, respectively, based on the heterogeneity of the group sizes across treatment arms. 
Definitions \ref{def:uniform-design} and \ref{def:non-uniform-design} below make the intuition more precise.

\begin{definition}[Nearly uniform design]\label{def:uniform-design} There exists a positive integer $N_0 > 0$ and absolute constants $\underline{c} \le \overline{c}$, such that $ N_q = c_q  {N}_0$ with  $\underline{c}\le c_q\le \overline{c}$, for all $q=1,\ldots, Q$. 
\end{definition}

Definition \ref{def:uniform-design} is a finite-sample characterization. It can allow $N_0$ to grow with $N$ as in \ref{regime:R1} and \ref{regime:R2}; it can also allow $N_0$ to be fixed as in  \ref{regime:R3} and \ref{regime:R4} with a growing number of treatment levels $Q$. It is a unified description of \ref{regime:R1}--\ref{regime:R4} where each arm contains a similar number of replications. 

\begin{definition}[General design]\label{def:non-uniform-design}
Partition the treatment arms as $\{ 1,\ldots, Q \}= \cQ_{\textsc{s}}\cup\cQ_{\textsc{l}}$ with detailed descriptions below.

(i) $\cQ_{\textsc{l}}$ contains the arms with large sample sizes.  There exists a positive integer $N_0$ and absolute constants $\underline{c} \le \overline{c}$, such that  $N_q = c_q N_0$ with  $\underline{c}\le c_q\le \overline{c}$, for all $q\in\cQ_{\textsc{l}}$.

(ii) $\cQ_{\textsc{s}}$ contains the arms with small sample sizes. There exists a fixed integer $\overline{n}  $ such that  
    $N_q \le \overline{n} $ for all $q\in\cQ_{\textsc{s}}$. Further partition $\cQ_{\textsc{s}}$ as $\cQ_{\textsc{s}} = \cQ_{\textsc{u}} \cup \cQ_{\textsc{r}}$ where
    \begin{itemize}
        \item $\cQ_{\textsc{r}}$ contains the arms with replications, that is,  $ 2\le N_q \le \overline{n}$ for all $q\in\cQ_{\textsc{r}}$;
                \item $\cQ_{\textsc{u}}$ contains the arms without  replications, that is, $N_q = 1$ for all $q\in\cQ_{\textsc{u}}$.
    \end{itemize}
\end{definition}

For simplicity, we will use $|\cQ_{\star}|$ and $N_\star = \sum_{q\in\cQ_\star} N_q$ to denote  the number of arms and the sample size in $\cQ_\star$, respectively, where $\star \in \{ \textsc{s},\textsc{u},\textsc{r},\textsc{l}\}$. As a special case of Definition \ref{def:non-uniform-design}, $|\cQ_{\textsc{r}}| = |\cQ_{\textsc{l}}| = 0$ corresponds to \textit{unreplicated designs} in which each treatment level has only one observation. 

Definition \ref{def:non-uniform-design} gives a partition of the treatment levels. It is a finite-sample characterization and does not impose any restrictions on the magnitude of $N_0$ and $\overline{n}$. Nevertheless, it is indeed motivated by the regime in which the arms in $\cQ_\textsc{l}$ contain many replications and the arms in $\cQ_\textsc{s}$ contain small numbers of replications. In other words, Definition \ref{def:non-uniform-design} is more interesting for the regime in which $N_0$ is much larger than $ \overline{n} $.  More generally, it is possible for some arm in $\cQ_{\textsc{l}}$ to converge faster than others. Our way of defining $\cQ_{\textsc{l}}$ provides a simple way to decipher the finite-sample implications of the derived BEBs. In the main theoretical results developed later, we will impose further restrictions in the theorems and assume conditions on $N_0$ and $\overline{n}$.

In practice, for most experiments, the partition is naturally dictated by the design protocol and can therefore be fixed \emph{a priori}. For example, in the partially nested experiment discussed in Example \ref{exp:provider} later, treated and control units are naturally categorized into large and small arms. In general, the partition should always be determined case by case, taking the specifics of the design into account.

Table \ref{tab:many-settings} summarizes the important regimes and reviews the established and missing theoretical results.  The overarching  goal of the paper is to provide BEBs for all regimes in Table \ref{tab:many-settings}. 
\begin{table}[h]
\centering
\caption{Theoretical results for multi-armed experiments under the randomization model. The regimes
\ref{regime:R1}--\ref{regime:R4} correspond to nearly uniform designs by Definition \ref{def:uniform-design}, whereas the regime \ref{regime:R5} corresponds to general designs by Definition \ref{def:non-uniform-design}. 
}
\label{tab:many-settings}
\begin{tabular}{M{1.0cm}M{1.0cm}M{3cm}M{6cm}}
\toprule
 Regime & $Q$   & $N_q$                    & CLT, variance estimation, and BEB              \\ \hline
   \ref{regime:R1} &  Small &   Large & CLT and variance estimation; no BEB  \\  
\ref{regime:R2}        & Large & Large                    & Seems similar to \ref{regime:R1} but not studied \\  
 \ref{regime:R3}        & Large & Small but $N_q\ge$ 2 & Not studied    \\  
   \ref{regime:R4} &  Large &  $N_q = 1$ & Not studied; variance estimation is nontrivial \\ \hline
   \ref{regime:R5} &   \multicolumn{2}{c}{{Mixture of the above}} &  Not studied \\ \bottomrule
\end{tabular}%
\end{table}

\subsection{Motivating examples}\label{sec:motivating-examples}
There are many practical experimental settings that are relevant to our regimes. We will also use the $2^K$ factorial design as a canonical example for many theoretical results throughout. We review the basic setup of the $2^K$ factorial design in Example \ref{eg::factorial-design}  below \citep{dasgupta2015causal, lu2016covariate, zhao2021regression}.

\begin{example}[Factorial design] 
\label{eg::factorial-design}
A $2^K$ factorial design has $K$ binary factors which generate $Q=2^K$ treatment levels. Index the potential outcomes $Y_i(q)$'s also as $Y_i(z_1, \ldots, z_K)$'s, where $q = 1,\ldots, Q$ and $z_1,\ldots, z_K = 0,1$. The parameter of interest $    \gamma = F^\top \overline{Y}  $ may consist of a subset of the factorial effects.  The contrast matrix $F$ has orthogonal columns and entries of $\pm Q^{-1}$. For example, when $K = 2$, the three factorial effects are characterized by the following contrast matrix:
\[
{F} = 
\bordermatrix{
~ & \tau_{\{1\}} & \tau_{\{2\}}  & \tau_{\{1,2\}}  \cr
(00)  & -1 & -1 &  1 \cr
(01)  & -1 & 1  & -1  \cr
(10)  & 1 & -1  & -1 \cr
(11)  & 1 &  1  &  1 
}.
\]
See \citet{dasgupta2015causal} for precise definitions of main effects and interactions in $2^K$ factorial experiments.  
\end{example}

By definition, the factorial design can have a large number of treatment levels and varying treatment group sizes. Previous asymptotic results only covered factorial designs under \ref{regime:R1} with fixed $K$ and large sample sizes for all treatment levels. This asymptotic regime can be a poor approximation to finite-sample properties of factorial designs with even a moderate $K$ (for example, if $K=10$ then $Q=2^K > 1000$). Such regimes are getting popular in conjoint survey experiments in political science \citep{hainmueller2014causal, caughey2019item, zhirkov2022estimating, shi2025forward}.  Based on simulation, \citet[][Appendix D]{zhao2021regression} showed that CLTs are likely to hold even with diverging $K$ and small sample sizes for all treatment levels.  Allowing for a diverging $K$, \citet[][Theorem A1]{li2017general} derived the CLT for a single factorial effect under the sharp null hypothesis of no treatment effects for any units whatsoever, i.e., $Y_i(1) = \cdots = Y_i(Q)$ for all $i=1,\ldots, N$. However, deriving general asymptotic results for the factorial design has been an open problem in the literature.

As another motivating example, we consider the following partially nested experiment with provider effects, which necessitates the study of general designs given by Definition \ref{def:non-uniform-design}.  
\begin{example}[Partially nested experiment with provider effects]\label{exp:provider}
Let $q=1,\ldots ,Q-1$ index the treated arms in which only limited units are recruited; for simplicity assume $N_q\le \overline{n}$ with a bounded $\overline{n}$ for $q\le Q-1$. Finally, let $q=Q$ index the control arm with many units so that $N_Q\to\infty$. Unlike the classical treatment-control experiments where only two arms are involved, in this case the treatment has many versions, due to the fact that different physicians are involved or different types/dosages of drugs are administered, etc. As an example, \cite{bauer2008evaluating} studied an effectiveness trial of the Reconnecting Youth preventive intervention program, in which high-risk participants in the intervention arm received the Reconnecting Youth treatment administered in groups, whereas high-risk participants assigned to the control arm were left ungrouped. Such experiments are called ``partially nested experiments'' because the treatment allocations are nested in small groups while the control arm is not. The effect of interest is 
\begin{align}\label{eqn:provider-effect}
    \gamma = \sum_{q=1}^{Q-1} f_q\overline{Y}(q) - \overline{Y}(Q),
\end{align}
where $f_q$'s are pre-specified coefficients, for example, $f_q = (Q-1)^{-1}$ for $1 \le q\le Q-1$.  The coefficient matrix in this example is a contrast vector:
$F = (f_1,\dots,f_{Q-1}, -1)^\top$  with $\sum_{q=1}^{Q-1}f_q = 1$.
\end{example}


\subsection{Our contributions}\label{sec:contribution}

Section \ref{sec::open-questions} has reviewed various designs and the associated open problems. In this paper, we will give a unified study of all the designs in Table \ref{tab:many-settings}. We further the literature in the following ways. 

First, we formulate the inference problem under the randomization model in terms of linear permutation statistics of the form $  \Gamma = \left(\Gamma_1,\ldots, \Gamma_H\right)^\top$ with  
$$
\Gamma_h = \sum_{i=1}^N M_h(i,\pi(i)), \quad h=1,\ldots,H,
$$ 
where $M_1,\ldots,M_H$ are $H$ deterministic $N \times N$ matrices and $\pi : [N] \rightarrow [N]$ is a random permutation on the set of integers $[N] = \{ 1,\ldots, N \}$. This formulation is intuitive because the treatment assignment in Definition \ref{def:cr} follows from a random permutation of the treatment levels. In our analysis, different estimators correspond to different specifications of the matrices $M_1,\ldots,M_H$, which depend on the potential outcomes.
This formulation allows us to build upon the existing results in probability theory \citep{Bolthausen1984AnEO, chatterjee2008multivariate} to derive BEBs on the point estimator of the causal effect. In particular, our analysis emphasizes the dependence on the number of treatment levels and the dimension of the causal effects of interest. Importantly, we derive BEBs that can deal with general designs with varying group sizes.

Second, we establish a novel BEB on quadratic forms of the linear estimator under the randomization model. Importantly, this BEB allows the number of treatment levels  to diverge, the sample sizes across treatment levels to vary, and the dimension of the causal effects of interest to diverge. It serves as the basis for the $\chi^2$ approximation for large-sample Wald-type inference. Moreover, in Appendix \ref{sec:general-BE}, we present general BEBs on multivariate permutation statistics that serve as the basis of these results, along with a thorough discussion on their connection to the existing results in the literature \citep[e.g.,][]{hoeffding1951combinatorial, Bolthausen1984AnEO, chatterjee2008multivariate, fang2015rates, wang2022rerandomization}, which is of independent interest. 

Third, we propose variance estimators for unreplicated designs and mixture designs that allow for the group size to be one in many treatment levels. To the best of our knowledge, the variance estimators are new in the literature of design-based causal inference, although they share some features with those in finely stratified survey sampling \citep[e.g.,][]{cochran1977sampling, wolter2007introduction, breidt2016nonparametric} and experiments \citep[e.g.,][]{abadie2008estimation, fogarty2018mitigating}. However, the theoretical analysis of the new variance estimators is much more challenging because of the dependence of the treatment indicators under the randomization model. We also study their probability limits and establish the theory for large-sample Wald-type inference.

Fourth, in the process of achieving the above three sets of results, we established some immediate theoretical results that are potentially useful for other problems. For instance, we prove a novel BEB for linear permutation statistics over convex sets, building upon a recent result based on Stein's method \citep{fang2015rates}.  We also obtain fine results on the sample moments under the randomization model. Due to the space limit, we relegate them to Appendices \ref{sec:general-BE} and \ref{sec:additional} in the supplementary material.

\subsection{Notation}
We use $C$ to denote generic constants that may vary. Let $\Phi(t)$ denote the cumulative distribution function of a standard Normal distribution. For two sequences of numbers, $a_N$ and $b_N$, let $a_N = O(b_N)$ denote $ a_N \le Cb_N$ for some positive constant $C>0$, and let $a_N = o(b_N) $ denote $a_N/b_N \to 0 $ as $N\to\infty$.  
Let $0_N$ and $ 1_N$ denote, respectively, vectors of all zeros and ones in $\bbR^N$. For two random variables $X$ and $X'$, we use $X\lesssim X'$ or $X'\gtrsim X$ to represent that $X'$  stochastically dominates $X$, i.e., $\bbP\{X' \le t\} \le \bbP\{X \le t\}$ for all $t\in\bbR$. For any covariance matrix $V$, let $V^\star$ denote the corresponding correlation matrix.

Consider a matrix $M=(M(h,l))\in\bbR^{H\times H}$. Let $M(\cdot,l)\in \bbR^{H\times 1}$ and $M(h, \cdot)\in\bbR^{1\times H}$ denote its $l$-th column and $h$-th row, respectively.
Let $\varrho_k(M)$ denote its $k$-th largest singular value. Specially, let $\varrho_{\max}(M)$ and $\varrho_{\min}(M)$ denote the largest and smallest singular values, respectively. Define its condition number as the ratio of its largest and smallest singular values: $
\kappa(M) = {\varrho_{\max}(M)}/{\varrho_{\min}(M)}
$.   Let $\|M\|_\textsc{f} = (\sum_{h=1}^H\sum_{l=1}^H M(h,l)^2)^{1/2}$, $\|M\|_{\operatorname{op}} = \{\varrho_{\max}(M^\top M)\}^{1/2}$, $\|M\|_{p,r}= (\sum_{l=1}^H \|M(\cdot,l)\|_p^r)^{1/r} = \{\sum_{l=1}^H(\sum_{h=1}^H   |M(h,l)|^p)^{r/p}\}^{1/r}$ ($1\le p < \infty$ and $1\le r < \infty$), $\|M\|_\infty = \max_{h,l\in [H]} |M(h,l)|$ be, respectively, the Frobenius norm, the operator norm, the $L_{p,r}$ norm and the vectorized $\ell_\infty$ norm. 

Design-based results rely on conditions on 
$$
M_N(q) = \max_{i\in[N]}|Y_i(q)-\overline{Y}(q)|, \quad (q=1,\ldots, Q)
$$
which is the maximum absolute deviation from the mean for potential outcome $Y_i(q)$'s. \citet{hajek1960limiting} used it in proving the CLT for simple random sampling, and \citet{li2017general} used it in proving CLTs for design-based causal inference. It will also appear frequently in our presentation below.

\section{BEBs for the moment estimator under completely randomized experiments}\label{sec:PCLT-projection}
This section presents the BEBs for the moment estimator $\hgamma$ in \eqref{eqn:estimates} under completely randomized experiments. 
Section \ref{sec::BEBs-linear-estimators} presents general BEBs for linear projections of $\hgamma$. Section \ref{sec::BEB-explain} further provides more discussion to facilitate the understanding of the established BEBs. Sections \ref{sec::BEB-nearly-uniform-design} and \ref{sec::BEB-non-uniform-design} then apply them to derive useful BEBs for nearly uniform designs and general designs, respectively.

\subsection{BEBs on the moment estimator}\label{sec::BEBs-linear-estimators}

To simplify the presentation, standardize $ \hgamma = F^\top \hY$:
\begin{align}\label{eqn:tilde-gamma}
    \tilde{\gamma} = V_\hgamma^{-1/2}(\hgamma - \gamma) \quad \text{ with } \quad \E{\tilde{\gamma}} = 0 \text{ and } \COV{\tilde{\gamma}} = I_H.
\end{align}
The standardization \eqref{eqn:tilde-gamma} assumes that the covariance matrix $V_\hgamma$ is not singular. We assume it for convenience and it holds for most interesting cases. Nevertheless, there are special cases under which $V_\hgamma$ becomes degenerate. One possible reason is that the coefficient matrix $F$ has linearly dependent columns. For such $F$, we can find a subset of contrast vectors to establish the BEB, and the rest is just a linear combination. Another reason for a singular $V_\hgamma$ is that the covariance structure of the potential outcomes might be degenerate and the contrast matrix $F$ aligns with the null space of $V_\hY$, which are usually extreme cases that are less of interest. For estimating average treatment effect in the two-arm randomized experiments, a degenerating estimator corresponds to the setting where the potential outcomes are perfectly negatively correlated (Example \ref{exp:two-arm}), which is a rare setup in reality. 

Our key results are BEBs on linear projections of $\tilde{\gamma}$. Theorem \ref{thm:be-proj-standard} below gives a general BEB for $\tilde{\gamma}$. 

\begin{theorem}[BEBs for linear projections of $\tilde{\gamma}$]\label{thm:be-proj-standard}
Assume complete randomization.

(i) There exists a universal constant $C>0$, such that for any $b\in\bbR^H$ 
with $ \|b\|_2 = 1$, we have
\begin{align*}
\sup_{t\in\bbR}\lt|\bbP\{b^\top \tilde{\gamma} \le t\} - \Phi(t)\rt| \le C\lt\|b^\top V_\hgamma^{-1/2} F^\top\rt\|_\infty \cdot 
\max_{q\in [Q]} N_q^{-1} M_N(q) . 
\end{align*}

(ii) Further assume that there exists $\sigma_F \ge 1$ such that the covariance matrix $V_\hgamma$ satisfies
\begin{align}\label{eqn:well-conditioned}
     V_\hgamma = F^\top V_\hY F \succeq  \sigma^{-2}_F F^\top  \diag{N_q^{-1}S(q,q)} F . 
\end{align}
Then there exists a universal constant $C>0$, such that 
\begin{align}\label{eqn:uniform-be}
    \sup_{b\in\bbR^H,\|b\|_2 =1}\sup_{t\in\bbR}\lt|\bbP\{b^\top \tilde{\gamma}  \le t\} - \Phi(t)\rt| 
    \le C \max_{i\in[N],q\in[Q]} \min\lt\{\mathrm{I}(i,q), \mathrm{II}(i,q)\rt\} 
\end{align}
where 
\begin{align}\label{eqn::terms1and2-be-proj-standard}
    \mathrm{I}(i,q) = {\sigma_F} \lt|\frac{ Y_i(q)-\overline{Y}(q)}{\sqrt{N_q S(q,q)}}\rt|,\quad \mathrm{II}(i,q) = {\sigma_F} \frac{  \|F(q, \cdot)\|_2 \cdot N_q^{-1}|Y_i(q)-\overline{Y}(q)|}{\sqrt{\varrho_{\min}(F^\top  \diag{N_q^{-1}S(q,q)} F)}}.
\end{align}
\end{theorem}

We relegate the proof of Theorem \ref{thm:be-proj-standard} in Appendix \ref{sec:pf-be-proj-standard} of the Supplementary Material. To prove Theorem \ref{thm:be-proj-standard}(i), we formulate $\tilde{\gamma}$ as a multivariate linear permutation statistic and apply an existing BEB by \citet{Bolthausen1984AnEO} to obtain the BEB for $\tilde{\gamma}$. To prove Theorem \ref{thm:be-proj-standard}(ii), we need to further derive upper bounds in terms of I and II in \eqref{eqn::terms1and2-be-proj-standard} from two different perspectives, which is non-trivial to the best of our knowledge. More concretely, Term I is motivated by a careful analysis of each parts of the BEB bound in Theorem \ref{thm:be-proj-standard}(i) for the large arms, which summarizes an overall $N_q^{-1/2}$ rate, while the rest pieces mainly depend on the size of $\sigma_F$ and the scale of the standardized potential outcomes. Term II is motivated by the case where there are many small arms with small $N_q$'s but a large $Q$. In this case, the inverse square root variance matrix $V_\hgamma^{-1/2}$ is small due to the existence of many such small arms, and the rest pieces depend on the scale of the design matrix and potential outcomes. Theorem \ref{thm:be-proj-standard}(ii) is the key result that is applicable in a wide range of designs.


\subsection{Understanding Theorem \ref{thm:be-proj-standard}}\label{sec::BEB-explain}

In this subsection we discuss Theorem \ref{thm:be-proj-standard} from several aspects. 

First, 
we emphasize the applicability of Theorem \ref{thm:be-proj-standard} in a wide range of settings. The upper bound in Theorem \ref{thm:be-proj-standard}(i) depends on the choice of $b$, whereas the upper bound in Theorem \ref{thm:be-proj-standard}(ii) is uniform over all $b$.  Theorem \ref{thm:be-proj-standard}(i) is useful when either the infinity norm of the scaled inverse variance $\|b^\top V_\hgamma^{-1/2} F^\top\|_\infty$ or the maximum term $\max_{q\in [Q]} N_q^{-1} M_N(q)$ is small, even though the other part does not shrink to zero. 
Moreover, Theorem \ref{thm:be-proj-standard}(ii) covers a wide range of design regimes. Technically, the upper bound in Theorem \ref{thm:be-proj-standard}(ii) depends on the minimum value of two terms. It is convenient to apply these two terms to different treatment arms based on the structure of the design. We elaborate this idea by revisiting \ref{regime:R1} to \ref{regime:R5}.
\begin{itemize}
    \item For \ref{regime:R1} and \ref{regime:R2}, because the $N_q$'s are large, we can use term I in \eqref{eqn::terms1and2-be-proj-standard} and obtain a sufficient condition for a vanishing upper bound.

    \item For \ref{regime:R3} and \ref{regime:R4}, the $N_q$'s are bounded and term I in \eqref{eqn::terms1and2-be-proj-standard} has constant order. However, term II in \eqref{eqn::terms1and2-be-proj-standard} is small under mild conditions on $F$. For instance, in the factorial design in Example \ref{eg::factorial-design}, the following algebraic facts hold:
\begin{equation}
\label{eq::contrast-matrix}
\|F\|_{\infty}  = Q^{-1},\quad \|  F(q, \cdot) \|_2 = Q^{-1} \sqrt{H}, \quad \varrho_{\min}(F^\top  F)  = Q^{-1}.
\end{equation}
Combining \eqref{eqn:well-conditioned} and  \eqref{eq::contrast-matrix}, we have
\begin{align}
    \varrho_{\min}(F^\top V_\hY F) &\ge \sigma_F^{-2} \varrho_{\min}(F^\top \diag{N_q^{-1}S(q,q)} F) \notag\\
    &\ge \sigma_F^{-2}\min_{q\in[Q]} \{N_q^{-1}S(q,q)\}\cdot \varrho_{\min}(F^\top  F) \notag\\
    &= Q^{-1} \sigma_F^{-2} \min_{q\in[Q]} \{N_q^{-1}S(q,q)\}.\label{eqn:factorial-F}
\end{align}
Assume $\max_{q\in[Q]}M_N(q)^2/\min_{q\in[Q]}S(q,q)$ is of constant order. Because the $N_q$'s are bounded, term II in \eqref{eqn::terms1and2-be-proj-standard} has order $O(\sqrt{H/Q})$, which is small if $H/ Q \rightarrow 0$. 

\item For \ref{regime:R5}, we can partition the treatment arms based on the sizes of $N_q$'s to achieve a trade-off between terms I and II in \eqref{eqn::terms1and2-be-proj-standard}.  In particular, for general designs in Definition \ref{def:non-uniform-design}, a natural partition is $ [Q] = \cQ_{\textsc{l}} \cup \cQ_{\textsc{S}}$. On the one hand, the arms in $\cQ_{\textsc{l}}$ contain many units, so term I in \eqref{eqn::terms1and2-be-proj-standard} vanishes asymptotically. On the other hand, $\cQ_{\textsc{s}}$ contains many arms which makes term II in \eqref{eqn::terms1and2-be-proj-standard} small under mild conditions on $F$.
\end{itemize}

We will provide rigorous results in the next two sections by applying Theorem \ref{thm:be-proj-standard} to obtain useful BEBs for different designs.


Second, we highlight some important theoretical insights of Theorem \ref{thm:be-proj-standard}(ii) regarding a trade-off between the non-uniformity of the design and the estimand regularity. In order to translate Theorem \ref{thm:be-proj-standard} into useful asymptotic results, we need additional regularity conditions that reflect and respect the trade-off between the key quantities ($N_q$'s, $Q$ and $F$) so that term I or term II vanish asymptotically. For those arms with large $N_q$'s, the scale of $F(q,\cdot)$ and $Q$ does not impact the estimation to a great amount, because term I in \eqref{eqn:uniform-be} suggests a vanishing upper bound regardless the choice of $Q$ and $F$. In other words, we can handle more general coefficients when more replications are available. For those arms with smaller $N_q$'s, we have to resort to term II of \eqref{eqn:uniform-be}, which requires the ratio $\|F(q,\cdot)\|_2/(\varrho_{\min}\{F^\top V_\hY F\})^{1/2} $ to vanish. This implies that most of the rows of the coefficient matrix (i.e., $F(q,\cdot)$'s) corresponding to the small arms should be nonzero and close in scale. We will formalize these intuitions in Sections \ref{sec::BEB-nearly-uniform-design} and \ref{sec::BEB-non-uniform-design}. 

Third, we add some discussion of the addtional condition \eqref{eqn:well-conditioned} that appeared in Theorem \ref{thm:be-proj-standard}(ii). Condition \eqref{eqn:well-conditioned} involves the potential outcomes and the linear coefficient jointly. In the specific designs, we can impose conditions separately on the potential outcomes and the linear contrast while keeping all our theoretical results valid. The role of Condition \eqref{eqn:well-conditioned} is to help with the derivation of the upper bound in Theorem \ref{thm:be-proj-standard}(ii). Condition \eqref{eqn:well-conditioned} is useful for both terms I and II in \eqref{eqn:uniform-be}. For term I, the parameter $\sigma_F^2$ directly appears in the expression of $\mathrm{I}(i,q)$ in \eqref{eqn::terms1and2-be-proj-standard}. For term II, it is used to further bound the denominator of $\mathrm{II}(i,q)$ in the corollaries in Section \ref{sec::BEB-nearly-uniform-design}.  
Condition \eqref{eqn:well-conditioned} requires the covariance matrix $V_\hgamma$ to be ``well-conditioned'' in the sense that the positive definite part of $V_\hgamma$ plays the dominant role. Recall that 
the covariance matrix formula of $V_\hgamma$  has two parts: a positive definite part, $ F^\top\text{Diag}\{N_q^{-1}S(q,q)\}_{q\in[Q]}F $, and a negative definite part, $ - N^{-1}F^\top S F $.  
Condition \eqref{eqn:well-conditioned}, coupled with the covariance formula in \eqref{eqn:mean-var-Gamma}, implies that 
\begin{align*}
    \sigma_F^{-2} F^\top \diag{N_q^{-1}S(q,q)}_{q\in[Q]} F \preceq V_\hgamma \preceq F^\top \diag{N_q^{-1}S(q,q)}_{q\in[Q]} F,
\end{align*}
i.e., the covariance matrix $V_\hgamma$ is upper and lower bounded by $F^\top \text{Diag}\{N_q^{-1}S(q,q)\}_{q\in[Q]} F$, up to constants. Condition \eqref{eqn:well-conditioned} is a regularity condition that rules out those counterexamples, which involve extreme choices of $F$ and $S$ and lead to an ill-conditioned covariance structure. See the counterexample given in Example \ref{exp:bad-unreplicated-exp} below. 
In general, violation of \eqref{eqn:well-conditioned} occurs when the potential outcomes are highly correlated and lead to an ill-conditioned covariance $S$, which we hope to rule out.  
Condition \eqref{eqn:well-conditioned} holds in a wide range of practically interesting settings.
Below we give some canonical examples, as well as two general sufficient conditions for \eqref{eqn:well-conditioned}. 
\begin{example}[Two-arm randomized experiments]\label{exp:two-arm}
    In   treatment-control experiments, we are interested in estimating the average treatment effect $\gamma = \overline{Y}(1) - \overline{Y}(0)$, with contrast vector $ F = (1,-1)^\top $. The difference-in-means  estimator is $\hgamma = \hY_1 - \hY_0$. We can compute
    \begin{align*}
        V_\hgamma 
        &= \frac{p_0}{N_1}S(1,1) + \frac{p_1}{N_0} S(0,0) + \frac{2}{N}S(1,0),
    \end{align*}
    where $p_1 = N_1/N$ and $p_0 = N_0/N$. We can verify that \eqref{eqn:well-conditioned} is equivalent to ruling out the scenario where the potential outcomes are perfectly negatively correlated (i.e., there exists a constant $c>0$ such that $Y_i(0) = -cY_i(1)$ for all $i\in[N]$). Similar conditions also appear in existing literature, for example, Assumption 3 in \citet{lei2021regression}. See Section \ref{sec:pf-two-arm} for detailed justification. 
\end{example} 

\begin{example}[A counterexample in unreplicated $2^K$ factorial designs]\label{exp:bad-unreplicated-exp}
    We give a counterexample in unreplicated $2^K$ factorial designs, which have $N_q = 1$ for all $q\in[Q]$. Consider the one-dimensional contrast:
    \begin{align}
        F = Q^{-1}(1_{2^{K-1}}^\top, -1_{2^{K-1}}^\top)^\top.
    \end{align}
    Let $S$ be the following positive semidefinite matrix:
    \begin{align}
        S = 
        \begin{pmatrix}
            1_{2^{K-1} \times 2^{K-1}} & -1_{2^{K-1} \times 2^{K-1}} \\
            -1_{2^{K-1} \times 2^{K-1}} & 1_{2^{K-1} \times 2^{K-1}}
        \end{pmatrix}.
    \end{align}
    Then we can verify that $V_\hgamma = 0$ but $F^\top \diag{S(q,q)}_{q\in[Q]} F = Q^{-1} > 0$ is positive.  This is a degenerate case with no uncertainty in $\hgamma$.
\end{example}





We conclude this section with Lemma \ref{lem:suff-conds} below, which gives two sufficient conditions for \eqref{eqn:well-conditioned} to aid the understanding. 
\begin{lemma}[Sufficient conditions for \eqref{eqn:well-conditioned}]\label{lem:suff-conds} 
(i) {\em Additive effects}. Condition \eqref{eqn:well-conditioned} holds with $\sigma_F = 1$ if the individual causal effects are constant, that is, $  F^\top (Y_i(q)  - \overline{Y}(q))_{q=1}^Q = 0$ for all $i\in[N]$. 
(ii) {\em Well-conditioned correlation.} Condition \eqref{eqn:well-conditioned} holds with $\sigma_F = c\sigma$ if $\max_{q\in[Q]}N_q \le (1-c)N$ for some $0<c<1$ and the condition number of the correlation matrix corresponding to $V_\hY$ is upper bounded by $\sigma^2$. 
\end{lemma}




The sufficient conditions in Lemma \ref{lem:suff-conds} are somewhat standard in the literature, especially under \ref{regime:R1}. \citet[][Corollary 2]{li2017general} gives a CLT under the assumption of constant  individual causal effects, which is a special case of Lemma \ref{lem:suff-conds}(i). 
Lemma \ref{lem:suff-conds}(i) ensures that under the sharp null hypothesis with $Y_i(q) = Y_i$ for $q\in[Q]$, Condition \eqref{eqn:well-conditioned} holds for all of \ref{regime:R1}--\ref{regime:R5}. 
Lemma \ref{lem:suff-conds}(ii) also generalizes the classical results under \ref{regime:R1}. \citet[][Theorem 5]{li2017general} proves a CLT under the assumption that $S$ has a finite limiting value. When the limit is positive definite, $V_\hY$ also converges to a positive definite matrix, which becomes a special case of  Lemma \ref{lem:suff-conds}(ii). In general, Lemma \ref{lem:suff-conds}(ii) can cover many other interesting scenarios. For example, with Lemma \ref{lem:suff-conds}(ii) we can verify that if the potential outcomes from different treatment arms are uncorrelated with $S(q,q') = 0$ for all $q\neq q'$ and the sample sizes from different arms satisfy $\max_{q\in[Q]}N_q \le (1-c)N$ for some $0<c<1$, then Condition \eqref{eqn:well-conditioned} holds with $\sigma_F^{-2} = c$. 
The general forms of conditions in \eqref{eqn:well-conditioned} and  Lemma \ref{lem:suff-conds} are useful for all of \ref{regime:R1}--\ref{regime:R5}.

\subsection{A BEB with a proper coefficient matrix in nearly uniform designs}\label{sec::BEB-nearly-uniform-design}

In \ref{regime:R1} with a fixed $Q$ and large $N_q$'s, it is intuitive to have CLTs for linear transformations of $\hY$ because $\hY$ itself has a CLT. In other regimes, for instance, \ref{regime:R4}, CLTs for linear transformations of $\hY$ are less intuitive. Consider a diverging $Q$ and bounded $N_q$'s. If   $F = (1,-1,\dots,0)^\top \in \bbR^Q$, 
then the CLT for $F^\top \hY = \hY_1 - \hY_2$ does not hold due to the bounded sample size in treatment arms 1 and 2. As another toy example, if 
\begin{align}
    F = 
    \begin{pmatrix}
        1 & -1 & 0 & \cdots & 0\\
        1 & -1 & 0 & \cdots & 0
    \end{pmatrix}^\top \in\bbR^{Q\times 2}, 
\end{align}
then $F^\top\hY$ has degenerate covariance structure and Theorem \ref{thm:be-proj-standard} cannot be directly applied. Therefore, CLTs should be established for proper coefficient matrices. Corollary \ref{cor:uniform-design-be} below gives a BEB for proper coefficient matrices. We first introduce Condition \ref{condition::proper} below on the coefficient matrix $F$. 



%
\begin{condition}[Proper coefficient matrix in nearly uniform designs]
\label{condition::proper}
The coefficient matrix $F$ satisfies $ \|F\|_\infty \le cQ^{-1}$ and $\varrho_{\min}\{F^\top F\} \ge c'Q^{-1}$ for some constants $c,c'>0$. 
\end{condition} 
 
Condition \ref{condition::proper} depends on the scale of $F$, although the BEB should not depend on the scale of $F$ due to the standardization of $\hgamma$. We present the above form of Condition \ref{condition::proper} to facilitate the discussion of the factorial design in Example \ref{eg::factorial-design}, in which the scale of $F$ is motivated by scientific questions of interest.
When $Q$ is fixed, Condition \ref{condition::proper} holds if $F$ has full column rank. So in \ref{regime:R1}, Condition \ref{condition::proper} does not impose any additional assumptions beyond the standard ones. When $Q$ diverges, Condition \ref{condition::proper}  rules out sparse $F$ that only results in a linear combination of $\hY$ over a small number of treatment arms. Also, the minimum eigenvalue condition in Condition \ref{condition::proper} ensures the non-degenerate covariance structure of the estimator $\hgamma$.  
 
 We then give Corollary \ref{cor:uniform-design-be} below. 

\begin{corollary}[BEB for nearly uniform designs]\label{cor:uniform-design-be}
Assume complete randomization that satisfies Definition \ref{def:uniform-design} and Condition \ref{condition::proper}. Also assume \eqref{eqn:well-conditioned}. 
There exists a universal constant $C>0$, such that 
\begin{align}\label{eqn:uniform-design-be}
 \sup_{b\in\bbR^H,\|b\|_2 =1}\sup_{t\in\bbR}\lt|\bbP\{b^\top \tilde{\gamma}  \le t\} - \Phi(t)\rt|  
 \le  
 C\sigma_F    \frac{  \max_{ q\in[Q]} M_N(q)  }{  \{  \min_{q\in[Q]} S(q,q) \}^{1/2} }  
\sqrt{ \frac{H}{N} } 
 .
\end{align}
\end{corollary}

We relegate the proof of Corollary \ref{cor:uniform-design-be} to Section \ref{sec:uniform-design-be} in the Supplementary Material.  Technically, we derive the upper bound in Corollary \ref{cor:uniform-design-be} based on the upper bound from term II in \eqref{eqn::terms1and2-be-proj-standard} in Theorem \ref{thm:be-proj-standard}(ii). We first upper bound the numerator of term II using the condition that $\|F\|_\infty$ is upper bounded by $cQ^{-1}$. We then lower bound the denominator using \eqref{eqn:well-conditioned} and the fact that $\varrho_{\min}\{F^\top F\} \ge c' Q^{-1}$ by Condition \ref{condition::proper}.

We make several further comments on Corollary \ref{cor:uniform-design-be}. 
First, Theorem \ref{thm:be-proj-standard}(ii) is uniform over $b$. Therefore, the upper bound in Corollary \ref{cor:uniform-design-be} preserves the uniformity and does not depend on $b$. 
Second, the upper bound in \eqref{eqn:uniform-design-be} reveals the interplay of several quantities: the number of parameters $H$, the number of units $N$, the scale of the potential outcomes $M_N(q)$, the minimum second moments $\min_{q\in[Q]} S(q,q)$  as well as the structure of $F$. 
Third, the upper bound in \eqref{eqn:uniform-design-be} decreases at the rate of $(H / N)^{1/2}$. Under regimes {\ref{regime:R1}--\ref{regime:R4}}, $QN_0$ and $N$ have the same order as $N\rightarrow \infty$.  
To ensure convergence in distribution in \eqref{eqn:uniform-design-be}, we only require $H$ to be small compared with $N$, or, equivalently, $H$ to be small compared with $QN_0$.  Importantly, there is no further restriction on $Q$ or $N_0$, as long as  $H / (QN_0)$ converges to $0$.  Therefore, Corollary \ref{cor:uniform-design-be} is applicable for regimes {\ref{regime:R1}--\ref{regime:R4}}. 
Fourth, the denominator of \eqref{eqn:uniform-design-be} depends on $\min_{q\in[Q]} S(q,q)$, which is useful when the variances of the potential outcomes are lower bounded.  For ease of presentation, we do not discuss more complicated cases such as some $S(q,q)$'s are small. We can slightly modify the proof of Corollary \ref{cor:uniform-design-be} to cover scenarios where some $S(q,q)$'s are close or equal to zero. See Remark \ref{rmk:relax-minimal-S} in Section \ref{sec:uniform-design-be} of the Supplementary Material for more detailed discussion. 


Example \ref{eg::factorial-proper-F} below gives a more detailed discussion of Condition \ref{condition::proper} in the nearly uniform factorial design. 

\begin{example}[Nearly uniform factorial design]\label{eg::factorial-proper-F}
Recall Example \ref{eg::factorial-design} and assume it satisfies Definition \ref{def:uniform-design}. Let $F\in\bbR^{Q\times H}$ with $H = K+K(K-1)/2={K(K+1)}/{2}$ be the coefficient matrix for all main effects and two-way interactions. 
Assume \eqref{eqn:well-conditioned} and recall \eqref{eq::contrast-matrix}.  Corollary \ref{cor:uniform-design-be} implies 
\begin{align} \label{eqn:factorial-BE}
 \sup_{b\in\bbR^H,\|b\|_2 =1}\sup_{t\in\bbR}\lt|\bbP\{b^\top \tilde{\gamma}  \le t\} - \Phi(t)\rt|  
 \le  
 C\sigma_F     \frac{ \max_{q\in[Q]} M_N(q) }{ \{  \min_{q\in[Q]} S(q,q) \}^{1/2} }  \sqrt{\frac{K^2}{N}}   .
\end{align}
From \eqref{eqn:factorial-BE}, we can obtain a sufficient condition for the upper bound to converge to 0, which implies a CLT of $\tilde{\gamma} $. 
\end{example}

\subsection{A BEB for general designs}\label{sec::BEB-non-uniform-design}

Now consider general designs in Definition \ref{def:non-uniform-design}. Analogous to our discussion in nearly uniform designs, we impose a condition on the coefficient matrix. Partition the coefficient matrix $F$ into $    F_{\textsc{s}}$ and $    F_{\textsc{l}}$, and further partition $   F_{\textsc{s}} $ into  $    F_{\textsc{u}}$ and $    F_{\textsc{r}}$. So we have
\begin{align}\label{eqn:F-partition}
    F = 
    \begin{pmatrix}
    F_{\textsc{s}}\\
    F_{\textsc{l}}
    \end{pmatrix}
    \quad \text{ where }
      F_{\textsc{s}} = 
    \begin{pmatrix}
    F_{\textsc{u}}\\
    F_{\textsc{r}}
    \end{pmatrix}.
\end{align}
Here $F_{\textsc{s}}, F_{\textsc{l}}, F_{\textsc{u}}, F_{\textsc{r}}$ are submatrices of $F$ corresponding to the columns indexed by treatment arms in $\cQ_\textsc{s}, \cQ_{\textsc{l}}, \cQ_{\textsc{u}}, \cQ_{\textsc{r}}$, respectively.

\begin{condition}[Proper coefficient matrices in general designs]
\label{condition::proper-non-uniform}
The submatrix $F_\textsc{s}$ of the coefficient matrix $F$ satisfies $ \|F_\textsc{s}\|_\infty \le c|\cQ_\textsc{s}|^{-1}$ and $\varrho_{\min}\{F_\textsc{s}^\top F_\textsc{s}\} \ge c'|\cQ_\textsc{s}|^{-1}$ for some constants $c,c'>0$. 
\end{condition}

Condition \ref{condition::proper-non-uniform} is similar to Condition \ref{condition::proper}. However, Condition \ref{condition::proper-non-uniform} imposes restrictions on the submatrix $F_\textsc{s}$, whereas  Condition \ref{condition::proper} imposes restrictions on the whole matrix $F$. Importantly, Condition \ref{condition::proper-non-uniform} does not impose the restrictions on $F_\textsc{l}$, which corresponds to the treatment arms with enough replications.

We can apply Theorem \ref{thm:be-proj-standard} to establish the following BEB for general designs:

\begin{corollary}[BEB for general designs]\label{cor:non-uniform-design-be}
Assume complete randomization  that satisfies Definition \ref{def:non-uniform-design} and  Condition \ref{condition::proper-non-uniform}. Also assume \eqref{eqn:well-conditioned}. 
There exists a universal constant $C>0$, such that 
\begin{align}\label{eqn:non-uniform-design-be}
 &\sup_{b\in\bbR^H,\|b\|_2 =1}\sup_{t\in\bbR}\lt|\bbP\{b^\top \tilde{\gamma}  \le t\} - \Phi(t)\rt|  \\
 \le&  C \sigma_F  \max\lt\{ \max_{q\in\cQ_{\textsc{l}}}\frac{  M_N(q)}{\sqrt{N_q S(q,q)}}, \frac{   \max_{q\in \cQ_{\textsc{s}}} M_N(q)}{   \{   \min_{q\in \cQ_{\textsc{s}}} S(q,q) \}^{1/2}  }\cdot \sqrt{\frac{H}{N_{\textsc{s}}}}\rt\}. \notag
\end{align}
\end{corollary}

We relegate the proof of  Corollary \ref{cor:non-uniform-design-be} to Section \ref{sec:pf-non-uniform-design-be} of the Supplementary Material. To prove Corollary \ref{cor:non-uniform-design-be}, we apply Theorem \ref{thm:be-proj-standard} in several key steps. We first partition $\cQ$ into $\cQ_{\textsc{s}} \cup \cQ_{\textsc{l}}$  based on the size of the treatment arms. With the key bound in \eqref{eqn:uniform-be}, we  then apply term I to $\cQ_{\textsc{l}}$ and term II to $ \cQ_{\textsc{s}} $ with some further simplifications of the denominator of term II.

The upper bound  \eqref{eqn:non-uniform-design-be} is uniform over all $b$. It depends on the sizes of the treatment arms in a subtle way.  On the one hand, for $q\in\cQ_{\textsc{l}}$, the $N_q$'s are large, so the first part of \eqref{eqn:non-uniform-design-be} converges to zero if the following ``local'' condition holds for all $q\in\cQ_{\textsc{l}}$:
$$
\frac{  M_N(q)^2  }{ S(q,q) } = o(N_q)  .
$$
On the other hand, for $q\in\cQ_{\textsc{s}}$, the $N_q$'s are small, but the second part of \eqref{eqn:non-uniform-design-be} still converges to zero if the following ``global'' condition holds:
$$
\frac{   \max_{q\in \cQ_{\textsc{s}}} M_N(q)^2 }{     \min_{q\in \cQ_{\textsc{s}}} S(q,q)   } = o\left(  \frac{H}{ N_{\textsc{s}} } \right).
$$

%

We first apply Corollary \ref{cor:non-uniform-design-be} to the general factorial design. 

\begin{example}[An example of general factorial designs]\label{exp:general-factorial-BEB}
Recall Example \ref{eg::factorial-design}.  Assume the baseline arm $q=1$ contains a large number of units possibly due to lower cost while the other arms have $N_q \le \overline{n}$ for some fixed $\overline{n}$. This gives a general design by Definition \ref{def:non-uniform-design} with $\cQ_{\textsc{l}} = \{1\}$ and $\cQ_{\textsc{s}} = \{2, \ldots, Q\}$. 
Let $F\in\bbR^{Q\times H}$ with $H={K(K+1)}/{2}$ be the contrast matrix for all main effects as well as two-way interactions. 
Assume \eqref{eqn:well-conditioned} and recall \eqref{eq::contrast-matrix}. Condition \ref{condition::proper-non-uniform} holds naturally for large $Q$, because we have $\|F_\textsc{s}\|_\infty = Q^{-1} \le |\cQ_\textsc{s}|^{-1}$ in factorial designs and the eigenvalue of $F_\textsc{s}^\top F_\textsc{s}$ can be lower bounded as follows:
\begin{align*}
    \varrho_{\min}\{F_\textsc{s}^\top F_\textsc{s}\} &\ge \varrho_{\min}\{F^\top F\} - \varrho_{\max}\{F_\textsc{l}^\top F_\textsc{l}\} \\
    &= Q^{-1} - Q^{-2} H = Q^{-1} (1 - HQ^{-1}) = O(Q^{-1}) = O(|\cQ_\textsc{s}|^{-1}).
\end{align*}
Applying Corollary \ref{cor:non-uniform-design-be}, we have
\begin{align}\label{eqn:non-uniform-factorial-be}
 &\sup_{b\in\bbR^H,\|b\|_2 =1}\sup_{t\in\bbR}\lt|\bbP\{b^\top \tilde{\gamma}  \le t\} - \Phi(t)\rt|  \\
 \le&  C \sigma_F  \max\lt\{\frac{ M_N(1)}{\sqrt{N_1 S(1,1)}}, 
 \frac{  \max_{ q \geq 2}   M_N(q)}{    \{  \min_{q\geq 2} S(q,q) \}^{1/2} }  \sqrt{  \frac{K^2}{N_\textsc{s}} }  \rt\}. \notag
\end{align}
From \eqref{eqn:non-uniform-factorial-be}, if $K\to\infty, N_1\to\infty$, and
$$
\frac{M_N(1)}{\sqrt{S(1,1)}} = o(N_1^{1/2}) ,\quad 
\frac{  \max_{q\geq 2} M_N(q)}{ \{  \min_{q\geq 2} S(q,q)\}^{1/2} } = o( {  N_\textsc{s}^{1/2} } / K),
$$ 
then the upper bound in \eqref{eqn:non-uniform-factorial-be} vanishes asymptotically.  
\end{example}

We then apply Corollary \ref{cor:non-uniform-design-be} to the partially nested experiment with provider effects (Example \ref{exp:provider}). 
\begin{example}[Revisit Example \ref{exp:provider}]\label{exp:revisit-provider}
Recall Example \ref{exp:provider} with the contrast vector:
\begin{align*}
    F = (\underbrace{f_1,\dots,f_{Q-1}}_{F^\top_\textsc{s}}, \underbrace{-1 \vphantom{f_1,\dots,f_{Q-1}}}_{F^\top_\textsc{l}})^\top.
\end{align*}
Assume that there exists some $c,c'>0$ such that
\begin{gather}\label{eqn:good-cq}
    \|F_\textsc{s}\|_\infty = \max_{q\le Q-1}|f_q| \le c(Q-1)^{-1},\quad \varrho_{\min}\{ F_\textsc{s}^\top F_\textsc{s} \} = \sum_{q=1}^{Q-1} f_q^2 \ge c' (Q-1)^{-1},
\end{gather}
which holds for the special case with $f_q = (Q-1)^{-1}$. For general $f_q$'s, 
Condition \eqref{eqn:good-cq} guarantees Condition \ref{condition::proper-non-uniform}. Intuitively, Condition \eqref{eqn:good-cq} requires a dense number of $c_q$'s to be of the order $O((Q-1)^{-1})$ so that the target parameter \eqref{eqn:provider-effect} is the contrast between the means of a weighted average of the treated arms and the control arm. Meanwhile, Condition \eqref{eqn:well-conditioned} holds under many settings (see for example the sufficient conditions in Lemma \ref{lem:suff-conds}). 
The point estimator $\hgamma = \sum_{q=1}^{Q-1} f_q \hY_q - \hY_Q $ satisfies
\begin{align}\label{eqn:provider-effect-be}
 & \sup_{t\in\bbR}\lt|\Prob{\frac{\hat{\gamma} - \gamma}{\sqrt{\Var{\hgamma}}} \le t} - \Phi(t)\rt|  \notag\\
 \le&  C\sigma_F  \max\lt\{ \frac{M_N(Q)}{\sqrt{N_Q S(Q,Q)}}, ~ \frac{  \max_{ q \le Q-1}   M_N(q)}{    \{  \min_{q\le Q-1} S(q,q) \}^{1/2} }  \sqrt{  \frac{1}{N_\textsc{s}} } \rt\}.
\end{align}
As $N_Q\to\infty$ and $N_\textsc{s}\to\infty$, the upper bound in \eqref{eqn:provider-effect-be} converges to zero if
\begin{align*}
    \frac{M_N(Q)}{\sqrt{S(Q,Q)}} = o(N_Q^{1/2}) ,\quad 
\frac{  \max_{q\le Q-1} M_N(q)}{ \{  \min_{q\le Q-1} S(q,q)\}^{1/2} } = o( {  N_\textsc{s}^{1/2} }).
\end{align*}
\end{example}

\section{Design-based causal inference}\label{sec:inference}

Now we turn to the central task of design-based causal inference under complete randomization. We focus on the large-sample Wald-type inference based on the quadratic form 
$$
\widehat{T} = (\hgamma - \gamma)^\top \hV_{\hgamma}^{-1} (\hgamma - \gamma),
$$
recalling the point estimator  $\hgamma$ and  the variance estimator $\hV_{\hgamma}$ in \eqref{eqn:estimates}. In \ref{regime:R1} with fixed $(Q, H)$ and large $N_q$'s, the standard asymptotic argument suggests that we can use $q_{H, \alpha}$, the upper $\alpha$-quantile of $\chi^2_H$, as the critical value for the quadratic form. For simplicity, we say that the corresponding confidence set is asymptotically valid if  $ \lim_{N\to\infty} \mathbb{P}\{  \widehat{T}  \le q_{H, \alpha} \} \ge 1-\alpha$.

The rigorous theoretical justification for the above Wald-type inference procedure typically follows from two steps:
\begin{enumerate}
\item 
[(Step 1)]\label{step1}
First, analyze the asymptotic distribution of the corresponding quadratic form with the true covariance matrix
\begin{align}\label{eqn:quad-form}
   T = (\hgamma - \gamma)^\top V_{\hgamma}^{-1} (\hgamma - \gamma).
\end{align}

\item
[(Step 2)]\label{step2}
Second, construct a consistent or conservative estimator $\hV_{\hgamma}$ for the true covariance matrix    $V_{\hgamma}$. 
\end{enumerate}

Under regime \ref{regime:R1}, both Steps 1 and 2 have rigorous theoretical justification ensured by \eqref{eq::basic-asymptotic-inference}. Beyond \ref{regime:R1}, it is challenging to derive the asymptotic distribution of the quadratic form in Step 1 especially when $H$ and thus the degrees of freedom of $T$ diverge. To achieve the requirement in Step 1, we use results based on Stein's method to derive BEBs on quadratic forms of linear permutation statistics. To avoid excessive notation, we present the results that are most relevant to our inference problem in the main paper and relegate more general yet more complicated results to Appendices \ref{sec:general-BE} and \ref{sec:additional}. Moreover, the sample variances $\hS(q,q)$'s and thus the variance estimator $\hV_{\hgamma}$ in \eqref{eqn:estimates} are not even well defined when some treatment arms do not have replications of the outcome. Without replications in all arms, we must find an alternative form of  $\hV_{\hgamma}$ to estimate $V_{\hgamma}$. This is a salient problem for \ref{regime:R4} and \ref{regime:R5}. Finally, in all regimes \ref{regime:R1}--\ref{regime:R5}, we need to study the properties of $\hV_{\hgamma}$ to achieve the requirement in Step 2.

Due to the different levels of technical complexities, we divide this section into three subsections.   Section \ref{sec:var-uniform} discusses nearly uniform designs with replications in all arms. Section \ref{sec:var-unreplicate} discusses unreplicated designs. Section \ref{sec::non-uniform-design-inference} discusses the general designs. In every subsection, we first present a BEB on the quadratic form in Step 1, then present the properties of the covariance estimator $\hV_{\hgamma}$, and finally present the formal result to justify the Wald-type inference.

To facilitate the discussion, we introduce the following notation 
\begin{align}\label{eqn:T0}
T_0 = \xi_H^\top \xi_H \quad \text{ where }
\xi_H\sim\cN(0,I_H)
\end{align}
for a $\chi^2_H$ random variable with possibly diverging degrees of freedom. The $T_0$ in \eqref{eqn:T0} has mean $H$ and variance $2H$.  We will show that asymptotically with large $N$, the distribution of $T$ is approximately equal to $T_0$, whereas the distribution of $\hat{T}$ is stochastically dominated by that of $T_0$ due to the conservativeness of the variance estimation. 

 We introduce the following moment condition on the potential outcomes for our theoretical analysis below. 
 
 \begin{condition}[Bounded fourth moment of the potential outcomes]\label{cond:moments}
There exists an absolute constant $\Delta>0$ such that 
$
    \max_{q\in[Q]}  N^{-1} \sum_{i=1}^N \{Y_i(q) - \overline{Y}(q)\}^4 \le \Delta^4.
$
\end{condition}

As a technical comment, we can allow $\Delta$ to grow in theory, but to keep the presentation more elegant, we assume $\Delta$ to be a constant in Condition \ref{cond:moments}. Moreover, it is possible to replace the fourth moment condition by a $\max_{q\in[Q]} N^{-1} \max_{i=1,\dots,N} |Y_i(q) - \overline{Y}(q)|^2 = o(1)$ as \cite{li2017general} when we focus on inference on a fixed set of contrasts (i.e., a fixed $H$). More general results are in the supplementary material (see Sections \ref{sec:pf-uniform-var}, \ref{pf:lem-mean-var-grouping-thm-unrep-var} and \ref{sec:pf-non-uniform-var}).


\subsection{Nearly uniform design with replications in all arms}\label{sec:var-uniform}

In this subsection, we study the Wald-type inference for nearly uniform designs given by Definition \ref{def:uniform-design}.
First, we present a BEB for $T$ in \eqref{eqn:quad-form} in Theorem \ref{thm:quad-be-nearly-uniform} below.

\begin{theorem}[BEB for  the quadratic form $T$ for  nearly uniform designs with replications]\label{thm:quad-be-nearly-uniform}
Assume complete randomization that satisfies Definition \ref{def:uniform-design} and Condition \ref{condition::proper}. Further assume  \eqref{eqn:well-conditioned}. There exists a universal constant $C>0$, such that
\begin{align}\label{eqn:quad-be-nearly-uniform}
    \sup_{t\in\bbR} |\bbP(T\le t)-\bbP(T_0 \le t)| \le \frac{C\max_{q\in[Q]}M_N(q)^3 }{ \{ \min_{q\in[Q]}S(q,q) \}^{3/2}}\cdot \frac{H^{19/4}}{N^{1/2}}.
\end{align}
\end{theorem}

We relegate the proof of Theorem \ref{thm:quad-be-nearly-uniform} to Section \ref{sec:fine-BE} in the Supplementary Material. To prove Theorem \ref{thm:quad-be-nearly-uniform}, we first establish a general BEB over convex sets for multivariate linear permutation statistics based on \cite{fang2015rates}. This involves constructing an ``exchangeable pair'' and carrying out delicate moment calculations under complete randomization for applying \cite{fang2015rates}  based on Stein's method.  Theorem \ref{thm:be-bounded} in Section \ref{sec:fine-BE} presents this general BEB, which is of independent interest beyond our setting. We then apply Theorem \ref{thm:be-bounded} to derive the BEB for the quadratic form $T$ in \eqref{eqn:quad-be-nearly-uniform}.


Theorem \ref{thm:quad-be-nearly-uniform} bounds the difference between the distribution of $T$ and $T_0$ with possibly diverging $H$. Its upper bound is more useful when $H^{19/2} / N \rightarrow 0$, which restricts the number of parameters of interest. The condition $H^{19/2} / N \rightarrow 0$ holds naturally in the factorial design in Example \ref{eg::factorial-design} under regime \ref{regime:R4} if only the main effects and two-way interactions are of interest which gives $H = O(K^2) = O( (\log N)^2 ) $.

Second, we discuss variance estimation. Recall $\hS(q,q)$ and $\hV_\hY$ be defined as in  \eqref{eqn:hS} and \eqref{eqn:hY-hV}. Consider the point estimator $\hgamma$ and covariance estimator $\hV_\hgamma$ in \eqref{eqn:estimates}.
We have  Theorem \ref{thm:uniform-var} below. 

\begin{theorem}[Variance estimation in nearly uniform designs]\label{thm:uniform-var} 
Consider designs that satisfies Definition \ref{def:uniform-design} with $\min_{q\in[Q]} N_q \ge 2$. Assume Condition \ref{cond:moments}. 
\begin{enumerate}[label = (\roman*)]
    \item\label{thm:uniform-var-1} 
$\bbE\{\hV_{\hgamma}\} \succeq  V_\hgamma$.
    
 \item\label{thm:uniform-var-2}
 $ \|\hV_{\hgamma}-\bbE\{\hV_{\hgamma}\}\|^2_{\infty} = O_{\bbP}\lt({\|F\|_\infty^4 Q^4 N^{-3}H^2}\rt).$
 
 \item \label{thm:uniform-var-3}
 $ \|\hV_{\hgamma}-\bbE\{\hV_{\hgamma}\}\|^2_{{\operatorname{op}}} = O_{\bbP}\lt({ \|F\|_\infty^4 Q^4 N^{-3} H^4}\rt).$

%
%
%
\end{enumerate}

\end{theorem}

We relegate the proof to Section \ref{sec:pf-wald-uniform}  in the Supplementary Material. Theorem \ref{thm:uniform-var}\ref{thm:uniform-var-1} reveals that the covariance estimator $\hV_{\hgamma}$ is conservative, which is well-known in design-based causal inference \citep{neyman1923application, imbens15, li2017general}. We prove stronger results than Theorem \ref{thm:uniform-var}\ref{thm:uniform-var-2} and \ref{thm:uniform-var-3} by establishing finite-sample tail bounds on $\hV_{\hgamma}$ based on  Chebyshev's inequality and detailed calculations of the moments under complete randomization. 


 Theorem \ref{thm:uniform-var}(ii) and (iii) are novel results on the stochastic orders of the estimation error of the covariance estimator in $L_{\infty}$ norm and operator norm, respectively. In Example \ref{eg::factorial-design} of the factorial design with $\|F\|_\infty = O(Q^{-1})$, Theorem \ref{thm:uniform-var} simplifies to 
$$
   N\|\hV_{\hgamma}-\bbE\{\hV_{\hgamma}\}\|_{\infty} =O_{\bbP}\lt(  H/N^{1/2}  \rt),\quad
    N\|\hV_{\hgamma}-\bbE\{\hV_{\hgamma}\}\|_{\operatorname{op}}  = O_{\bbP}\lt( H^2/N^{1/2}   \rt).
$$
The estimation error shrinks to zero if only the main effects and two-way interactions are of interest. 
The results in Theorem \ref{thm:uniform-var} suffice for inference, and we relegate the finer probability tail bound for  $\hV_{\hgamma}$ to the supplementary material.

Third, we present formal results on inference. To simplify the presentation, we impose Condition \ref{cond:easy-spec} below. 

%

%

\begin{condition}\label{cond:easy-spec}
(i) There exists a universal constant $\nu>0$ that does not depend on $N$ and $Q$ such that  $ \max_{q\in[Q]} M_N(q) \le \nu$. (ii) There exists a universal constant $\underline{S} > 0$ that does not depend on $N$ and $Q$ such that 
$ \min_{q\in[Q]} S(q,q) \ge \underline{S}.$
\end{condition}

We present Condition \ref{cond:easy-spec} to simplify the presentation of the theory in the main paper. More generally,
we can relax Condition \ref{cond:easy-spec}(i) on the universal upper bound on $M_N(q)$ and Condition \ref{cond:easy-spec}(ii) on the  universal lower bound on $S(q,q)$ by imposing conditions on the tail behavior of the potential outcomes. 
See Section \ref{sec:extend-conditions} in the Supplementary Materials for more discussions.


Theorem \ref{thm:wald-uniform} below justifies the Wald-type inference under the nearly uniform design with replications, where $H$ can be either fixed or diverging.
 
\begin{theorem}[Validity of Wald-type inference under nearly uniform designs with replications]\label{thm:wald-uniform} 
Consider the nearly uniform design given by Definition \ref{def:uniform-design} that satisfies $\min_{q\in[Q]}N_q\ge 2$ and Condition \ref{condition::proper}. Also assume \eqref{eqn:well-conditioned}, Conditions \ref{cond:moments} and \ref{cond:easy-spec}. 
Let $N\rightarrow \infty$. 
If $H^{19/2} / N \to 0$, then
the Wald-type confidence set
$\{  \bar \gamma :  (\hgamma - \bar \gamma)^\top \hV_{\hgamma}^{-1} (\hgamma - \bar \gamma)  \leq q_{H, \alpha}  \}$
for $\gamma$ is asymptotically valid. 


\end{theorem}

We relegate the proof of Theorem \ref{thm:wald-uniform} to Section \ref{sec:pf-wald-uniform} in the Supplementary Material. To prove Theorem \ref{thm:wald-uniform}, we establish the limiting distribution of $\hat T$ under both regimes with a fixed $H$ and a diverging $H$. Theorem \ref{thm:wald-uniform-general} in the Supplementary Material summarizes the precise results on the limiting distributions of $\hat{T}$, which is of independent interest. The proof of Theorem \ref{thm:wald-uniform-general} involves translating the finite sample bounds in Corollary \ref{cor:uniform-design-be} and Theorem \ref{thm:quad-be-nearly-uniform} into asymptotic results.
With a fixed $H$, Theorem \ref{thm:wald-uniform-general} shows that $\hat{T} \rightsquigarrow \cL$ for some distribution $\cL$ that is stochastically dominated by $\chi^2_H$. With a diverging $H$, Theorem \ref{thm:wald-uniform-general} shows that the standardized $\hat T$ converges to the standard normal distribution.


Theorem \ref{thm:wald-uniform} extends the known result for \ref{regime:R1} with a fixed $H$ and provides a novel result that allows for diverging $Q$ and $H$. It relies crucially on the BEB on the quadratic form $T$ in Theorem \ref{thm:quad-be-nearly-uniform} and the stochastic properties of $\hV_\hgamma$ in Theorem \ref{thm:uniform-var}.


\subsection{Unreplicated design}\label{sec:var-unreplicate}

In this subsection, we study inference for unreplicated designs with $N_q = 1$ for $q=1,\ldots, Q$. The BEB on $T$ is identical to that in Theorem \ref{thm:quad-be-nearly-uniform}. We give the formal result in  Theorem \ref{thm:quad-be-unreplicated} below for completeness.

\begin{theorem}[BEB for the quadratic form in unreplicated designs]\label{thm:quad-be-unreplicated}
Assume complete randomization that satisfies $N_q = 1$ for $q=1,\ldots, Q$ and Condition \ref{condition::proper}. Also assume \eqref{eqn:well-conditioned}. The BEB \eqref{eqn:quad-be-nearly-uniform} holds.
\end{theorem}

However, covariance estimation without replications is a fundamentally challenging problem that is not unique to the design-based framework, as reviewed in Section \ref{sec:motivation}. The commonly-used  covariance estimator $\hV_\hgamma$ in \eqref{eqn:estimates} is not well defined. We must construct a new estimator. Related variance estimation problems appeared in stratified survey sampling and stratified randomized experiments. For example, \citet[][Section 5A.12]{cochran1977sampling} proposed a variance estimation in stratified survey sampling with one unit per stratum by grouping strata into pairs. Similarly, \citet[][Section 2.5]{wolter2007introduction} also discussed the use of the collapsed stratum estimator with a general extension to more than two strata per group. For stratified experiments, there are also similar problems. \cite{fogarty2018mitigating} discussed several conservative estimators which also give the form of variance estimators based on collapsed strata or incorporating covariate information. However, these theoretical analyses cannot easily generalize to a permutation distribution where observations are correlated. Many works \citep[e.g.][]{HansenHurwitzMadow1953a, HansenHurwitzMadow1953b, breidt2016nonparametric, abadie2008estimation} also discussed using predictors to assist variance estimation, which is beyond the scope of our discussion. 

Below we first consider a naive variance estimator, which is intuitive and easy for implementation. However, we will show that it is almost always strictly conservative for the true variance. As a remedy, we also propose a grouping strategy that is similar to collapsing strata to form conservative variance estimators in finely stratified survey sampling and experiments. 

We use slightly simplified notation for unreplicated designs, where the observed allocation $Z_i$ and the arm $q$ have a one-to-one correspondence. We can denote the single observed outcome in arm $ q$ by $Y_q$. The point estimator still has the form $\hgamma = F^\top \hY$ where $\hY = (Y_1,\ldots, Y_Q)^\top$ is simply the observed outcome vector. Without replications, we cannot calculate $\hS(q,q)$ based on only the single observation within arm $q \in \cQ_\textsc{u} = \{ 1,\ldots, Q\} $.

\subsubsection{First strategy for variance estimation}\label{sec:first-strategy}
A strategy for constructing a variance estimator in unreplicated designs is based on the fact that $\hgamma$ is the average of the random vectors:
\begin{align*}
    \hgamma = F^\top \hY = \sum_{q\in[Q]} F(q,\cdot)^\top Y_q = Q^{-1}\sum_{q\in[Q]} Q{F(q,\cdot)^\top Y_q},
\end{align*}
which motivates us to construct the variance estimator: 
\begin{align}\label{eqn:hVo}
    \hV_{\hgamma} 
    = \mu_Q\sum_{q\in[Q]} \lt(Q{F(q,\cdot)^\top Y_q} - \hgamma  \rt) \lt(Q {F(q,\cdot) Y_q} - \hgamma^\top  \rt).
\end{align}
In \eqref{eqn:hVo}, $\mu_Q = \{Q(Q-2)\}^{-1}$ is a correction factor, which is motivated by moments calculation by the proof of Theorem \ref{thm:hVo} below:
\begin{theorem}[First strategy for variance estimation for unreplicated designs]\label{thm:hVo}
Consider designs that satisfy Definition \ref{def:non-uniform-design} with $|\cQ_{\textsc{r}}| = |\cQ_{\textsc{l}}| = 0$ and the covariance estimator in \eqref{eqn:hVo}.
\begin{enumerate}[label = (\roman*)]
    \item\label{thm:hVo-1} 
We have  
    \begin{align}\label{eqn:E-hVo}
        \bbE\{\hV_{\hgamma}\} 
        = & V_\hgamma  + \frac{Q-1}{Q(Q-2)}F^\top S F  \notag\\
    & + \frac{1}{Q(Q-2)} \sum_{q\in[Q]} \lt(\gamma - {Q F(q,\cdot)^\top\overline{Y}(q)}\rt) \lt(\gamma^\top - {Q F(q,\cdot)\overline{Y}(q)}\rt). 
    \end{align}
    Therefore,     $\bbE\{\hV_{\hgamma}\} \succeq V_\hgamma $. 
    
    \item \label{thm:hVo-2} 
Assume Conditions \ref{cond:moments} and \ref{cond:easy-spec}. We have 
    $\|\hV_{\hgamma} - \bbE\{\hV_{\hgamma}\}\|^2_{\infty}  = O_\bbP\lt(\|F\|_\infty^4
    QH^2\rt)$.
    \item \label{thm:hVo-3}
    Assume Conditions \ref{cond:moments} and    \ref{cond:easy-spec}. We have 
    $\|\hV_{\hgamma} - \bbE\{\hV_{\hgamma}\}\|^2_{\text{\em op}}  = O_\bbP\lt(\|F\|_\infty^4
    QH^4\rt)$.
\end{enumerate}

\end{theorem}

Theorem \ref{thm:hVo} is established by moment calculation and the use of Chebyshev's inequality (see Section \ref{sec:pf-hVo}).  Theorem \ref{thm:hVo}(i) suggests that the variance estimator \eqref{eqn:hVo} is conservative. The bias vanishes if and only if both of the following conditions hold:
\begin{gather*}
    F^\top (Y_i(q))_{q\in[Q]} = \gamma, \quad \text{ for all } i \in [N] \text{ and } \\
    F(q,\cdot)^\top\overline{Y}(q) = Q^{-1}\gamma, \quad \text{ for all } q\in[Q]. 
\end{gather*}
This is a very stringent condition and imposes restrictive conditions on the coefficient matrix $F$ and the potential outcomes. Therefore,  Theorem \ref{thm:hVo} suggests a trade-off for the use of the variance estimator \eqref{eqn:hVo}: it is easy for implementation but almost always strictly conservative.




\subsubsection{Second strategy for variance estimation}\label{sec:second-strategy}

Due to the above limitation of the variance estimator \eqref{eqn:hVo}, we consider a new strategy based on grouping the outcomes.
With a little abuse of notation, we still consider the covariance estimator of the form:
\begin{align}\label{eqn:hV-hY-abuse}
    \hV_{\hgamma} =  F^\top \hV_\hY  F,
\end{align}
where $\hV_\hY$ is a $Q\times Q$ diagonal matrix. The key is to construct its diagonal elements $\hV_\hY(q,q)$ for all $q$'s because we cannot define $\hV_\hY(q,q)$ as \eqref{eqn:hY-hV}.

To obtain substitutes for $\hS(q,q)$, we must borrow information across treatment arms. This motivates us to consider the following grouping strategy.

\begin{definition}[Grouping]
\label{def::Grouping}
Partition $\cQ_\textsc{u}$ as $   \cQ_{\textsc{u}} = \cup_{g=1}^G \cQ_{\textsc{u},g}$ where $\cQ_{\textsc{u},g} \cap \cQ_{\textsc{u},g'}  = \varnothing$ for all $g\neq g'$ and $ | \cQ_{\textsc{u},g}| \ge 2$ for all $g \in [G]$. The partition does not depend on the observed data. 
\end{definition}

Definition \ref{def::Grouping} does not allow for data-dependent grouping, which can cause theoretical complications due to double-dipping into the data. Examples \ref{exp:pairing} and \ref{exp:regression} below are special cases of Definition \ref{def::Grouping}. 
By the construction in Definition \ref{def::Grouping}, the $\cQ_{\textsc{u},g}$'s have no overlap, so we can also use $\langle g \rangle$  to denote $\cQ_{\textsc{u},g}$ and $\cG = \{ \langle g \rangle\}_{g=1}^G$ to denote the grouping strategy without causing confusions.  Moreover, $|\langle g \rangle|$ must be larger than or equal to two so that there are at least two treatment levels in each $\langle g \rangle$. In general, we use $ \langle g \rangle_q$ to indicate the group $\langle g \rangle$ that contains arm $q$, but when no confusion arises, we also simplify the notation to $ \langle g \rangle$ if the corresponding $q$ is clear from the context.

Define 
\begin{align*}
    \hY_{\langle g \rangle} = \frac{1}{|\langle g \rangle|} \sum_{q\in\langle g \rangle} Y_q,
\end{align*}
 as the group-specific average, and construct 
 \begin{align}\label{eqn:hV-QU}
\hV_\hY(q,q) = \mu_{\langle g \rangle}(Y_q - \hY_{{\langle g \rangle}})^2 ,\quad \text{ if } q\in{\langle g \rangle}
\end{align}
as the $q$th diagonal element of $\hV_\hY$, where 
\begin{align}\label{eqn:mu-bg}
     \mu_{\langle g \rangle} = (1-2N^{-1})^{-1}(1-|{\langle g \rangle}|^{-1})^{-2}
\end{align}
is a correction factor that is motivated by the theory below. Although the mean of $    \hY_{\langle g \rangle} $ has a simple formula, the mean of $\hV_\hY(q,q) $ has a cumbersome form. We present a lower bound of $    \bbE\{\hV_\hY(q,q)\} $ below and relegate the complete formula to Section \ref{pf:lem-mean-var-grouping-thm-unrep-var} the supplementary material. The results require Condition \ref{cond:cond-N} below on the largest eigenvalue of the population correlation matrix of the potential outcomes in group $\langle g \rangle$, defined as
\begin{align}\label{eqn:rho-max-S}
    \varrho_{\langle g \rangle} = \varrho_{\max}\{(S^\star(q,q'))_{q,q'\in\langle g\rangle}\},
\end{align}
where $S^\star$ is the population correlation matrix. 

\begin{condition}[Bound on $  \varrho_{\langle g \rangle} $]\label{cond:cond-N}   
$
    N- \varrho_{\langle g \rangle} -  (|{\langle g \rangle}|-1) \ge 0 
$
for all $g\in\cG$.
\end{condition}

Condition \ref{cond:cond-N} reflects a trade-off between $N$, $\langle g\rangle$ and $\varrho_{\langle g \rangle}$. It is more likely to hold with smaller correlations between arms within the same group and smaller subgroup sizes. By a natural bound $\rho_{\langle g \rangle} \le |{\langle g \rangle}|$, Condition \ref{cond:cond-N} holds if $|{\langle g \rangle}| \le {(N + 1)}/{2}$ for all $g\in[G]$. Examples \ref{exp:pairing} and \ref{exp:regression} below satisfy Condition \ref{cond:cond-N} automatically. 
With Condition \ref{cond:cond-N}, we can present Lemma \ref{lemma::mean-variance-grouping} below.

\begin{lemma}
[Sample mean and variance under grouping]
\label{lemma::mean-variance-grouping}
Assume grouping $\cG$ according to Definition \ref{def::Grouping}. We have 
\begin{align*}
\E{\hY_{{\langle g \rangle}}}  =    \overline{Y}_{\langle g \rangle} ,\quad \text{ where }   \overline{Y}_{\langle g \rangle} = \frac{1}{|{\langle g \rangle}|N} \sum_{q\in{\langle g \rangle}} \sum_{i=1}^NY_i(q) = \frac{1}{|{\langle g \rangle}|} \sum_{q\in{\langle g \rangle}} \overline{Y}(q).
\end{align*}
Further assume Condition \ref{cond:cond-N}. We have
\begin{align*}
     \bbE\{\hV_\hY(q,q)\} \ge S(q,q) + \underbrace{{\Omega}(q,q)}_{\text{\em term III}} +  \underbrace{\mu_{\langle g \rangle}  (\overline{Y}(q) - \overline{Y}_{\langle g \rangle})^2}_{\text{\em term IV}},
\end{align*}
where
\begin{align} \label{eq::Omega-qq}
    {\Omega}(q,q) &= \mu_{\langle g \rangle}|g|^{-2}\lt(1-\frac{\varrho_{\langle g \rangle}}{N} - \frac{|g|-1}{N}\rt)\sum_{q'\in{\langle g \rangle},q'\neq q}S(q',q')  \geq 0.
\end{align}
\end{lemma}

By Lemma \ref{lemma::mean-variance-grouping}, $\hV_\hY(q,q)$, as an estimator for $S(q,q) $, is conservative, and the conservativeness depends on the variation of other arms $q'$ that belong to ${\langle g \rangle}_q$ (term III) and the between-arm heterogeneity in means within ${\langle g \rangle}_q$ (term IV). We comment on some special cases below. 
\begin{itemize}
    \item If we assume homogeneity in means within subgroups, i.e.,
\begin{align}\label{eqn:weak-means}
    \overline{Y}(q) = \overline{Y}_{\langle g \rangle},~ \text{ for all } q\in\langle g \rangle,
\end{align}
then term IV vanishes.

   \item If we assume homoskedasticity across treatment arms within the same subgroup, i.e.,
\begin{align}\label{eqn:weak-vars}
    S(q,q) = S(q',q'),~ \text{ for all } q,q'\in \langle g \rangle,
\end{align}
then term III becomes 
\begin{align}\label{eqn:strong-po}
    {\Omega}(q,q) = \mu_{\langle g \rangle}(|g|-1)|g|^{-2}\lt(1-\frac{\varrho_{\langle g \rangle}}{N} - \frac{|g|-1}{N}\rt)S(q,q) . 
\end{align}
Then we can combine \eqref{eqn:strong-po} with $S(q,q)$ and use a smaller correction factor  
\begin{align*}
    \mu'_{\langle g \rangle} = (1-|g|^{-1})^{-1}\{(1-|g|^{-1})(1-2N^{-1}) + |g|^{-1}(1 - (2|g| - 1)/N) \}^{-1} \le \mu_{\langle g \rangle}
\end{align*}
to reduce the conservativeness of variance estimation. 

  \item If we assume the strong null hypothesis within subgroups, i.e.,
  \begin{align*}
      Y_i(q) = Y_i(q'), \text{ for all }i\in[N] \text{ and } q,q'\in {\langle g \rangle},
  \end{align*}
  then both \eqref{eqn:weak-means} and \eqref{eqn:weak-vars} hold.  Applying the correction factor $\mu'_{\langle g \rangle}$, we can show 
$
      \bbE\{\hV_\hY(q,q)\} = S(q,q).
$ 
\end{itemize}

Lemma \ref{lemma::mean-variance-grouping} suggests that ideally, we should group treatment arms based on the prior knowledge of the  means and variances of the potential outcomes. While more general grouping strategies are possible, we give two examples for their simplicity of implementation. Both target the factorial design in Example \ref{eg::factorial-design}.

\begin{example}[Pairing by the lexicographic order]\label{exp:pairing}
Recall Example \ref{eg::factorial-design}. 
We order the observations based on the lexicographical order of their treatment levels, then group the $(2k-1)$-th level with the $(2k)$-th level $(1\le k\le 2^{K-1})$. When $K=3$, the grouping reduces to
 \begin{align*}
   \langle 1 \rangle=  \{ (000), (001) \}, \quad
      \langle 2 \rangle   = \{ (010), (011) \}, \quad
        \langle 3 \rangle   = \{    (100), (101) \} , \quad 
          \langle 4 \rangle   =    \{   (110), (111)\} . 
 \end{align*}
 If the last factor has a small effect on the outcome, then we expect small differences in the mean potential outcomes within groups. 
As a sanity check, Condition \ref{cond:cond-N} holds under this grouping strategy.
\end{example}

\begin{example}[Grouping based on a subset of the factors]\label{exp:regression}
Recall Example \ref{eg::factorial-design} again. If we have the prior knowledge that $K_0 < K$ factors are the most important ones, we can group the treatment levels based on these factors. Without loss of generality, assume that the first $K_0$ factors are the important ones. In particular, we can create $ G = 2^{K_0} < Q$ groups, with each group $     {\langle g \rangle}$ corresponding to treatment levels with the same important factors. Example \ref{exp:pairing} above is a special case with the first $K-1$ factors as the important ones. 
Also, Condition \ref{cond:cond-N} holds under this grouping strategy.
\end{example}


\begin{remark}[Practical grouping strategies]\label{rmk:grouping-strategy}
    We have included two strategies for covariance estimation. On the one hand, they may seem ad hoc from a theoretical perspective. On the other hand, they are intuitive methods for covariance estimation. Covariance estimation without replications is a challenging problem in general. Therefore, the proposals can be viewed as a first attempt for variance estimation in unreplicated designs. Indeed, more research efforts should be put into this problem. For instance, how do we compare different covariance estimation strategies? What is the ``optimal'' strategy for covariance estimation? We believe this is another independent project that goes beyond the scope of the current paper. In the current work, we focused on the factorial design example and proposed to apply two grouping strategies that respect the structure of the design and are easy to implement.  In particular, it is interesting to consider borrowing additional information such as pre-treatment covariates to form better groups.  Again, these directions require more technical work and go beyond the scope of the current paper.
\end{remark}




Now we turn to the theoretical analysis of \eqref{eqn:hV-QU}.  Its properties depend on how successful the grouping $\cG$ is, quantified by Condition \ref{cond:bounded-bgv} below.

\begin{condition}[Bound on the within-group variation in potential outcome means]\label{cond:bounded-bgv}
There exists a $\zeta>0$, such that  
$
    \max_{g\in [G]} \max_{q\in{\langle g \rangle}} |\overline{Y}(q) - \overline{Y}_{\langle g \rangle}| \le \zeta.
$
\end{condition}

The $\zeta$ in Condition \ref{cond:bounded-bgv} bounds the between-arm distance of the mean potential outcomes under grouping $\cG$. It plays a key role in Theorem \ref{thm:unreplicated-var} below.

\begin{theorem}[Variance estimation for unreplicated designs]\label{thm:unreplicated-var}
Consider designs that satisfy Definition \ref{def:non-uniform-design} with $|\cQ_\textsc{r}| = |\cQ_\textsc{l}| = 0$ and the covariance estimator in \eqref{eqn:hV-QU}.
\begin{enumerate}[label = (\roman*)]
    \item\label{thm:unreplicated-var-1} 
Assume Condition \ref{cond:cond-N}.    We have   
    \begin{align*}
        \bbE\{\hV_\hY\} 
        =  V_\hY  +  \Omega  +  \diag{\mu_{\langle g \rangle}(\overline{Y}(q) - \overline{Y}_{\langle g \rangle})^2}_{q\in\cQ_{\textsc{u}}} + N^{-1} (\Theta + S)  
    \end{align*}
    with $\Omega = \diag{ \Omega(q,q)  }_{q\in\cQ_{\textsc{u}}}$ and $ \Theta = \diag{\Theta(q,q)}_{q\in\cQ_\textsc{U}}$, where the $\Omega(q,q) $'s are defined in \eqref{eq::Omega-qq} and the $\Theta(q,q)$'s are defined in \eqref{eqn:Theta-qq} in the supplementary material, satisfying $0\le \Theta(q,q) \le 5\mu_{\langle g \rangle}\max_{q'\in \langle g \rangle} S(q',q').$ 
    Therefore,     $\bbE\{F^\top\hV_\hY F\} \succeq V_\hgamma $. 
    
    \item \label{thm:var-unreplicated-2} 
Assume Conditions \ref{cond:moments} and    \ref{cond:bounded-bgv}. We have 
    \begin{align*}
        \|\hV_{\hgamma}-\bbE\{\hV_{\hgamma}\}\|^2_{\infty} & = O_\bbP\lt\{(\max_{g\in[G]}\mu_{\langle g \rangle})^2 \|F\|_\infty^4 NH^2\rt\}.
    \end{align*}
    \item \label{thm:var-unreplicated-3}
    Assume Conditions \ref{cond:moments} and    \ref{cond:bounded-bgv}. We have 
    \begin{align*}
        \|\hV_{\hgamma}-\bbE\{\hV_{\hgamma}\}\|^2_{\operatorname{op}} & = O_\bbP\lt\{(\max_{g\in[G]}\mu_{\langle g \rangle})^2 \|F\|_\infty^4 NH^4\rt\}.
    \end{align*}
\end{enumerate}

\end{theorem}

Theorem \ref{thm:unreplicated-var}\ref{thm:unreplicated-var-1} demonstrates that based on $\eqref{eqn:hV-QU}$, the covariance estimator  $\hV_\hY$ is conservative for $V_\hY$, which implies that $F^\top\hV_\hY F$ is conservative for the true covariance matrix of $\hgamma$. The conservativeness, however, has a more complex pattern compared with the setting with replications within all arms \citep{neyman1923application, imbens15, li2017general}.  
Theorem \ref{thm:unreplicated-var}(i) shows three sources of conservativeness. The first part, captured by $\Omega $, is due to the between-arm heteroskedasticity within each subgroup. However, it is fundamentally difficult to estimate each $S(q,q)$ without replications. The second part, captured by $\diag{\mu_{\langle g \rangle}(\overline{Y}(q) - \overline{Y}_{\langle g \rangle})^2}_{q\in[Q]}$,  is due to the between-arm heterogeneity in means within each subgroup. The part will be small if the grouping strategy ensures that the grouped arms have similar population averages of potential outcomes.
The third part, captured by $N^{-1} (\Theta + S)$,  is due to the difficulty of estimating $S$ and in particular, the off-diagonal terms of $S$. The difficulty of estimating $S$ has been well documented ever since \citet{neyman1923application} even in experiments with replications in each arm. It is possible to reduce this part but it requires additional assumptions, for example, the individual causal effects are constant.

Theorem \ref{thm:unreplicated-var}\ref{thm:var-unreplicated-2} and \ref{thm:var-unreplicated-3} give the stochastic order of the estimation error of the covariance estimator $\hV_{\hgamma}$ under the $L_\infty$ norm and operator norm, respectively. If $\max_{g\in[G]}\mu_{\langle g \rangle}$, $\Delta$ and $\zeta$ are all constants, then 
$$
    N\|\hV_{\hgamma}-\bbE\{\hV_{\hgamma}\}\|_{\infty}  = O_\bbP (H  / {N}^{1/2}  ) , \quad
     N  \|\hV_{\hgamma}-\bbE\{\hV_{\hgamma}\}\|_{\operatorname{op}}   = O_\bbP (H^2  / {N}^{1/2}  ),
$$
which gives sufficient conditions on $H$ to ensure the convergence of $\hV_{\hgamma}$ in $L_\infty$ norm and operator norm, respectively.

Finally, equipped with the BEB on the quadratic form $T$ in \eqref{eqn:quad-form} and the conservative variance estimator studied in Theorem \ref{thm:unreplicated-var}, it is immediate to establish Theorem \ref{thm:wald-unreplicated} below for inference, which parallels Theorem \ref{thm:wald-uniform}.

\begin{theorem}[Wald-type inference under unreplicated design]\label{thm:wald-unreplicated} 
Consider the unreplicated design that satisfies $N_q =1$ for all $q = 1,\ldots, Q$ and Condition \ref{condition::proper}. Also assume \eqref{eqn:well-conditioned} and Conditions \ref{cond:moments}--\ref{cond:bounded-bgv}.  Let $N\rightarrow \infty$. If $H^{19/4}N^{-1/2} \to 0$, the Wald-type confidence set is asymptotically valid.
\end{theorem}

\subsection{General design}\label{sec::non-uniform-design-inference}

In this section, we consider general designs in Definition \ref{def:non-uniform-design}. First, we show a BEB on $T$ in \eqref{eqn:quad-form} in Theorem \ref{thm:quad-be-non-uniform} below. 

\begin{theorem}[Quadratic form BEB for general designs]\label{thm:quad-be-non-uniform}
Consider the general design in Definition \ref{def:non-uniform-design} that satisfies Condition \ref{condition::proper-non-uniform} together with $\|F_\textsc{l}\|_\infty = O(Q^{-1})$ and $N=O(|\cQ_\textsc{s}|)$. Also assume \eqref{eqn:well-conditioned}. There exists a universal constant $C>0$, such that
\begin{align}\label{eqn:quad-be-non-uniform}
    \sup_{t\in\bbR} |\bbP(T\le t)-\bbP(T_0\le t)| \le C \frac{\max_{q\in[Q]}M_N(q)^3 }{\{ \min_{q\in\cQ_\textsc{S}}S(q,q) \}^{3/2}}\cdot \frac{H^{19/4}}{N^{1/2}}.
\end{align}
\end{theorem}

The proof of Theorem \ref{thm:quad-be-non-uniform} relies on the general BEB Theorem \ref{thm:be-bounded}. See Section \ref{sec:pf-wald-non-uniform} for more details.
Theorem \ref{thm:quad-be-non-uniform} assumes $\|F_\textsc{l}\|_\infty $ and $N$ has the same order as $Q^{-1}$ and $|\cQ_\textsc{s}|$, respectively, which is helpful to establish the root-$N$ convergence of BEB. We can relax the assumptions if we only need the CLT rather than the BEB. See Remark \ref{rmk:relax-non-uniform-BEB} in Section \ref{sec:quad-be-non-unifor} of the the supplementary material.  For ease of presentation, we omit the general results. A subtle feature of the upper bound in \eqref{eqn:quad-be-non-uniform} is that $\max_{q\in[Q]}M_N(q)$ is the  maximum value of the $M_N(q)$'s over all treatment arms whereas $\min_{q\in\cQ_\textsc{S}}S(q,q)$ is the minimum value of the $S(q,q)$'s  over treatment arms in $\cQ_\textsc{S}$ only.

Second, we construct a covariance estimator. It is a combination of the covariance estimators discussed in Sections \ref{sec:var-uniform} and \ref{sec:var-unreplicate}. For the treatment arms with replications, we can calculate sample variances of the potential outcomes based on the observed data. For the treatment arms without replications $\cQ_\textsc{u}$, we need the grouping strategy in Definition \ref{def::Grouping}. Therefore, we construct a diagonal covariance estimator $\hV_\hY$ with the $q$-th diagonal term
\begin{align*}
    \hV_\hY(q,q) = \left\{
    \begin{array}{cc}
        \mu_{\langle g \rangle} (Y_q - \hY_{\langle g \rangle})^2, &\quad  q\in\cQ_{\textsc{u}} \\
        \widehat{S}(q,q), & \quad  q\in\cQ_{\textsc{r}}\cup\cQ_{\textsc{l}}.
    \end{array}
    \right.
\end{align*}
In a matrix form, it is equivalent to
\begin{align}\label{eqn:composite-var}
    \hV_\hY = \begin{pmatrix}
    \hV_{\hY,\textsc{u}}& 0& 0,\\
    0& \hV_{\hY,\textsc{r}}& 0,\\
    0& 0& \hV_{\hY,\textsc{l}}
    \end{pmatrix},
\end{align}
where $\hV_{\hY,\textsc{u}}, \hV_{\hY,\textsc{r}}, \hV_{\hY,\textsc{l}}$
correspond to the diagonal covariance estimators for treatment arms 
$\cQ_{\textsc{u}}, \cQ_{\textsc{r}}, \cQ_{\textsc{l}}$, respectively. Recall the partitioning of $F$ given in \eqref{eqn:F-partition}.
Construct the final covariance estimator below: 
\begin{align}\label{eqn:composite-var-2}
    \hV_\hgamma = F^\top \hV_{\hY} F = F_{\textsc{u}}^\top \hV_{\hY,\textsc{u}} F_{\textsc{u}} + F_{\textsc{r}}^\top \hV_{\hY,\textsc{r}} F_{\textsc{r}} + F_{\textsc{l}}^\top \hV_{\hY,\textsc{l}} F_{\textsc{l}}. 
\end{align}

\begin{remark}
    The variance estimator \eqref{eqn:composite-var-2} uses the second variance estimation strategy in Section \ref{sec:second-strategy} for the unreplicated design component. An extension of the first strategy in Section \ref{sec:first-strategy} is also feasible based on the following partition:
    \begin{align}
        &F^\top \diag{S(q,q)}_{q\in[Q]} F \label{eqn:decompose-FSF} \\
        = F_{\textsc{l}}^\top & \diag{S(q,q)}_{q\in\cQ_{\textsc{l}}} F_{\textsc{l}} + F_{\textsc{r}}^\top \diag{S(q,q)}_{q\in\cQ_{\textsc{r}}} F_{\textsc{r}} + F_{\textsc{u}}^\top \diag{S(q,q)}_{q\in\cQ_{\textsc{u}}} F_{\textsc{u}}.
    \end{align}
    The classical variance estimator for the first two terms of \eqref{eqn:decompose-FSF} is well-defined because each arm in $\cQ_\textsc{l}$ and $\cQ_\textsc{r}$ still contains at least two units. For the last part, the variance estimator can be constructed in a similar way to \eqref{eqn:hVo}.
\end{remark}

The decomposition in \eqref{eqn:composite-var-2} allows us to characterize the statistical properties of $ \hV_\hY $ by combining the results from Sections \ref{sec:var-uniform} and \ref{sec:var-unreplicate}.

\begin{theorem}[Covariance estimation for general designs]\label{thm:non-uniform-var} 
Consider the designs in Definition \ref{def:non-uniform-design} and the covariance estimator in \eqref{eqn:composite-var-2}.
Assume Conditions \ref{cond:moments}, \ref{cond:cond-N}, and \ref{cond:bounded-bgv}. 
Assume $\max_{g\in[G]}\mu_{\langle g \rangle}, \Delta$ and $\zeta$ are constants.
\begin{enumerate}[label = (\roman*)]
    \item\label{thm:non-uniform-var-1}
    $\bbE\{\hV_\hgamma\}  \succeq V_\hgamma$.
    
    \item\label{thm:non-uniform-var-2}
    We have
    $$
            \|\hV_{\hgamma}-\bbE\{\hV_{\hgamma}\}\|^2_{\infty} = 
          O_\bbP (   \|F_{\textsc{u}}\|_\infty^4 
 |\cQ_{\textsc{u}}|H^2 + 
  \|F_{\textsc{r}}\|_\infty^4 |\cQ_{\textsc{r}}| H^2  + { \|F_{\textsc{l}}\|_\infty^4 |\cQ_{\textsc{l}}|^4 N_\textsc{l}^{-3}  H^2} ).
     $$
    
    \item\label{thm:non-uniform-var-3}
    We have 
    $$
            \|\hV_{\hgamma}-\bbE\{\hV_{\hgamma}\}\|^2_{\operatorname{op}} = 
          O_\bbP (   \|F_{\textsc{u}}\|_\infty^4 
   |\cQ_{\textsc{u}}|H^4 +  
    \|F_{\textsc{r}}\|_\infty^4 |\cQ_{\textsc{r}}|  H^4 + { \|F_{\textsc{l}}\|_\infty^4 |\cQ_{\textsc{l}}|^4 N_\textsc{l}^{-3}  H^4}  )  . 
     $$ 
\end{enumerate}
\end{theorem}

In Theorem \ref{thm:non-uniform-var}, we assume $\max_{g\in[G]}\mu_{\langle g \rangle}, \Delta$ and $\zeta$ to be constants to simplify the presentation. Without this assumption, we can derive results similar to those in Theorem \ref{thm:unreplicated-var} but relegate finer results to the supplementary material. Theorem \ref{thm:non-uniform-var}\ref{thm:non-uniform-var-1} shows the conservativeness of $\hV_\hgamma$ as a direct consequence of Theorems \ref{thm:uniform-var}\ref{thm:uniform-var-1} and  \ref{thm:unreplicated-var}\ref{thm:unreplicated-var-1}. Theorem \ref{thm:non-uniform-var}\ref{thm:non-uniform-var-2} and \ref{thm:non-uniform-var-3} show the stochastic order of the estimation error of $\hV_\hgamma$ in $L_\infty$ norm and operator norm, respectively. We only discuss Theorem \ref{thm:non-uniform-var}\ref{thm:non-uniform-var-2} below. 
If $\| F \|_{\infty} = O(Q^{-1})$ as in the factorial design in Example \ref{eg::factorial-design}, it reduces to 
\begin{align}\label{eqn:V-infty}
   \lt\|\hV_{\hgamma} - \bbE\{\hV_\hgamma\}\rt\|_\infty = O_\bbP\lt\{ Q^{-2}H(|\cQ_{\textsc{u}}|^{1/2} + |\cQ_{\textsc{r}}|^{1/2} + |\cQ_{\textsc{l}}|^{2}N_{\textsc{l}}^{-3/2})\rt\}.
\end{align}
Therefore, if $N$ and $Q$ are of the same order, then $N\|\hV_{\hgamma}-\bbE\{\hV_{\hgamma}\}\|_{\infty} = O_\bbP ( HN^{-1/2} ) $.
Besides, when one or two of $\cQ_{\textsc{u}},\cQ_{\textsc{r}},\cQ_{\textsc{l}}$ are small or absent, the stochastic orders in Theorem \ref{thm:non-uniform-var} still hold because the large terms in \eqref{eqn:V-infty} will dominate the rest. In particular, if $|\cQ_\textsc{u}| = |\cQ_\textsc{r}| = 0$, then 
$
    \|\hV_{\hgamma} - \bbE\{\hV_\hgamma\}\|_\infty = O_\bbP ( HN_{\textsc{l}}^{-3/2}) ,
$ 
which gives the same rate as Theorem \ref{thm:uniform-var}. If $|\cQ_\textsc{l}| = 0$, then we should interpret $0\cdot \infty = 0$ in Theorem \ref{thm:non-uniform-var}\ref{thm:non-uniform-var-2} to obtain
$
    \|\hV_{\hgamma} - \bbE\{\hV_\hgamma\} \|_\infty = O_\bbP ( HQ^{-3/2} )  = O_\bbP ( HN_{\textsc{s}}^{-3/2} ) ,
$
which also agrees with Theorem \ref{thm:uniform-var}. 

Finally, the BEB on the quadratic form $T$ and the conservativeness of the covariance estimator ensure Theorem \ref{thm:wald-non-uniform} below for inference.  

\begin{theorem}[Wald-type inference under general designs]\label{thm:wald-non-uniform} 
Consider the general design in Definition \ref{def:non-uniform-design} that satisfies Condition \ref{condition::proper-non-uniform} together with $\|F_\textsc{l}\|_\infty = O(Q^{-1})$ and $N=O(|\cQ_\textsc{s}|)$. Also assume  \eqref{eqn:well-conditioned} and Conditions \ref{cond:moments}--\ref{cond:bounded-bgv}. Let $N\rightarrow \infty$. If  $H^{19/4}N^{-1/2} \to 0$, the Wald-type confidence set is asymptotically valid.
\end{theorem}

This concludes our discussion of design-based causal inference with possibly a diverging number of treatment levels and varying group sizes across treatment levels.

\section{Discussion}\label{sec::discussion}

We provide general BEBs for design-based causal inference that can accommodate possibly diverging treatment levels and varying group sizes. They serve as the theoretical foundation for causal inference in modern complex experiments. When we were polishing the paper, \citet{shi2025forward} and \citet{masoero2024multiple} used our BEBs to analyze factorial designs and multiple randomization designs, respectively. We look forward to see more applications of our theoretical results in future research.  

Our paper is mainly theoretical. Nevertheless, we use simulation studies to evaluate finite-sample properties of the estimators as well as their variance estimators. We relegate the details to Appendix \ref{sec:simulation} in the supplementary material. It is of interest to see more comprehensive simulation studies and concrete empirical applications in the future.

We focused on scalar outcomes. Results for vector outcomes are also important in both theory and practice. \citet{li2017general} reviewed CLTs and many applications with vector outcomes under the regime of a fixed number of treatment levels and large sample sizes within all treatment levels. We include an extension of the BEB for vector outcomes under a general regime; see Section \ref{sec:vec-outcome} in the supplementary material.


Asymptotic results for design-based inference are often criticized because the population of interest is finite but the asymptotic theory requires a growing sample size. Establishing BEBs is an important theoretical step to characterize the finite-sample performance of the statistics. Alternatively, it is also desirable to derive non-asymptotic concentration inequalities for the estimators under the randomization model. This requires a deeper understanding of sampling without replacement and permutation statistics. We leave it to future research. 


\section*{Funding}
The authors were partially supported by the U.S. National Science Foundation (\# 1945136).


\section*{Supplement}
The supplementary material contains additional results on general linear permutational statistics, randomization-based inference, and all the technical proofs.

\bibliographystyle{asa.bst}
\bibliography{ref}


\newpage 

\begin{appendix}

\setcounter{page}{1}
\renewcommand{\thepage}{S\arabic{page}}

\setcounter{equation}{0}
\renewcommand{\theequation}{S\arabic{equation}}

\setcounter{theorem}{0}
\renewcommand{\thetheorem}{S\arabic{theorem}}

\setcounter{lemma}{0}
\renewcommand{\thelemma}{S\arabic{lemma}}

\setcounter{proposition}{0}
\renewcommand{\theproposition}{S\arabic{proposition}}

\setcounter{corollary}{0}
\renewcommand{\thecorollary}{S\arabic{corollary}}

\setcounter{table}{0}
\renewcommand{\thetable}{S\arabic{table}}

\setcounter{figure}{0}
\renewcommand{\thefigure}{S\arabic{figure}}

\setcounter{example}{0}
\renewcommand{\theexample}{S\arabic{example}}

\setcounter{condition}{0}
\renewcommand{\thecondition}{S\arabic{condition}}

\setcounter{definition}{0}
\renewcommand{\thedefinition}{S\arabic{definition}}

\setcounter{remark}{0}
\renewcommand{\theremark}{S\arabic{remark}}

\begin{center}
\Huge Supplementary materials
\end{center}

Appendix \ref{sec:simulation} provides simulation results that evaluate the finite-sample properties of the point and variance estimators under a non-uniform design.  

Appendix \ref{sec:general-BE} reviews existing and develops new BEBs for linear permutation statistics. 

Appendix \ref{sec::proof-linear-permutational} gives the proofs of the results in  Appendix \ref{sec:general-BE}.

Appendix \ref{sec:additional} presents additional results for design-based causal inference. 

Appendix \ref{sec:main-proof} gives the proofs of the results in the main paper and Appendix \ref{sec:additional}.

\medskip 

\section{Simulation}\label{sec:simulation}


In this section, we will evaluate the finite-sample properties of the point estimates and the proposed variance estimator in factorial experiments. We consider general designs because there have been extensive numerical studies for nearly uniform designs before. 

\subsection{Practical implementation}\label{section::simulation-implementation}

For illustration purposes, we focus on conducting inference for the main effects in general factorial designs.
To do this, we need grouping strategies to implement the proposed variance estimator \eqref{eqn:composite-var}. As we discussed in Section \ref{sec:var-unreplicate}, the structure of factorial designs can provide some practical guidance on the choice of grouping strategy. In addition, our theoretical results in Theorem \ref{thm:unreplicated-var} also provide insight into reducing the conservativeness of the variance estimator. In our simulation, we will compare three variance estimation strategies:
\begin{enumerate}[label = (\roman*)]
    \item \textit{Pairing according to the lexicographical order.} This corresponds to our discussion in Example \ref{exp:pairing}. If arms with similar factor combinations have close means, pairing based on the lexicographical order can guarantee small between-arm discrepancy in means and reduce the conservativeness. 
    
    Moreover, pairing strategies have another benefit in factorial experiments. We can use a smaller correction factor $\tilde{\mu}_{\langle g\rangle}$ for variance estimation if our goal is to conduct inference marginally (i.e. build confidence intervals on each of $\gamma_h$ separately). The reason is that, while it is hard to control the $\Omega$ matrix in Theorem \ref{thm:unreplicated-var}\ref{thm:unreplicated-var-1} in general, we can control the diagonals of $F_{\textsc{u}}^\top \Omega F_{\textsc{u}}$ because $F_\textsc{u}$ has element $\pm Q^{-1}$. We can get more intuition by noticing that 
    $
    \sum_{q'\in{\langle g \rangle_q}, q'\neq q} S(q',q')
    $
    is the core of $\Omega(q,q)$ and that the following algebraic fact holds under pairing:
    \begin{align}\label{eqn:fact-on-Omega}
        \sum_{q\in\cQ_{\textsc{u}}}\sum_{q'\in{\langle g \rangle_q}, q'\neq q} S(q',q') 
        &= \sum_{q\in\cQ_{\textsc{u}}} S(q,q).
    \end{align}
The identity    \eqref{eqn:fact-on-Omega} enables us to transform the diagonals of $F_{\textsc{u}}^\top \Omega F_{\textsc{u}}$ from a source of conservativeness to the part of the true variance. Hence it allows us to choose a smaller correction factor:
\begin{align*}
    \tilde{\mu}_{\langle g \rangle} = (1-|{\langle g \rangle}|^{-1})^{-1}(1-3N^{-1})^{-1} = 2(1-3N^{-1})^{-1} ,
\end{align*}
which is approximately one half of $\mu_{\langle g \rangle}=4(1-2N^{-1})^{-1}$ in \eqref{eqn:mu-bg} when $N$ is large.
 
    \item \textit{Regression-based variance estimation with the target factors as regressors.} Regression-based approach is a commonly used strategy for analyzing factorial experiments.  For general designs, \cite{zhao2021regression} pointed out that ordinary least squares (OLS) with unsaturated model specifications can give biased point estimates and variance estimators. Instead, one should apply weighted least squares (WLS) and the sandwich variance estimation.
    
    \item \textit{Regression-based variance estimation with the target factors and their high-order interactions as regressors.} This strategy differs from strategy (ii) in whether the interactions are included. If all possible two-way interactions of the target factors are specified in the regression model and the true $k$-way ($k\ge 3$) interactions are zero, then this strategy is equivalent to the general factor-based grouping strategy introduced in Example \ref{exp:regression}.
\end{enumerate}

In the next section, we will provide more details on implementing the above strategies in simulation. 

\subsection{Simulation settings}
In this section, we use more simulations to check the performance of the point estimator and the proposed variance estimation strategies for different sample sizes by varying the number of factors $K$. We set up $2^K$ experiments with $K = 7,8,9,10$ and $Q = 2^K$ as follows: 
    \begin{itemize}
        \item unreplicated arms: $|\cQ_{\textsc{u}}| =  0.65 \cdot Q$ (rounded to integer) and $N_q = 1$ for each $q\in\cQ_{\textsc{u}}$.
        \item replicated small arms: $|\cQ_{\textsc{r}}| = 0.33\cdot Q$ (rounded to integer) and $N_q = 2$ for each $q\in\cQ_{\textsc{r}}$.
        \item large arms: $|\cQ_{\textsc{l}}| = 0.02\cdot Q$ (rounded to integer) and $N_q = 30$ for each $q\in\cQ_{\textsc{l}}$.
    \end{itemize}
    In this setup, the sample sizes for $\cQ_{\textsc{u}}$, $\cQ_{\textsc{r}}$ and $\cQ_{\textsc{l}}$ are close, and the total sample size $N = 1.91\cdot Q$ (rounded to integer). With $K = 7,8,9,10$, we have $N = 257, 514, 1028, 1999$. We generate the potential outcomes independently from a shifted exponential distribution:
    \begin{align*}
        Y_i(q) \sim \text{EXP}(\lambda_q) - 1/\lambda_q + \mu_q,
    \end{align*}
    where $\lambda_q$ are randomly set as $1$ or $2$ with equal probability to induce heteroskedasticity. We generate the $\mu_q$'s such that: (i) the main effects of factor $F_k$ with $k=1,4,7,10$ are set as zero; (ii) a random subset of two-way interactions is set as zero as well; (iii) all the $k$-way ($k\ge 3$) interactions are zero; (iv) the nonzero main effects and two-way interactions from a uniform distribution: $ \text{Unif}([-0.25,-0.05]\cup[0.05,0.25])$. 

    We focus on estimating the main factorial effects for factor $F_{l}$ for $l=1,\dots,5$. We apply the point estimates $\hgamma$ in \eqref{eqn:estimates} and compare three variance estimation strategies discussed in Section \ref{section::simulation-implementation}: 
    \begin{enumerate}
    \item
    LEX: We use the grouping strategy based on pairing by the lexicographical order. 
    \item
    $\text{EHW}_0$: We use the sandwich variance estimators based on 
    WLS with the target factors:
    \begin{align*}
        Y \sim F_1 + F_2 + F_3 + F_4 + F_5, \text{ with weights } w_i = N_{Z_i}^{-1}. 
    \end{align*}
    \item 
    $\text{EHW}_1$: We use the sandwich variance estimators based on WLS with the target factors and their two-way interactions:
    \begin{align*}
        Y \sim F_1 + F_2 + F_3 + F_4 + F_5 + \text{Interaction}_2(F_1, F_2, F_3, F_4, F_5), \text{ with weights } w_i = N_{Z_i}^{-1}. 
    \end{align*}
    
    \end{enumerate}

    \subsection{Simulation results}

\begin{figure}[ht!]
    \centering
    \includegraphics[width=0.7\linewidth]{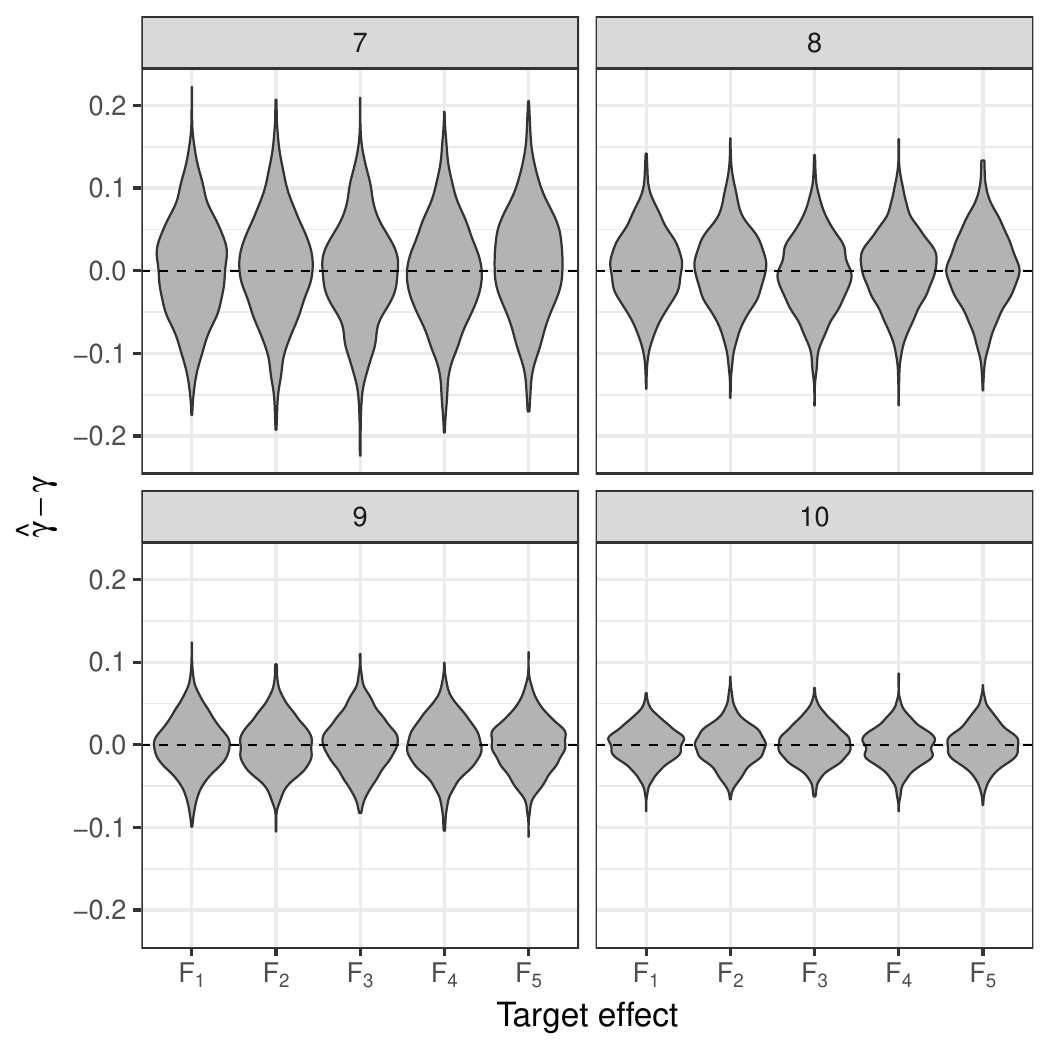}
    \caption{Violin plots of the differences between the estimators and true parameters for the five
    target effects for the four experiments with $K = 7,8,9,10$. }
    \label{fig:small-effects-K}
\end{figure}

    The Monte Carlo simulations are all repeated $1,000$ times. Figure \ref{fig:small-effects-K} shows violin plots of the differences between the point estimates and the true parameters. Table \ref{tab:small-effects-K} compares the aforementioned variance estimators based on two criteria: coverage rate of $95\%$ confidence intervals and rejection rate of the null hypothese that the main effects are zero, which corresponds to the ``Coverage'' column and the ``Power'' column, respectively. 
    
    
     
    From Figure \ref{fig:small-effects-K}, we can see that, even in a general design where the treatment group sizes vary greatly, the point estimates are centered around the truth and asymptotic normality holds. Besides, the variance shrinks as the sample size increases. 
    
    From Table \ref{tab:small-effects-K}, we can see that all three variance estimation strategies lead to valid type I error control and have increasing power as the sample size grows. When $K=7$, the EHW strategies give less conservative variance estimators and higher power because there is only one nonzero effect ($F_6$) that the regression does not capture. When $K$ grows and more model misspecification occurs, the pairing strategy guarantees sharper coverage and better power while WLS becomes slightly more conservative. This is because the between-group variation induced by grouping tends to be smaller with finer groups (see Theorem \ref{thm:unreplicated-var} and the relevant discussion). For the sandwich variance estimator, including more terms in the regression can mitigate the conservativeness.


\begin{table}[ht!]
\centering
\caption{Coverage and power results based on the variance estimation strategies with different number of factors}
\label{tab:small-effects-K}
\begin{tabular}{cccccccc}
\toprule 
\multirow{2}{*}{K}  & \multirow{2}{*}{Effect} & \multicolumn{3}{c}{Coverage}            & \multicolumn{3}{c}{Power}               \\ \cmidrule{3-8} 
                    &                         & LEX   & $\text{EHW}_0$ & $\text{EHW}_1$ & LEX   & $\text{EHW}_0$ & $\text{EHW}_1$ \\ \midrule
\multirow{5}{*}{7}  & $F_1$                      & 0.981 & 0.978          & 0.968          & 0.019 & 0.022          & 0.032          \\
                    & $F_2$                      & 0.967 & 0.963          & 0.951          & 0.591 & 0.604          & 0.636          \\
                    & $F_3$                      & 0.971 & 0.970          & 0.957          & 0.864 & 0.866          & 0.886          \\
                    & $F_4$                      & 0.969 & 0.969          & 0.959          & 0.031 & 0.031          & 0.041          \\
                    & $F_5$                      & 0.966 & 0.965          & 0.956          & 0.907 & 0.910          & 0.920          \\ \midrule
\multirow{5}{*}{8}  & $F_1$                      & 0.975 & 0.985          & 0.981          & 0.025 & 0.015          & 0.019          \\
                    & $F_2$                      & 0.966 & 0.982          & 0.976          & 0.889 & 0.838          & 0.863          \\
                    & $F_3$                      & 0.971 & 0.982          & 0.980          & 0.987 & 0.978          & 0.981          \\
                    & $F_4$                      & 0.971 & 0.986          & 0.979          & 0.029 & 0.014          & 0.021          \\
                    & $F_5$                      & 0.963 & 0.979          & 0.975          & 0.995 & 0.990          & 0.990          \\ \midrule
\multirow{5}{*}{9}  & $F_1$                      & 0.960 & 0.978          & 0.967          & 0.040 & 0.022          & 0.033          \\
                    & $F_2$                      & 0.976 & 0.986          & 0.983          & 0.998 & 0.996          & 0.997          \\
                    & $F_3$                      & 0.975 & 0.984          & 0.983          & 1.000 & 1.000          & 1.000          \\
                    & $F_4$                      & 0.967 & 0.979          & 0.972          & 0.033 & 0.021          & 0.028          \\
                    & $F_5$                      & 0.975 & 0.988          & 0.985          & 1.000 & 1.000          & 1.000          \\ \midrule 
\multirow{5}{*}{10} & $F_1$                      & 0.981 & 0.993          & 0.990          & 0.019 & 0.007          & 0.010          \\
                    & $F_2$                      & 0.970 & 0.984          & 0.982          & 1.000 & 1.000          & 1.000          \\
                    & $F_3$                      & 0.974 & 0.989          & 0.986          & 1.000 & 1.000          & 1.000          \\
                    & $F_4$                      & 0.974 & 0.985          & 0.982          & 0.026 & 0.015          & 0.018          \\
                    & $F_5$                      & 0.969 & 0.986          & 0.984          & 1.000 & 1.000          & 1.000          \\ \bottomrule 
\end{tabular}%
\end{table}

\section{General combinatorial Berry--Esseen bounds for linear permutation statistics}\label{sec:general-BE}

Appendix \ref{sec:general-BE} presents general BEBs on multivariate linear permutation statistics. Section \ref{sec:formulation} provides a unified formulation for linear permutation statistics, which includes the point estimates in the main paper as a special case. Section \ref{sec:linear-projection} discusses BEBs for linear projections of multivariate permutation statistics. Section \ref{sec:fine-BE} provides dimension-dependent BEBs over convex sets, which are the basic tools for proving the BEBs for the quadratic forms of linear permutation statistics.

In addition to the notation used in the main paper, we need additional notation for the rest of the supplementary material. 
For a  positive integer $N$, let $\bbS_N$ denote the set of permutations over $[N]$. We use $\pi \in \bbS_N$ to denote a permutation, which is a bijection from $[N]$ to $[N]$ with $\pi(i)$ denoting the integer on index $i$ after permutation. We also use the same notation $\pi$ to denote a random permutation, which is uniformly distributed over $\bbS_N$.

For a matrix $M=(M(h,l))\in\bbR^{H\times H}$, define its column, row and all-entry sums as  
$$
{M}(+, l) = \sum_{h=1}^H M(h, l),\quad
{M}(h,+) = \sum_{l=1}^H M(h,l),\quad 
M(+,+) = \sum_{h=1}^H\sum_{l=1}^H M(h,l),
$$
respectively.
For two matrices $M,M'\in\bbR^{H\times H}$, define the trace inner product as 
$$
\tr{M}{M'} = \operatorname{trace}(M^\top M') = \sum_{h=1}^H\sum_{l=1}^H M(h,l)M'(h,l).
$$
Vectorize $M$ as $\myvec{M}$ by stacking its column vectors. We will use the following basic result on matrix norms: 
\begin{align}\label{eqn:op-inf}
    \|M\|_{\operatorname{op}} \le {H}\|M\|_{\infty}.
\end{align}

\subsection{Multivariate permutation statistics}\label{sec:formulation}

To analyze estimates of the form \eqref{eqn:estimates}, we need a general formulation of multivariate permutation statistics. Let $P\in\bbR^{N}$ be a random permutation matrix, which is obtained by randomly permuting the columns (or rows) of the identity matrix $I_N$. Also define $M_1,\ldots,M_H$ as $H$ deterministic $N \times N$ matrices.
We want to study the random vector \citep{chatterjee2008multivariate} 
\begin{align}\label{eqn:rv-target}
    \Gamma = \left(\trace{M_1 P},\ldots, \trace{M_H P}\right)^\top.
\end{align}

Each random permutation matrix $P$ can also be represented by a random permutation $\pi$.
Then  
\begin{align*}
    \trace{M_h P} = \sum_{i=1}^N M_h(i,\pi(i)), \qquad (h=1,\ldots,H).
\end{align*}

Example \ref{exp:revisit-CR} below  revisits complete randomization.

\begin{example}[Revisiting complete randomization]\label{exp:revisit-CR}
Under complete randomization, the treatment vector ${Z} = (Z_1,\cdots, Z_N)$ has a correspondence with $P$. As a toy example, consider an experiment with $Q = 2$, $N_1 = 1$ and $N_2 = 2$. One can label the rows and columns of $P$ as follows:
\begin{align*}
    \bordermatrix{~  & i=1 & i=2 & i=3 \cr
     q=1 & 0   & 1   & 0   \cr
     q=2 & 1   & 0   & 0   \cr
     q=2 & 0   & 0   & 1}.
\end{align*}
The pattern of $1$'s indicates exactly the treatment allocation.
Generally, if we let the rows of $P$ represent the treatment arms and view the columns as indicator vectors of individuals, a permutation over the columns means a pattern of treatment allocation for all units. We can use \eqref{eqn:rv-target} to reformulate the sample mean vector $\hY$ as $ \Gamma  = (\Gamma_1, \ldots, \Gamma_Q)^\top$, where $\Gamma_q = \trace{M_q P}$ with
\begin{align}\label{eqn:pop}
M_q = 
\bordermatrix{~ &         Z = 1               & \cdots    &  Z = q                                      & { \cdots }           &  Z=Q    \cr
1        & 0^\top_{N_{1}}  & \cdots  & N_{q}^{-1}Y_1(q)\cdot 1^\top_{N_{q}}  &  \cdots  &  0^\top_{N_{Q}}  \cr
2        & 0^\top_{N_{1}}  &  \cdots   &  N_{q}^{-1}Y_2(q)\cdot 1^\top_{N_{q}}  &   \cdots  &  0^\top_{N_{Q}}  \cr 
\cdots &  \cdots                   &  \cdots                                      &  \cdots  &  \cdots  &  \cdots \cr 
N        & 0^\top_{N_{1}}  &  \cdots   &  N_{q}^{-1}Y_N(q)\cdot 1^\top_{N_{q}}  &   \cdots  &  0^\top_{N_{Q}}  \cr} .
\end{align}
\end{example}

Lemma \ref{lem:mean-var} below gives the mean and covariance of $\Gamma$:
\begin{lemma}[Mean and covariance of $\Gamma$]\label{lem:mean-var} 
\quad
\begin{enumerate}[label = (\roman*)]
\item For random permutation matrix $P$, we have
\begin{gather}
    \bbE\{P(\cdot, i)\} = \frac{1}{N}  1_N,\quad \bbE\{P(\cdot, i) P(\cdot, i)^\top\} = \frac{1}{N}I_N\label{eqn:EP}
    \end{gather}
for all $i$,    and
    \begin{gather}
\bbE\{P(\cdot, i)P(\cdot, j)^\top\} = \frac{1}{N(N-1)}( 1_{N\times N} - I_N) \label{eqn:EPiPj}
\end{gather}
for $  i\neq j$. 
\item For the random vector $\Gamma$ defined in \eqref{eqn:rv-target}, we have
\begin{align}
    \bbE\{\Gamma_h\} = \frac{1}{N}\sum_{i=1}^N\sum_{j=1}^NM_h(i,j)\label{eqn:EGamma}
    \end{align}
for all $h$,    and
    \begin{align}
  \bbE\{\Gamma_h\Gamma_l\} &= \frac{1}{N-1}\tr{M_h}{M_l} + \frac{1}{N(N-1)}{M}_h(+,+){M}_l(+,+)\notag\\
    & - \frac{1}{N(N-1)} \sum_{k=1}^N M_h(+,k) {M}_l(+,k) - \frac{1}{N(N-1)} \sum_{k=1}^N {M}_h(k,+) {M}_l(k,+).\label{eqn:EGammakl}
\end{align}
for $h\neq l$.
\end{enumerate}
\end{lemma}

Special cases of Lemma \ref{lem:mean-var} have appeared in some previous works under certain simplifications. For example, \cite{hoeffding1951combinatorial} computed the mean and variance for scalar $\Gamma$ with $H=1$. \cite{chatterjee2008multivariate} did the calculation under the conditions of zero row and column sums as well as orthogonality of the population matrices. \cite{bolthausen1993rate} relaxed the constraints of orthogonality and presented the covariance formula only under the zero column sum condition. 

As an application, we can obtain the mean and covariance of $\hY$:
\begin{example}[Mean and covariance matrix of $\hY$]\label{exp:multi-avg}
Based on \eqref{eqn:pop}, we can  verify  that 
\begin{align*}
    \tr{M_q}{M_l} = 0, \text{ if } q\neq l.
\end{align*}
Using \eqref{eqn:EGamma}, we can compute 
\begin{align*}
    \bbE\{\Gamma_q\} = \frac{1}{N}\sum_{i=1}^N Y_i(q),
\end{align*}
and
\begin{gather*}
    \bbE\{(\Gamma_q - \bbE\Gamma_q)^2\} = \lt(\frac{1}{N_q} - \frac{1}{N}\rt)S(q,q), \\
    \bbE\{(\Gamma_q - \bbE\Gamma_q)(\Gamma_l - \bbE\Gamma_l)\} =  - \frac{1}{N}S(q,l).
\end{gather*}

\end{example}

From now on, for ease of discussion, we assume Condition \ref{cond:str-Mk} below:
\begin{condition}[Standardized orthogonal structure of $M_h$'s]\label{cond:str-Mk}
For each $h\in[H]$, the row and column sums of $M_h$ are zero and 
\begin{align*}
    \operatorname{Tr}(M_h^\top M_h) = N - 1.
\end{align*}
The $M_h$'s are mutually orthogonal with respect to the trace inner product:
\begin{align*}
    \operatorname{Tr}(M_h^\top M_l) = 0, \text{ for } h\neq l.
\end{align*}
\end{condition}

Lemma \ref{lem:reformulate} below ensures that imposing Condition \ref{cond:str-Mk} causes no loss of generality.

\begin{lemma}[Reformulation of the multivariate permutation statistics]\label{lem:reformulate}

Let $P$ be a random $N\times N$ permutation matrix, and $M_1,\ldots,M_H$ be $H$ deterministic $N\times N$ matrices. 
Let $\bbE\{\Gamma\}, V = \COV{\Gamma}, V^\star = \operatorname{Corr}(\Gamma)$ be respectively the expectation, covariance and correlation of $\Gamma$ defined in \eqref{eqn:rv-target}. Let $\tilde{V} = V^{-1/2}$. Define the $\{M'_h\}_{h=1}^H$ as
$$
  M'_h(i,j) = {M}_h(i,j) - N^{-1}{M_h}(i,+) - N^{-1}{M_h}(+,j) + N^{-2} {M_h}(+,+),
  $$
  and then define the $\{M''_h\}_{h=1}^H$ as 
  $$
     M''_h(i,j) = \sum_{l=1}^H \tilde{V}_{hl}M'_l(i,j) .
     $$
%

(i) ${M}''_1, \ldots,  {M}''_H$
satisfy Condition \ref{cond:str-Mk} and
\begin{align*}
V^{-1/2}(\Gamma - \bbE\{\Gamma\}) = \lt(\operatorname{Tr}({M}''_1P), \ldots, \operatorname{Tr}({M}''_HP)\rt)^\top.
\end{align*}

(ii) We have 
\begin{align}\label{eqn:standardM-bd}
    \max_{h\in[H]}\max_{i,j\in[N]} | {M}''_h(i,j)| \le \varrho_{\min}(V)^{-1/2} \sqrt{H} \max_{h\in[H]}\max_{i,j\in[N]} |M'_h(i,j)|.
\end{align}

\end{lemma}

\subsection{BEBs for linear projections of multivariate linear permutation statistics}\label{sec:linear-projection}

In this subsection, we establish BEBs for linear permutation statistics. 
\cite{Bolthausen1984AnEO} established a BEB for univariate permutation statistics, which is a basic tool for our proofs.

\begin{lemma}[Main theorem of \cite{Bolthausen1984AnEO}]\label{lem:bolthausen-1984}
There exists an absolute constant $C>0$, such that
\begin{align*}
    \sup_{t\in\bbR} |\bbP\{\Gamma_1 \le t\} - \Phi(t)| \le \frac{C}{N}\sum_{i,j\in[N]} |M_1(i,j)|^3.
\end{align*}
\end{lemma}

We can use Lemma \ref{lem:bolthausen-1984} to prove Theorem \ref{thm:linear-projection} below. 

\begin{theorem}\label{thm:linear-projection}
Assume  Condition \ref{cond:str-Mk}. Let $b\in\bbR^H$ be a vector with $\|b\|_2 = 1$. Then there exists an absolute constant $C > 0$, such that
\begin{align*}
    \sup_{t\in\bbR}|\bbP\{b^\top\Gamma \le t\} - \Phi(t)| \le C{\max_{i,j\in[N]} \lt|\sum_{h=1}^Hb_hM_h(i,j)\rt|}.
\end{align*}

\end{theorem}

The proof of Theorem \ref{thm:linear-projection} is straightforward based on Lemma \ref{lem:bolthausen-1984}. It is more interesting  to compute the upper bound in specific examples, which we will do in Appendix \ref{sec:additional}.  Theorem \ref{thm:linear-projection} is a finite-sample result. It implies a CLT when the upper bound vanishes:
\begin{align}\label{eqn:BE-to-clt}
    \max_{i,j\in[N]} \lt|\sum_{h=1}^Hb_hM_h(i,j)\rt| \to 0, \text{ as } N\to\infty.
\end{align}
We can further upper bound the left hand side of \eqref{eqn:BE-to-clt}:
\begin{align*}
    \max_{i,j\in[N]} \lt|\sum_{h=1}^Hb_hM_h(i,j)\rt| \le \max_{i,j\in[N],h\in[H]} |M_h(i,j)| \cdot \|b\|_1 \le \sqrt{H}\max_{i,j\in[N],h\in[H]} |M_h(i,j)|.
\end{align*}
Hence  Theorem \ref{thm:linear-projection} reveals a trade-off between $H$ and $\max_{i,j\in[N],h\in[H]} |M_h(i,j)|$. Alternatively, we can use the Cauchy-Schwarz inequality to obtain another bound:
\begin{align*}
    \max_{i,j\in[N]} \lt|\sum_{h=1}^Hb_hM_h(i,j)\rt| \le \max_{i,j\in[N]} \lt\{\sum_{h=1}^H|M_h(i,j)|^2\rt\}^{1/2} \cdot \|b\|_2 \le \max_{i,j\in[N]} \lt\{\sum_{h=1}^H|M_h(i,j)|^2\rt\}^{1/2}.
\end{align*}
which can be a better bound for some $M_q$'s.

Besides, the combinatorial CLT of \citet[][Theorem 3]{hoeffding1951combinatorial} establishes the following sufficient condition for $b^\top\Gamma$ converging to a standard Normal distribution: 
\begin{lemma}[Combinatorial CLT by Theorem 3 of \cite{hoeffding1951combinatorial}]
$b^\top\Gamma$ is asymptotically Normal if 
\begin{align}\label{eqn:hoeffding-cond}
    \frac{\max_{i,j\in[N]} \lt\{\sum_{h=1}^Hb_hM_h(i,j)\rt\}^2}{N^{-1}\sum_{i,j\in[N]}\lt\{\sum_{h=1}^Hb_hM_h(i,j)\rt\}^2} \to 0.
\end{align}
\end{lemma}

Under Condition \ref{cond:str-Mk}, we have
\begin{align*}
     \sum_{i,j\in[N]}\lt\{\sum_{h=1}^Hb_hM_h(i,j)\rt\}^2 = N-1.
\end{align*}
Hence \eqref{eqn:hoeffding-cond} is equivalent to \eqref{eqn:BE-to-clt}. But Theorem \ref{thm:linear-projection} is stronger because \eqref{eqn:hoeffding-cond} implies not only convergence in distribution but also an upper bound on the convergence rate in the Kolmogorov distance.

\subsection{A permutational BEB over convex sets} \label{sec:fine-BE}

With independent random variables, the BEBs over convex sets match the optimal rate $N^{1/2}$ \citep{nagaev1976estimate, bentkus2005lyapunov}. 
We achieve the same order for linear permutation statistics by using a result based on Stein's method \citep{fang2015rates}.

\begin{definition}[Exchangeable pair]
\label{def::exchangeable-pair}
$(\Gamma,\Gamma')$ is an exchangeable pair if  $(\Gamma,\Gamma')$ and $(\Gamma',\Gamma)$ have the same distribution.
\end{definition}

\begin{definition}[Stein coupling, Definition 2.1 of \cite{fang2015rates}]\label{def:stein-coupling}
A triple of square integrable $H$-dimensional random vectors $ (\Gamma,\Gamma',G) $ is called a $H$-dimensional Stein coupling if 
\begin{align*}
    \bbE\{G^\top f(\Gamma') - G^\top f(\Gamma)\} = \bbE\{\Gamma^\top f(\Gamma)\}
\end{align*}
for all $f: \bbR^H \to  \bbR^H$ provided that the expectations exist.
\end{definition}

\citet[][Remark 2.3]{fang2015rates} made a connection between Definitions \ref{def::exchangeable-pair} and \ref{def:stein-coupling}, shown below.

\begin{lemma}
[Remark 2.3 of \cite{fang2015rates}]
\label{lemma-coupling-1}
If $(\Gamma,\Gamma')$ is an exchangeable pair and
$
    \bbE(\Gamma' - \Gamma\mid \Gamma) = -\Lambda \Gamma
$
for some invertible $\Lambda$, then $(\Gamma,\Gamma',\frac{1}{2}\Lambda^{-1}(\Gamma'-\Gamma))$ is a Stein coupling. 
\end{lemma}


\cite{fang2015rates} established the following BEB based on multivariate Stein coupling.

\begin{lemma}[Theorem 2.1 of \cite{fang2015rates}]\label{lem:bounded-pair}
Let $(\Gamma, \Gamma', G)$ be a $H$-dimensional Stein coupling. Assume $\operatorname{Cov}(\Gamma) = {I}_H$. Let $\xi_H$ be an $H$-dimensional standard Normal random vector. With $D = \Gamma' - \Gamma$, suppose that there are positive constants $\alpha$ and $\beta$ such that
$\|G\|_2\le\alpha$ and $\|D\|_2 \le \beta$. Let $\cA$ be the collection of all Borel measurable convex sets. 
Then there exists a universal constant $C > 0$, such that 
\begin{align*}
    &\sup_{A\in\cA}|\bbP\{\Gamma\in A\} - \bbP\{\xi_H \in A\}| \\
    &\le C(H^{7/4} \alpha \mathbb{E}\|D\|_2^2 + H^{1/4}\beta + H^{7/8}\alpha^{1/2}B_1^{1/2} + H^{3/8}B_2 + H^{1/8}B_3^{1/2}),
\end{align*}
where 
\begin{gather*}
    B_1^2 = \COV{\bbE(\|D\|_2^2\mid \Gamma)}, \\
    B_2^2 = \sum_{h=1}^H\sum_{l=1}^H \COV{\bbE(G_hD_l\mid \Gamma)}, \\
    B_3^2 = \sum_{h=1}^H\sum_{l=1}^H\sum_{m=1}^H \COV{\bbE(G_hD_lD_m\mid \Gamma)}.
\end{gather*}
 
\end{lemma}

Our construction of exchangeable pairs for linear permutation statistics is motivated by \cite{chatterjee2007multivariate}.
For $\Gamma$, construct a coupling random vector $\Gamma'$ by performing a random transposition to the original pattern of permutation. Here a random transposition is defined as follows:
\begin{definition}[Random transposition]\label{def:transposition}
The set of transpositions $\bbT_N = \{(t_1t_2)\}$ is defined as the subset of permutations over $[N]$ which only switches two indices $t_1$ and $t_2$ among $\{1,\ldots,N\}$ while keeping the others fixed. 
A random transposition $\tau$  is a uniform distribution on $\bbT_N$.
\end{definition}

If a random transposition $\tau$ and a random permutation $\pi$ are independent, their composite $\pi' = \tau\circ \pi$ is also a random permutation over $[N]$. 
As we discussed in Section \ref{sec:formulation}, $\pi$ and $\pi'$ can be represented as random permutation matrices $P$ and $P'$. Let
\begin{align}\label{eqn:target-vec-prime}
    \Gamma' = \lt(\operatorname{Tr}(M_1 P'),\ldots, \operatorname{Tr}(M_H P') \rt)^\top.
\end{align}
Now $(\Gamma, \Gamma')$ is an exchangeable pair and has the following basic property.

\begin{lemma}
[Lemma 8 of \cite{chatterjee2007multivariate}] \label{lemma-coupling-2}
$ \bbE\{\Gamma'-\Gamma\mid \pi\} = -\frac{2}{N-1}\Gamma$.
\end{lemma} 

By Lemmas \ref{lemma-coupling-1} and \ref{lemma-coupling-2}, $(\Gamma,\Gamma',-\frac{N-1}{4}(\Gamma-\Gamma'))$ with $G = -\frac{N-1}{4}(\Gamma-\Gamma')$ is a Stein coupling.


We prove the following result based on Lemma \ref{lem:bounded-pair}:
\begin{theorem}[Permutational BEB over convex sets]\label{thm:be-bounded}
Assume $|M_h(i,j)|\le B_N$ for $h\in[H]$ and $i,j\in[N]$. Assume  Condition \ref{cond:str-Mk}. Then there exists a universal constant $C>0$, such that
\begin{align}\label{eqn:be-bounded}
    &\sup_{A\in\cA}|\bbP\{\Gamma\in A\} - \bbP\{\xi_H \in A\}| \\
    &\le CH^{13/4}NB_N(B_N^2 + N^{-1}) + C H^{3/4}B_N \\
    &+ CH^{13/8}N^{1/4}B_N^{3/2} 
    + CH^{11/8}N^{1/2}B_N^2. \notag
\end{align}
When $B_N = O(N^{-1/2})$, the upper bound \eqref{eqn:be-bounded} becomes
\begin{align}\label{eqn:special-case}
\sup_{A\in\cA}|\bbP\{\Gamma\in A\} - \bbP\{\xi_H \in A\}| \le \frac{CH^{13/4}}{{N}^{1/2}}.
\end{align}
\end{theorem}

To end this subsection, we briefly comment on the literature of multivariate permutational BEBs and make a comparison between the existing results and Theorem \ref{thm:be-bounded}. \cite{bolthausen1993rate} proved a multivariate permutational BEB under some conditions, but their bound did not specify explicit dependence on the dimension ($H$ in our notation). \cite{chatterjee2007multivariate} proposed methods based on exchangeable pairs for multivariate normal approximation and applied them to permutation distributions. However, their methods only allow them to establish the following result:
\begin{align*}
    \sup_{g\in C^2(\bbR^H)}|\E{g(\Gamma)} - \E{g(\xi_H)}| \le \frac{CH^3}{N^{1/2}},
\end{align*}
where $C^2(\bbR^H)$ represents the collection of second-order continuously differentiable functions on $\bbR^H$. While the rate over $H$ is slightly better than \eqref{eqn:special-case}, the function class $C^2(\bbR^H)$ cannot cover the indicator functions. \cite{raic2015multivariate} conjectured the following result:
\begin{align}\label{eqn:raic}
    \sup_{A\in\cA}|\bbP\{\Gamma\in A\} - \bbP\{\xi_H\in A\}| \le C\frac{H^{1/4}}{N}\sum_{i\in[N]}\sum_{j\in[N]}\lt(\sum_{h\in[H]}M_h(i,j)^2\rt)^{3/2}.
\end{align}
When $B_N = O(N^{-1/2})$, \eqref{eqn:raic} has order $O(H^{7/4}N^{-1/2})$. However, \cite{raic2015multivariate} did not provide any proof for \eqref{eqn:raic}. \cite{wang2022rerandomization} proved a BEB for treatment-control randomized experiment using the coupling method, with the dependence on $N$ being slower than $N^{-1/2}$.  The dependence on $H$ may be further improved but it is beyond the scope of the current work.


\section{Proofs of the results in Appendix \ref{sec:general-BE}}\label{sec::proof-linear-permutational}

In this section, we prove the results in Appendix \ref{sec:general-BE}. Section \ref{sec:pre-lemma} presents several lemmas that are essential to the proofs. The main proofs start from Section \ref{sec:pf-linear-proj}.

\subsection{Lemmas}\label{sec:pre-lemma}

%

 Lemma \ref{lem:cond-check}  below gives the conditional moments of the exchangeable pair $(\Gamma,\Gamma')$ constructed in \eqref{eqn:rv-target} and \eqref{eqn:target-vec-prime}.

\begin{lemma}[Lemma 8 in \cite{chatterjee2007multivariate}]\label{lem:cond-check}
Construct an exchangeable pair $(\Gamma, \Gamma')$ based on \eqref{eqn:rv-target} and \eqref{eqn:target-vec-prime}.
\begin{enumerate}[label = (\roman*)]
\item We restate Lemma \ref{lemma-coupling-2}: 
\begin{align*}
    \bbE\{\Gamma'-\Gamma\mid \pi\} = -\frac{2}{N-1}\Gamma . 
\end{align*}

\item For the $h$-th coordinate $(\Gamma_h, \Gamma_h')$, we have
\begin{align*}
    \bbE\{(\Gamma_h - \Gamma_h')^2\mid\pi\} 
    &= \frac{2(N+1)}{N(N-1)}\sum_{i=1}^N M_h(i,\pi(i))^2 + \frac{2}{N} + \frac{2}{N(N-1)}\Gamma_h^2 \\
    &+ \frac{2}{N(N-1)}\sum_{i\neq j} M_h(i,\pi(j))M_h(j,\pi(i)).
\end{align*}

\item For the $h$-th coordinate $(\Gamma_h, \Gamma_h')$ and $l$-th coordinate  $(\Gamma_l, \Gamma_l')$, we have
\begin{align*}
    \bbE\{(\Gamma_h - \Gamma_h')(\Gamma_l - \Gamma_l')\mid\pi\} 
    &= \frac{2(N+1)}{N(N-1)}\sum_{i=1}^N M_h(i,\pi(i))M_l(i,\pi(i)) + \frac{2}{N(N-1)}\Gamma_h\Gamma_l \\
    &+ \frac{2}{N(N-1)}\sum_{i\neq j} M_h(i,\pi(j))M_l(j,\pi(i)).
\end{align*}
\end{enumerate}

\end{lemma}

 Lemma  \ref{lem:key-var-bds} below bounds the variances of linear permutation statistics.

\begin{lemma}\label{lem:key-var-bds} We have the following variance bounds for $N$ large enough. 
\begin{enumerate}[label = (\roman*)]
    \item $\Var{\sum_{h=1}^H  X_h}\le H\sum_{h=1}^H \COV{X_i}.$
    \item If $|M_0(i,j)|\le B_N$, then
    \begin{align}
        &\Var{\sum_{i=1}^N M_0(i,\pi(i))} \notag\\
        &= (N-1)^{-1}\sum_{i=1}^N\sum_{j=1}^N\{ M_0(i,\pi(i)) -  N^{-1}{M}_0(i,+) - N^{-1}{M}_0(+,j) + N^{-2}{M}_0(+,+)\}^2\notag\\
        &\le 32NB_N^2.\label{eqn:var-bd}
    \end{align}
    
    \item Suppose $M_1=(a_{ij})$ and $M_2=(b_{ij})$ have zero column and row sums.  If $|a_{ij}|\le B_N, |b_{ij}|\le B'_N$, then
    \begin{align*}
        \Var{\sum_{i\neq j}^N a_{i\pi(i)}b_{j\pi(j)}}\le 54 N^2B_N^2B_N'^2.
    \end{align*}
    
    \item Suppose $M_1=(a_{ij})$ and $M_2=(b_{ij})$ have zero column and row sums.  If $|a_{ij}|\le B_N, |b_{ij}|\le B'_N$, then  
    \begin{align*}
        \Var{\sum_{i\neq j}^N a_{i\pi(j)}b_{j\pi(i)} } \le 54 N^2B_N^2B_N'^2.
    \end{align*}
    
    \item Suppose $M_1=(a_{ij}), M_2=(b_{ij}), M_3=(c_{ij})$ all have zero column and row sums.  If $|a_{ij}|\le B_N, |b_{ij}|\le B'_N, |c_{ij}|\le B''_N$, then
    \begin{align*}
        \Var{\sum_{i\neq j}^N a_{i\pi(i)}b_{j\pi(j)}c_{i\pi(j)} }\le 15N^3B_N^2B_N'^2B_N''^2.
    \end{align*}
    
    \item Suppose $M_1=(a_{ij}), M_2=(b_{ij}), M_3=(c_{ij})$ all have zero column and row sums.  If $|a_{ij}|\le B_N, |b_{ij}|\le B'_N, |c_{ij}|\le B''_N$, then 
    \begin{align*}
        \Var{\sum_{i\neq j}^N a_{i\pi(i)}b_{j\pi(i)}c_{i\pi(j)} } \le 15N^3B_N^2B_N'^2B_N''^2.
    \end{align*}
\end{enumerate}
\end{lemma}

\begin{proof}[Proof of Lemma \ref{lem:key-var-bds}]
\begin{enumerate}[label = (\roman*)]
    \item This is a standard result by the Cauchy-Schwarz inequality.
    \item This is due to the variance formula of linear permutation statistics. See Lemma \ref{lem:mean-var}.
    
    \item We calculate 
    \begin{align*}
        &\bbE\lt\{\sum_{i\neq j}a_{i\pi(i)}b_{j\pi(j)}\rt\}^2\\
        &= \bbE\lt\{\sum_{i,k}\sum_{j\neq i, l\neq k}a_{i\pi(i)}b_{j\pi(j)}a_{k\pi(k)}b_{l\pi(l)}\rt\}\\
        &= \frac{1}{N(N-1)}\sum_{i\neq j,m\neq n} \lt\{a_{im}^2b_{jn}^2 + a_{in}b_{in}a_{jm}b_{jm}\rt\}\\
        & + \frac{1}{N(N-1)(N-2)}\sum_{i\neq j\neq k}\sum_{m\neq n\neq o} \lt\{a_{im}^2b_{jn}b_{ko} + a_{im}a_{jn}b^2_{ko} + a_{im}b_{im}a_{jn}b_{ko} +
        a_{im}b_{jn}a_{ko}b_{ko}\rt\}\\
        & + 
        \frac{1}{N(N-1)(N-2)(N-3)}\sum_{i\neq j\neq k \neq l}\sum_{m\neq n\neq o\neq p} \lt\{
        a_{im}b_{jn}a_{ko}b_{lp}\rt\}\\
        & = \text{I} + \text{II} + \text{III}.
    \end{align*}
    
    For I, we have
    \begin{align}\label{eqn:3.I-bd}
        N(N-1)\text{I} \le N^2(N-1)^2 \cdot 2B_N^2B_N'^2 = {2N^2(N-1)^2B_N^2B_N'^2}.
    \end{align}
    
    For II, using the property of zero column and row sums, we have
    \begin{align}\label{eqn:3.II-bd}
        N(N-1)(N-2)\text{II} \le 16N^2(N-1)^2B_N^2B_N'^2.
    \end{align}
    To see why \eqref{eqn:3.II-bd} is true, consider the first part of the summation:
    \begin{align*}
        &\lt|\sum_{i\neq j\neq k}\sum_{m\neq n\neq o} a_{im}^2b_{jn}b_{ko}\rt| \\
        =&\lt|\sum_{i\neq j\neq k}\sum_{m\neq n} a_{im}^2b_{jn}(-b_{km}-b_{kn})\rt|\\
        =&\lt|\sum_{i\neq j}\sum_{m\neq n} a_{im}^2b_{jn}(b_{im}+b_{jm}+b_{in}+b_{jn})\rt|\\
        \le& 4N^2(N-1)^2B_N^2B_N'^2. 
    \end{align*}
    Similar results hold for the other parts of the summation. Adding terms together, we obtain \eqref{eqn:3.II-bd}.
    
    For III, use the zero column and row sums property again, we have 
    \begin{align}\label{eqn:3.III-bd}
        N(N-1)(N-2)(N-3) \text{III} \le 36N^2(N-1)^2B_N^2B_N'^2.
    \end{align}
    
    Summing up \eqref{eqn:3.I-bd}--\eqref{eqn:3.III-bd} to obtain 
    \begin{align*}
        \Var{\sum_{i\neq j}^N a_{i\pi(i)}b_{j\pi(j)} } \le 54 N^2B_N^2B_N'^2.
    \end{align*}
    
    \item We calculate 
    \begin{align*}
        &\bbE\lt\{\sum_{i\neq j}a_{i\pi(j)}b_{j\pi(i)}\rt\}^2\\
        &= \bbE\lt\{\sum_{i,k}\sum_{j\neq i, l\neq k}a_{i\pi(j)}b_{j\pi(i)}a_{k\pi(l)}b_{l\pi(k)}\rt\}\\
        &= \frac{1}{N(N-1)}\sum_{i\neq j,m\neq n} \lt\{a_{im}^2b_{jn}^2 + a_{im}b_{im}a_{jn}b_{jn}\rt\}\\
        & + \frac{1}{N(N-1)(N-2)}\sum_{i\neq j\neq k}\sum_{m\neq n\neq o} \lt\{a_{in}b_{jm}a_{io}b_{km} +  a_{in}a_{jo}b_{jm}b_{kn} +  a_{in}a_{km}b_{jm}b_{io} +
         a_{in}a_{kn}b_{jm}b_{jo}\rt\}\\
        & + 
        \frac{1}{N(N-1)(N-2)(N-3)}\sum_{i\neq j\neq k \neq l}\sum_{m\neq n\neq o\neq p} \lt\{
        a_{in}a_{kp}b_{jm}b_{lo}\rt\}\\
        & = \text{I} + \text{II} + \text{III}.
    \end{align*}
    The rest of the analysis is nearly identical to part (iii). We omit the details.
    
    \item We calculate 
    \begin{align*}
        &\bbE\lt\{\sum_{i\neq j}^N a_{i\pi(i)}b_{j\pi(j)}c_{i\pi(j)}\rt\}^2\\
        =&\bbE\lt\{\sum_{i\neq j}\sum_{k\neq l} a_{i\pi(i)}b_{j\pi(j)}c_{i\pi(j)}a_{k\pi(k)}b_{l\pi(l)}c_{k\pi(l)}\rt\}\\
        =&\frac{1}{N(N-1)}\sum_{i\neq j}\sum_{m\neq n}\{a^2_{im}b^2_{jn}c^2_{in} + a_{im}b_{jn}c_{in}a_{jn}b_{im}c_{jm}\} \\
        +& \frac{1}{N(N-1)(N-2)}\sum_{i\neq j\neq k}\sum_{m\neq n\neq o}
        \{a_{im}^2b_{jn}c_{in}b_{ko}c_{io}+
        a_{im}b_{jn}c_{in}a_{ko}b_{im}c_{km}\\ &\phantom{\frac{1}{N(N-1)(N-2)}\sum_{i\neq j\neq k}\sum_{m\neq n\neq o}}+
        a_{im}b_{jn}c_{in}a_{jn}b_{ko}c_{jo} +
        a_{im}b_{jn}c_{in}a_{ko}b_{jn}c_{kn}
        \}\\
        +& \frac{1}{N(N-1)(N-2)(N-3)}\sum_{i\neq j\neq k \neq l}\sum_{m\neq n\neq o\neq p} a_{im}b_{jn}c_{in}a_{ko}b_{lp}c_{kp}\\
        =& \text{I} + \text{II} + \text{III}.
    \end{align*}
    For I, using the triangle inequality, we have
    \begin{align}\label{eqn:5.I-bd}
        N(N-1) \text{I} \le 2N^2(N-1)^2B_N^2B_N'^2B_N''^2.
    \end{align}
    For II, using the triangle inequality, we have
    \begin{align}\label{eqn:5.II-bd}
        N(N-1)(N-2) \text{II} \le 4N^2(N-1)^2(N-2)^2B_N^2B_N'^2B_N''^2.
    \end{align}
    For III, expanding along the indices $l$ and $o$, we have
    \begin{align}\label{eqn:5.III-bd}
        N(N-1)(N-2)(N-3) \text{III} \le 9N^6B_N^2B_N'^2B_N''^2.
    \end{align}
    
        Sum up \eqref{eqn:5.I-bd}--\eqref{eqn:5.III-bd} to get the final result.

    \item We calculate 
    \begin{align*}
        &\bbE\lt\{\sum_{i\neq j}^N a_{i\pi(i)}b_{j\pi(i)}c_{i\pi(j)}\rt\}^2\\
        =&\bbE\lt\{\sum_{i\neq j}\sum_{k\neq l} a_{i\pi(i)}b_{j\pi(i)}c_{i\pi(j)}a_{k\pi(k)}b_{l\pi(k)}c_{k\pi(l)}\rt\}\\
        =&\frac{1}{N(N-1)}\sum_{i\neq j}\sum_{m\neq n}\{a^2_{im}b^2_{jm}c^2_{in} + a_{im}b_{jm}c_{in}a_{jn}b_{in}c_{jm}\} \\
        +& \frac{1}{N(N-1)(N-2)}\sum_{i\neq j\neq k}\sum_{m\neq n\neq o}
        \{a_{im}^2b_{jm}c_{in}b_{km}c_{io}+
        a_{im}b_{jm}c_{in}a_{ko}b_{io}c_{km}\\ &\phantom{\frac{1}{N(N-1)(N-2)}\sum_{i\neq j\neq k}\sum_{m\neq n\neq o}}+
        a_{im}b_{jm}c_{in}a_{jn}b_{kn}c_{jo} +
        a_{im}b_{jm}c_{in}a_{ko}b_{jo}c_{kn}
        \}\\
        +& \frac{1}{N(N-1)(N-2)(N-3)}\sum_{i\neq j\neq k \neq l}\sum_{m\neq n\neq o\neq p} a_{im}b_{jm}c_{in}a_{ko}b_{lo}c_{kp}\\
        =& \text{I} + \text{II} + \text{III}.
    \end{align*}
    For I, using the triangle inequality, we have
    \begin{align}\label{eqn:6.I-bd}
        N(N-1) \text{I} \le 2N^2(N-1)^2B_N^2B_N'^2B_N''^2.
    \end{align}
    For II, using the triangle inequality, we have
    \begin{align}\label{eqn:6.II-bd}
        N(N-1)(N-2) \text{II} \le 4N^2(N-1)^2(N-2)^2B_N^2B_N'^2B_N''^2.
    \end{align}
    For III, expanding along the indices $l$ and $p$, we have
    \begin{align}\label{eqn:6.III-bd}
        N(N-1)(N-2)(N-3) \text{III} \le 9N^6B_N^2B_N'^2B_N''^2.
    \end{align}
    
    Sum up \eqref{eqn:6.I-bd}--\eqref{eqn:6.III-bd} to get the final result.
    
\end{enumerate}
\end{proof}

\subsection{Proof of Lemma \ref{lem:mean-var}}

The proof follows from combining the permutation distribution of $P$ with matrix algebra.
\begin{proof}[Proof of Lemma \ref{lem:mean-var}]

\textbf{Proof of \eqref{eqn:EP}.} Use the fact that each column of $P$ follows a uniform distribution over the canonical bases. 

\textbf{Proof of \eqref{eqn:EPiPj}.} Use the fact that $P(\cdot,i)P(\cdot,j)^\top$ is uniformly distributed over all the $N(N-1)$ off-diagonal positions.

\textbf{Proof of \eqref{eqn:EGamma}.} \eqref{eqn:EGamma} follows from Lemma \ref{lem:mean-var}(i) and the linearity of $\trace{\cdot}$.

\textbf{Proof of \eqref{eqn:EGammakl}.} For \eqref{eqn:EGammakl}, we have
\begingroup
\allowdisplaybreaks
\begin{align*}
    \bbE\{\Gamma_h\Gamma_l\} &= \bbE\lt\{\lt(\sum_{i=1}^N M_h(i,\cdot) P(\cdot, i)\rt)\lt(\sum_{i=1}^N M_l(i,\cdot) P(\cdot, i)\rt)\rt\} \\
    &= \bbE\lt\{\lt(\sum_{i=1}^N M_h(i,\cdot) P(\cdot, i)\rt)\lt(\sum_{i=1}^N M_l(i,\cdot) P(\cdot, i)\rt)\rt\} \\
    & = \bbE\lt\{ \sum_{i=1,j=1}^N M_h(i,\cdot)P(\cdot,i)P(\cdot,j)^\top M_l(j,\cdot)^\top \rt\} \\
    & = \frac{1}{N}\sum_{i=1}^N M_h(i,\cdot)M_l(i,\cdot)^\top + \frac{1}{N(N-1)} \sum_{i \neq j}^N M_h(i,\cdot) ( 1_{N\times N} - I_N) M_l(j,\cdot)^\top\\
    & = \frac{1}{N-1}\sum_{i=1}^NM_h(i,\cdot)M_l(i,\cdot)^\top + \frac{1}{N(N-1)}\sum_{i\neq j}^N M_h(i,\cdot) 1_{N\times N} M_l(j,\cdot)^\top\\
    & - \frac{1}{N(N-1)}\lt\{\sum_{i=1}^N M_h(i,\cdot)\rt\} \lt\{\sum_{i=1}^N M_l(i,\cdot)^\top\rt\} \\
    & = \frac{1}{N-1}\tr{M_h}{M_l} + \frac{1}{N(N-1)}\sum_{i\neq j} {M}_h(i,+){M}_l(j,+) - \frac{1}{N(N-1)} \sum_{k=1}^N{M}_h(+,k){M}_l(+,k)\\
    & = \frac{1}{N-1}\tr{M_h}{M_l} + \frac{1}{N(N-1)}\sum_{i=1,j=1}^N {M}_h(i,+){M}_l(j,+)\\
    & - \frac{1}{N(N-1)} \sum_{k=1}^N{M}_h(+,k){M}_l(+,k) - \frac{1}{N(N-1)} \sum_{k=1}^N{M}_h(k,+){M}_l(k,+)\\
    & = \frac{1}{N-1}\tr{M_h}{M_l} + \frac{1}{N(N-1)}{M}_h(+,+){M}_l(+,+)\\
    & - \frac{1}{N(N-1)} \sum_{k=1}^N{M}_h(+,k){M}_l(+,k) - \frac{1}{N(N-1)} \sum_{k=1}^N{M}_h(k,+){M}_l(k,+).
\end{align*}
\endgroup
\end{proof}

\subsection{Proof of Lemma \ref{lem:reformulate}}
\begin{proof}[Proof of Lemma \ref{lem:reformulate}]
(i)
By definition, 
\begin{align*}
{\Gamma_h - \bbE\{\Gamma_h\}} 
&= \sum_{i=1}^N {M}_h(i,\pi(i)) - N^{-1}\sum_{i=1}^N\sum_{j=1}^N {M}_h(i,j)\\
&= \sum_{i=1}^N V_{hh}^{-1/2}\lt\{{M}_h(i,\pi(i)) - N^{-1}{M_h}(i,+) - N^{-1}{M_h}(+,\pi(i)) + N^{-2} {M_h}(+,+)\rt\}.
\end{align*}
Now introduce a new matrix $M_h'$ with entries
\begin{align}\label{eqn:centering-1}
    {M}_h'(i,j) = {M}_h(i,j) - N^{-1}{M_h}(i,+) - N^{-1}{M_h}(+,j) + N^{-2} {M_h}(+,+).
\end{align}
Let $\tilde{V} = V^{-1/2}$ and let $\tilde{\Gamma} = \tilde{V}(\Gamma - \bbE\{\Gamma\})$  
with $\operatorname{Var}\{\tilde{\Gamma}\} = I_H, ~\bbE\{\tilde{\Gamma}\} = 0$.
Define $M''_h = \sum_{l=1}^H \tilde{V}_{hl}M'_l$. Because $M'_h$'s have zero row and column sums, we can verify that $M''_h$'s also satisfy:
\begin{align*}
    {M''_h}(i,+) = 0, ~{M''_h}(+,j) = 0, ~\forall~ i,j\in[N].
\end{align*}

Besides,
\begin{align*}
    \tilde{\Gamma}_h = \sum_{l=1}^H \tilde{V}_{hl}(\operatorname{Tr}(M'_lP) - \bbE\{\operatorname{Tr}(M'_lP)\}) = \operatorname{Tr}(M''_h P).
\end{align*}

Hence, combining Lemma \ref{lem:mean-var}, we have
\begin{align}\label{eqn:tGamma-hl-1}
    \bbE\{\tilde{\Gamma}_h\tilde{\Gamma}_l\} &= \frac{1}{N-1}\tr{M''_h}{M''_l} + \frac{1}{N(N-1)}{M''_h}(+,+){M''_l}(+,+)\notag\\
    & - \frac{1}{N(N-1)} \sum_{k=1}^N{M''_h}(+,k){M''_l}(+,k) - \frac{1}{N(N-1)} \sum_{k=1}^N{M''_h}(k,+){M''_l}(k,+)\\
    & = \frac{1}{N-1} \tr{M''_h}{M''_l}.
\end{align}
Recall 
\begin{align}\label{eqn:tGamma-hl-2}
    \E{\tilde{\Gamma}_h\tilde{\Gamma}_l} = \left\{
    \begin{array}{cc}
        1, & h=l; \\
        0, & h\neq l.
    \end{array}
    \right.
\end{align}
Combining \eqref{eqn:tGamma-hl-1} and \eqref{eqn:tGamma-hl-2}, we conclude
\begin{align*}
    \frac{1}{N-1} \tr{M''_h}{M''_l} = \left\{
    \begin{array}{cc}
        1, & h=l; \\
        0, & h\neq l.
    \end{array}
    \right.
\end{align*}
Therefore, Condition \ref{cond:str-Mk} holds for $M''_h$'s.

(ii)
For $i,j\in[N]$, define the vectors
\begin{align*}
    \boldsymbol{c}' = [M'_1(i,j),\ldots,M'_H(i,j)]^\top\in\bbR^H
\end{align*}
and
\begin{align*}
    \boldsymbol{c}'' = [M''_1(i,j),\ldots,M''_H(i,j)]^\top\in\bbR^H.
\end{align*}

We have
\begin{align*}
    \max_{h=1, \cdots, H} |M''_h(i,j)| \le \|\bc''\|_2 \le \varrho_{\min}(V)^{-1/2} \|\bc'\|_2 \le  \varrho_{\min}(V)^{-1/2} \sqrt{H} \max_{h=1,\ldots,H} |M'_h(i,j)|.
\end{align*}

\end{proof}

\subsection{Proof of Theorem \ref{thm:linear-projection}}\label{sec:pf-linear-proj}

Proving Theorem \ref{thm:linear-projection} reduces to checking the conditions of Lemma \ref{lem:bolthausen-1984}. 

\begin{proof}[Proof of Theorem \ref{thm:linear-projection}]
We have
\begin{align*}
    b^\top \Gamma = \sum_{h=1}^H b_h\trace{M_h P} = \trace{\lt(\sum_{h=1}^Hb_hM_h\rt) P}.
\end{align*}
Define
\begin{align*}
    M' = \sum_{h=1}^Hb_hM_h.
\end{align*}
We can verify  that the row sums and column sums of $M'$ are all zero. Also, using  Condition \ref{cond:str-Mk},
\begin{align}\label{eqn:tildeM-Fnorm}
    \innerprod{M'}{M'} = \sum_{h=1,l=1}^H b_hb_l\innerprod{M_h}{M_l} = (N-1)\sum_{h=1}^H b_h^2 = N-1.
\end{align}

Applying  Lemma \ref{lem:bolthausen-1984}, there exists an absolute constant $C>0$, such that
\begin{align*}
    \sup_{t\in\bbR}|\bbP\{b^\top\Gamma \le t\} - \Phi(t)| \le \frac{C}{N}  \sum_{i,j}|M'(i,j)|^3 \le \frac{C(N-1)}{N}\max_{i,j\in [N]}|M'(i,j)| \le C\max_{i,j\in[N]} |M'(i,j)|.
\end{align*}

\end{proof}

\subsection{Proof of Theorem \ref{thm:be-bounded}}

\begin{proof}[Proof of Theorem \ref{thm:be-bounded}]

We will apply Lemma \ref{lem:bounded-pair}.
The key step is to figure out the orders of $B_1,B_2,B_3$ in Lemma \ref{lem:bounded-pair}. One can upper bound $\Var{\bbE(\cdot\mid \Gamma)}$ by $\Var{\bbE(\cdot\mid \mathcal{F})}$
if $\sigma(\Gamma)\subset \mathcal{F}$.
This is a standard trick in Stein’s method and will be used without further mentioning.

Now we compute the quantities involved in Lemma \ref{lem:bounded-pair}. Recall we use the random transposition $\tau = (IJ)$ to construct exchangeable pairs. The $h$-th coordinate of $D = \Gamma' - \Gamma$ equals
\begin{align*}
    D_h = M_h(I,\pi(I)) + M_h(J,\pi(J)) - M_h(I,\pi(J)) - M_h(J,\pi(I)). 
\end{align*}
Hence 
\begin{align*}
   |D_h| \le 4 B_N, \quad |G_h| \le (N-1)B_N, \quad \|D\|_2 \le 4\sqrt{H}B_N, \quad \|G\|_2 \le (N-1)\sqrt{H}B_N.
\end{align*}

To apply Lemma \ref{lem:bounded-pair}, we need to bound the following quantities:
\begin{enumerate}[label = (\roman*)]
    \item $\bbE\{\|D\|_2^2\mid \pi\}$ and $\bbE\{\|D\|_2^2\}$.
    \item $B_1 = \sqrt{\Var{\bbE(\|D\|_2^2\mid \Gamma)}}$ and $B_2 = \sqrt{\sum_{k,l=1}^H \Var{\bbE(G_hD_l\mid \Gamma)}}$.
    \item $B_3 = \sqrt{\sum_{k,l,m=1}^H \Var{\bbE(G_hD_lD_m\mid \Gamma)}}$.
\end{enumerate}

\noindent\textbf{(i) Bound $\bbE\{\|D\|_2^2\mid \pi\}$ and $\bbE\{\|D\|_2^2\}$}.

By  Lemma \ref{lem:cond-check},
\begin{align*}
    \bbE\{\|D\|_2^2\mid \pi\} 
    &= \sum_{h=1}^H  \bbE\{D_h^2\mid\pi\} = \sum_{h=1}^H  \bbE\{(\Gamma_h - \Gamma_h')^2\mid\pi\}\\
    &= \frac{2(N+1)}{N(N-1)}\sum_{h=1}^H \sum_{i=1}^N M_h(i,\pi(i))^2 + \frac{2H}{N} + \frac{2}{N(N-1)}\sum_{h=1}^H \Gamma_h^2 \\
    &+ \frac{2}{N(N-1)}\sum_{h=1}^H \sum_{i\neq j} M_h(i,\pi(j))M_h(j,\pi(i))\\
    &\le \frac{2(N+1)HB_N^2}{N-1} + \frac{2H}{N} + \frac{2HNB_N^2}{(N-1)} + {2HB_N^2} \le 12HB_N^2 + \frac{2H}{N}. 
\end{align*}
This implies
\begin{align*}
    \bbE\{\|D\|_2^2\} = \bbE_{\pi}\bbE\{\|D\|_2^2\mid \pi\} \le 12HB^2_N + \frac{2H}{N}.
\end{align*}

\noindent\textbf{(ii) Bound $B_1$ and $B_2$}.

We prove the following result: there exists a universal constant $C>0$, such that
\begin{align*}
    B_1\le CHN^{-1/2}B_N^2, ~ B_2 \le CHN^{1/2}B_N^2.
\end{align*} 

By Lemma \ref{lem:cond-check}, 
\begin{align*}
    \bbE\{D_h^2\mid\pi\} 
    &= \frac{2(N+1)}{N(N-1)}\sum_{i=1}^N M_h(i,\pi(i))^2 + \frac{2}{N} + \frac{2}{N(N-1)}\Gamma_h^2 \\
    &+ \frac{2}{N(N-1)}\sum_{i\neq j} M_h(i,\pi(j))M_h(j,\pi(i))\\
    &= \frac{2(N+2)}{N(N-1)}\sum_{i=1}^N M_h(i,\pi(i))^2  + \frac{2}{N(N-1)}\sum_{i\neq j} M_h(i,\pi(j))M_h(i,\pi(j)) \\
    &+ \frac{2}{N(N-1)}\sum_{i\neq j} M_h(i,\pi(j))M_h(j,\pi(i))+ \frac{2}{N}\\
    & = \text{I} + \text{II} + \text{III} + \frac{2}{N}.
\end{align*}
For $h\neq l$,
\begin{align*}
    \bbE\{D_hD_l\mid\pi\} 
    &= \frac{2(N+1)}{N(N-1)}\sum_{i=1}^N M_h(i,\pi(i))M_l(i,\pi(i)) + \frac{2}{N(N-1)}\Gamma_h\Gamma_l \\
    &+ \frac{2}{N(N-1)}\sum_{i\neq j} M_h(i,\pi(j))M_l(j,\pi(i))\\
    & = \frac{2(N+2)}{N(N-1)}\sum_{i=1}^N M_h(i,\pi(i))M_l(i,\pi(i)) + \frac{2}{N(N-1)}\sum_{i\neq j} M_h(i,\pi(j))M_l(i,\pi(j)) \\
    &+ \frac{2}{N(N-1)}\sum_{i\neq j} M_h(i,\pi(j))M_l(j,\pi(i))\\
    & = \text{IV} + \text{V} + \text{VI}.
\end{align*}

For $B_1$, using Lemma \ref{lem:key-var-bds}(ii)--(iv), we have  
\begin{align*}
    \Var{ \text{I} } &\le \frac{4(N+2)^2}{N^2(N-1)^2}\cdot 32 N B_N^4 \le \frac{256B_N^4}{N},\\
    \Var{ \text{II} } &\le \frac{4}{N^2(N-1)^2}\cdot 54 N^2 B_N^4\le \frac{216B_N^4}{N},\\
    \Var{ \text{III} } &\le \frac{4}{N^2(N-1)^2}\cdot 54 N^2 B_N^4\le \frac{256B_N^4}{N}.\\
\end{align*}
Now apply Lemma \ref{lem:key-var-bds}(i) to obtain
\begin{align*}
    B_1^2 &= {\Var{\bbE(\|D\|_2^2\mid \Gamma)}}
    = {\Var{\bbE\lt(\sum_{h=1}^H  D_h^2\mid \Gamma\rt)}}
     \le {H\sum_{h=1}^H \Var{\bbE\lt( D_h^2\mid \Gamma\rt)}}
     \le CH^2N^{-1}B_N^4.
\end{align*}

Similarly, for $B_2$, we have
\begin{align*}
    B_2^2 = {\sum_{h,l=1}^H \Var{\bbE(G_hD_l\mid \Gamma)}}
    = \lt(\frac{N-1}{4}\rt)^2{\sum_{k,l=1}^H  \Var{\bbE\lt(D_hD_l\mid \Gamma\rt)}}
     \le CH^2NB_N^4.
\end{align*}

\noindent\textbf{(iii) Bound $B_3$}.

We prove the following result: there exists a universal constant $C>0$, such that
\begin{align*}
    B_3 \le C{H^{3/2}N^{1/2}B_N^3}
\end{align*}

For simplicity, we write $a_{ij} = M_h(i,j),b_{ij} = M_l(i,j)$ and $c_{ij} = M_m(i,j)$. Recall $G_h = (N-1)D_h/4 $. We have 
\begin{align}\label{eqn:DDD}
    \bbE\{D_hD_lD_m\mid\pi\} 
    = \bbE\{&
    (a_{I\pi(I)} + a_{J\pi(J)} -a_{I\pi(J)} -a_{J\pi(I)}) \\
    &\cdot (b_{I\pi(I)} + b_{J\pi(J)} -b_{I\pi(J)} -b_{J\pi(I)})\notag\\
    &\cdot (c_{I\pi(I)} + c_{J\pi(J)} -c_{I\pi(J)} -c_{J\pi(I)})\mid\pi\}.\notag
\end{align}
The expansion of \eqref{eqn:DDD} has $4^3=64$ terms, which can be characterized by the following categories of $a_{i_aj_a}b_{i_bj_b}c_{i_cj_c}$:
\begin{itemize}
    \item $a_{I\pi(I)}b_{I\pi(I)}c_{I\pi(I)}$. There are $2$ terms in total. We have
    \begin{align*}
        \bbE\{a_{I\pi(I)}b_{I\pi(I)}c_{I\pi(I)}\mid \pi\} = \frac{1}{N} \sum_{i=1}^N a_{i\pi(i)}b_{i\pi(i)}c_{i\pi(i)}.
    \end{align*}
    Because $|a_{i\pi(i)}b_{i\pi(i)}c_{i\pi(i)}|\le B_N^3$, by \eqref{eqn:var-bd}, we have
    \begin{align*}
        \Var{\bbE\{a_{I\pi(I)}b_{I\pi(I)}c_{I\pi(I)}\mid \pi\}} \le \frac{32NB_N^6}{N^2} = \frac{32B_N^6}{N}.
    \end{align*}

    \item $a_{I\pi(J)}b_{I\pi(J)}c_{I\pi(J)}$. There are $2$ terms in total. We have
    \begin{align}\label{eqn:second-ctg}
        \bbE\{a_{I\pi(J)}b_{I\pi(J)}c_{I\pi(J)}\mid \pi\} &= \frac{1}{N(N-1)} \sum_{i\neq j}^N a_{i\pi(j)}b_{i\pi(j)}c_{i\pi(j)} \notag\\
        &= \frac{1}{N(N-1)}\sum_{j=1}^N \sum_{i\neq j} a_{i\pi(j)}b_{i\pi(j)}c_{i\pi(j)}.
    \end{align}
    \eqref{eqn:second-ctg} can be viewed as univariate linear permutation statistics coming from a population matrix filled with entries that are identical on each row:
    \begin{align*}
       d_{kl} = \sum_{m\neq l} a_{ml}b_{ml}c_{ml}.
    \end{align*}
    Because $|d_{kl}| \le (N-1)B_N^3$, by Lemma \ref{lem:key-var-bds} (ii), we have
    \begin{align*}
        \Var{\bbE\{a_{I\pi(I)}b_{I\pi(I)}c_{I\pi(I)}\mid \pi\}} &\le \frac{16N \cdot \{(N-1)B_N^3\}^2}{N^2(N-1)^2} \le \frac{32B_N^6}{N}.
    \end{align*}

    \item $a_{I\pi(I)}b_{I\pi(I)}c_{J\pi(I)}$. There are $6$ terms in total. We have
    \begin{align*}
        \bbE\{a_{I\pi(I)}b_{I\pi(I)}c_{J\pi(I)}\mid \pi\} &= \frac{1}{N(N-1)} \sum_{i\neq j}^N a_{i\pi(i)}b_{i\pi(i)}c_{j\pi(i)}\\
        &= \frac{1}{N(N-1)} \sum_{i=1}^N\lt\{\sum_{j\neq i}  a_{i\pi(i)}b_{i\pi(i)}c_{j\pi(i)}\rt\}\\
        & = \frac{1}{N(N-1)} \sum_{i=1}^N   \{-a_{i\pi(i)}b_{i\pi(i)}c_{i\pi(i)}\} \\
        & \text{(since the column sums of $c_{ij}$ are all zero)}.
    \end{align*}
    Apply Lemma \ref{lem:key-var-bds} (ii) to obtain
    \begin{align*}
        \Var{\bbE\{a_{I\pi(I)}b_{I\pi(I)}c_{J\pi(I)}\mid \pi\}}\le \frac{16NB_N^6}{N^2(N-1)^2}\le\frac{32B_N^6}{N^3}.
    \end{align*}
    
    \item $a_{I\pi(I)}b_{I\pi(I)}c_{I\pi(J)}$. There are $6$ terms in total. This part is similar to the last one:
    \begin{align*}
        \Var{\bbE\{a_{I\pi(I)}b_{I\pi(I)}c_{I\pi(J)}\mid \pi\}}\le \frac{16NB_N^6}{N^2(N-1)^2}\le\frac{32B_N^6}{N^3}.
    \end{align*}

    \item $a_{I\pi(I)}b_{J\pi(I)}c_{J\pi(I)}$. There are $6$ terms in total. We have
    \begin{align}\label{eqn:third-ctg}
        \bbE\{a_{I\pi(I)}b_{J\pi(I)}c_{J\pi(I)}\mid \pi\} &= \frac{1}{N(N-1)} \sum_{i\neq j}^N a_{i\pi(i)}b_{j\pi(i)}c_{j\pi(i)}\notag\\
        &= \frac{1}{N(N-1)} \sum_{i=1}^N \lt\{\sum_{j\neq i}a_{i\pi(i)}b_{j\pi(i)}c_{j\pi(i)}\rt\}.
    \end{align}
    \eqref{eqn:third-ctg} can be viewed as a univariate linear permutation statistics from a population matrix with entries 
    \begin{align*}
        d_{kl} = a_{kl}\sum_{m\neq k}b_{ml}c_{ml}.
    \end{align*}
    Because $|d_{kl}| \le (N-1)B_N^3$,  we have
    \begin{align*}
        \Var{\bbE\{a_{I\pi(I)}b_{I\pi(I)}c_{I\pi(I)}\mid \pi\}} \le \frac{16N \cdot \{(N-1)B_N^3\}^2}{N^2(N-1)^2} \le \frac{16B_N^6}{N}.
    \end{align*}
    
    \item $a_{I\pi(I)}b_{I\pi(J)}c_{I\pi(J)}$. There  are $6$ terms in total. We can check (by using $\pi^{-1}$) that this term is similar to the last one:
    \begin{align*}
        \Var{\bbE\{a_{I\pi(I)}b_{I\pi(J)}c_{I\pi(J)}\mid \pi\}} \le \frac{16N \cdot \{(N-1)B_N^3\}^2}{N^2(N-1)^2} \le \frac{16B_N^6}{N}.
    \end{align*}
    
    \item $a_{I\pi(I)}b_{I\pi(I)}c_{J\pi(J)}$. There are $6$ terms in total. 
    Let $(d_{kl}) = (a_{kl}b_{kl})$ and  $d^\star_{kl} = d_{kl} - d_{\cdot l} - d_{k\cdot} + d_{\cdot\cdot}$ be the centered version with $|d^\star_{kl}|\le 4B_N^2$. We have
    \begin{align*}
        \bbE\{a_{I\pi(I)}b_{I\pi(I)}c_{J\pi(J)}\mid \pi\} &= \frac{1}{N(N-1)} \sum_{i\neq j}^N a_{i\pi(i)}b_{i\pi(i)}c_{j\pi(j)}\\
        &= \frac{1}{N(N-1)} \lt\{\sum_{i\neq j}^N d^\star_{i\pi(i)}c_{j\pi(j)} + \sum_{i\neq j}^N (d_{\cdot\pi(i)} - d_{\cdot\cdot} )c_{j\pi(j)} + \sum_{i\neq j}^N d_{i\cdot}c_{j\pi(j)}\rt\}\\
        & = \text{I} + \text{II} + \text{III}. 
    \end{align*}
    For I,  by Lemma \ref{lem:key-var-bds} (iii), we know
    \begin{align*}
        \Var{\text{I}}\le \frac{54N^2 \cdot (4B_N^2)^2 \cdot (B_N^2)}{N^2(N-1)^2} \le \frac{864B_N^6}{(N-1)^2}.
    \end{align*}
    For II, by re-indexing $h=\pi(i),l=\pi(j)$, we have
    \begin{align*}
        \text{II} = \frac{1}{N(N-1)}\sum_{l=1}^N\sum_{k\neq l}(d_{\cdot k} - d_{\cdot\cdot} )c_{\pi^{-1}(l)l}.
    \end{align*}
    Let $e_{kl} = \sum_{m\neq l}(d_{\cdot k} - d_{\cdot\cdot} )c_{kl}$. By Lemma \ref{lem:key-var-bds} (ii) and the fact that $|\sum_{k\neq l}(d_{\cdot k} - d_{\cdot\cdot} )c_{ml}|\le 2(N-1)B_N^3$, we have
    \begin{align*}
        \Var{\text{II}} \le \frac{32N\{2(N-1)B_N^3\}^2}{N^2(N-1)^2} \le \frac{128B_N^6}{N}.
    \end{align*}
    For III, the analysis is similar to II:
    \begin{align*}
        \Var{\text{III}} \le \frac{32N\{(N-1)B_N^3\}^2}{N^2(N-1)^2} \le \frac{32B_N^6}{N}.
    \end{align*}
    
    Since $\Var{\text{I}}$ is of lower order compared with that of II and III, we have
    \begin{align*}
        \Var{\bbE\{a_{I\pi(I)}b_{I\pi(I)}c_{J\pi(J)}\mid \pi\}} \le \frac{600B_N^6}{N}.
    \end{align*}
    
    \item $a_{I\pi(J)}b_{I\pi(J)}c_{J\pi(I)}$. There are $6$ terms in total.
    The analysis for this part is similar to the last part. The only difference is that we need to apply Lemma \ref{lem:key-var-bds} (iv) instead of (iii) to bound the variance for a term that looks like the I in the previous part. But the upper bound in (iii) and (iv) are the same. Hence  
    \begin{align*}
        \Var{\bbE\{a_{I\pi(J)}b_{I\pi(J)}c_{J\pi(I)}\mid \pi\}}\le \frac{600B_N^6}{N}.
    \end{align*}

    \item $a_{I\pi(I)}b_{J\pi(J)}c_{I\pi(J)}$. There are $12$ terms in total.  We have
    \begin{align*}
        \bbE\{a_{I\pi(I)}b_{J\pi(J)}c_{I\pi(J)}\mid \pi\} &= \frac{1}{N(N-1)} \sum_{i\neq j}^N a_{i\pi(i)}b_{j\pi(j)}c_{i\pi(j)}. 
    \end{align*}
    
    By Lemma \ref{lem:key-var-bds} (v), we have 
    \begin{align*}
        \Var{\bbE\{a_{I\pi(I)}b_{J\pi(J)}c_{I\pi(J)}\mid \pi\}} \le \frac{15N^3B_N^6}{N^2(N-1)^2} \le \frac{30B_N^6}{N}.
    \end{align*}
    
    \item $a_{I\pi(I)}b_{J\pi(I)}c_{I\pi(J)}$. There are $12$ terms in total. We have 
    \begin{align*}
        \bbE\{a_{I\pi(I)}b_{J\pi(I)}c_{I\pi(J)}\mid \pi\} &= \frac{1}{N(N-1)} \sum_{i\neq j}^N a_{i\pi(i)}b_{j\pi(i)}c_{i\pi(j)}. 
    \end{align*}
    
    By Lemma \ref{lem:key-var-bds} (vi), we have 
    \begin{align*}
        \Var{\bbE\{a_{I\pi(I)}b_{J\pi(J)}c_{I\pi(J)}\mid \pi\}} \le \frac{15N^3B_N^6}{N^2(N-1)^2} \le \frac{30B_N^6}{N}.
    \end{align*}
    
\end{itemize}

Now summing up the bullet points above, we have 
\begin{align*}
    \Var{\bbE\{D_hD_lD_m\mid\pi\}} \le \frac{CB_N^6}{N},
\end{align*}
for some absolute constant $C>0$.

\noindent\textbf{(iv)  Summarize (i) (ii) (iii) above.}

As a brief review, we have proved the following results: when $N$ is large,
\begin{enumerate}[label = (\arabic*)]
    \item $\|G\|_2\le \alpha = C(N-1)H^{1/2}B_N$, $\|D\|_2 \le \beta = CH^{1/2}B_N$.
    \item $\bbE\{\|D\|_2^2\} \le C(HB_N^2 + HN^{-1})$.
    \item $B_1 = \sqrt{\Var{\bbE(\|D\|_2^2\mid \Gamma)}} \le CHN^{-1/2}B_N^2$.
    \item $B_2 = \sqrt{\sum_{k,l=1}^H \Var{\bbE(G_hD_l\mid \Gamma)}}\le CHN^{1/2}B_N^2$.
    \item $B_3 = \sqrt{\sum_{k,l,m=1}^H \Var{\bbE(G_hD_lD_m\mid \Gamma)}} \le CH^{3/2}N^{1/2}B_N^3$.
\end{enumerate}
Using Lemma \ref{lem:bounded-pair} with (1) - (5), we have
\begin{align*}
    &\sup_{A\in\cA}|\bbP\{\Gamma\in A\} - \bbP\{\xi_H\in A\}| \\
    &\le C(H^{7/4} \alpha \mathbb{E}\|D\|_2^2 + H^{1/4}\beta + H^{7/8}\alpha^{1/2}B_1^{1/2} + H^{3/8}B_2 + H^{1/8}B_3^{1/2})\\
    &\le CH^{13/4}NB_N(B_N^2 + N^{-1}) + C H^{3/4}B_N + CH^{13/8}N^{1/4}B_N^{3/2} \\
    &+ CH^{11/8}N^{1/2}B_N^2 + CH^{7/8}N^{1/4}B_N^{3/2}.
\end{align*}

\end{proof}

\section{Additional results for randomization-based causal inference}\label{sec:additional}

Appendix \ref{sec:additional} presents some additional results for randomization-based causal inference. Section \ref{sec:BE-quad} presents general BEBs for quadratic forms, which are derived based on the results in Appendix \ref{sec:general-BE}. 
Section \ref{sec:high-moments} establishes some bounds on the high order moments of the sample averages. Section \ref{sec:tail-uniform} and \ref{sec:tail-non-uniform} provide more delicate tail bounds for linear combinations of sample variances.  Section \ref{sec:vec-outcome} extends the BEBs to vector potential outcomes.

\subsection{BEBs for quadratic forms}\label{sec:BE-quad}

In Section \ref{sec:PCLT-projection}, we proved the BEBs for linear projections of multivariate linear permutation statistics in randomized experiments.
In this subsection, we study a more general type of  distance:
\begin{align}\label{eqn:dcA}
    d_\cA(\tilde{\gamma}, \xi_H) = \sup_{A\in\cA} |\Prob{\tilde{\gamma} \in A} - \Prob{\xi_H \in A}|, 
\end{align}
recalling that $\cA$ is the collection of all Borel convex sets, $\tilde{\gamma}$ is defined as \eqref{eqn:standard-est}, and $\xi_H$ is a random vector in $\bbR^H$ with standard multivariate Normal distribution.  $\cA$ can cover many specific convex classes. For example, the set of ellipsoids defined as follows are a subset of $\cA$:
\begin{align}\label{eqn:ellipsoids}
    \cA_{2}(\lambda, t) = \lt\{\gamma\in\bbR^H: \sum_{h=1}^H\lambda_h^2\gamma_h^2 \le t \rt\}, ~\lambda_h > 0,~ t > 0.
\end{align}
\eqref{eqn:ellipsoids} is useful for deriving BEBs for quadratic forms of $\tilde{\gamma}$.

Recall $\tilde{\gamma}$ from \eqref{eqn:tilde-gamma}. We study the asymptotic distribution of the following random variable:
\begin{align*}
    T = \tilde{\gamma}^\top W \tilde{\gamma},
\end{align*}
where $W$ is a given positive definite matrix in $\bbR^{H\times H}$.
We also write $T_0$ as the counterpart with $\tilde{\gamma}$ replaced by Normal vectors $\xi_H$:
\begin{align}\label{eqn:E-V-TZ}
    T_0 = \xi_H^\top W \xi_H,~ \xi_H \sim \cN(0,I_H),
\end{align}
with 
\begin{equation}\label{eqn::moments-chi-squares}
\E{T_0} = \trace{W}, \quad  \Var{T_0} = 2\trace{W^2}.
\end{equation}
For the problems in the main paper, we need to deal  with the class of ellipsoids \eqref{eqn:ellipsoids}, which are convex. Applying Theorem \ref{thm:be-bounded}, we obtain the following result:
\begin{theorem}[Permutational BEBs for  quadratic forms]\label{thm:quad-clt}
We have 
\begin{align}
    & \sup_{t\in\bbR} |\bbP(T\le t)-\bbP(T_0\le t)|\notag\\
    & \le C(H^{13/4}NB_N(B_N^2 + N^{-1}) +  H^{3/4}B_N + H^{13/8}N^{1/4}B_N^{3/2} 
    + H^{11/8}N^{1/2}B_N^2),\label{eqn:quad-clt}
\end{align}
where 
\begin{align}\label{eqn:BN}
    B_N = \varrho_{\min}(V_{{\hgamma}})^{-1/2}\sqrt{H} \max_{h\in[H]}\max_{i\in[N],q\in[Q]}|f_{qh}N_{q}^{-1}(Y_i(q)- \overline{Y}(q))|.
\end{align}
When $B_N \le C H^{1/2}N^{-1/2}$, we have
\begin{align}\label{eqn:BN-special}
    \sup_{t\in\bbR} |\bbP(T\le t)-\bbP(T_0\le t)| \le \frac{CH^{19/4}}{N^{1/2}}.
\end{align}
\end{theorem}

\begin{remark}
The proof for Theorem \ref{thm:quad-clt} is an application of the bound over convex sets in Theorem \ref{thm:be-bounded}. In general, the bound \eqref{eqn:quad-clt} might not be sharp for quadratic forms.  While the BEB achieved the rate $N^{-1/2} $ which is analogous to the i.i.d. scenario, the rate in $H$ might not be optimal. However, we do not pursue the best possible bound here. In many  cases, \eqref{eqn:quad-clt} suffices for establishing asymptotics. For example, in factorial experiments, if we focus on lower order effects, $H$ is approximately the order of $\log(N)$.  Therefore, we can justify the asymptotic normality as $N\to\infty$ using \eqref{eqn:quad-clt}.
\end{remark}

We discuss how to bound $B_N$ to obtain a usable result from \eqref{eqn:quad-clt}.  The following lemma covers nearly uniform designs and, more broadly, general designs:

\begin{lemma}\label{lem:BN}
Assume \eqref{eqn:well-conditioned}.  
\begin{enumerate}[label = (\roman*)]
    \item For nearly uniform designs with either replicated or unreplicated arms, assume Condition \ref{condition::proper}. There exists a constant $C = C(c,c',\underline{c},\overline{c})$ that only depends on the constants in Definition \ref{def:uniform-design} and Condition \ref{condition::proper}, such that
    \begin{align*}
       \sup_{t\in\bbR} |\bbP(T\le t)-\bbP(T_0\le t)| \le  C\frac{\max_{ q\in[Q]} M_N(q)^3  }{ \{ \min_{q\in[Q]} S(q,q) \}^{3/2}}\cdot\frac{H^{19/4}}{N^{1/2}}.
    \end{align*}
    Moreover, under Condition \ref{cond:easy-spec}, we have
    \begin{align*}
    B_N \le \frac{2c^{1/2}\underline{c}^{-1}\nu}{(\overline{c}^{-1}\underline{S})^{1/2}} \cdot \lt(\frac{H}{QN_0}\rt)^{1/2}
    \end{align*}
    and the BEB \eqref{eqn:BN-special} holds.

    \item For general designs, assume Condition \ref{condition::proper-non-uniform}. There exists some constant $C = C(c,c',\underline{c},\overline{c})$ that only depends on the constants in Definition \ref{def:non-uniform-design} and Condition \ref{condition::proper-non-uniform}, such that the BEB \eqref{eqn:quad-clt} holds with 
    \begin{align*}
      B_N \le \frac{2C\max_{i\in[N],q\in[Q]}|Y_i(q) - \overline{Y}(q)|}{(\overline{n}^{-1}\min_{q\in\cQ_{\textsc{s}}}S(q,q))^{1/2}}\cdot \max\lt\{\frac{1}{|\cQ_\textsc{s}|^{1/2}}, \frac{\|F_\textsc{l}\|_\infty |\cQ_\textsc{s}|^{1/2}}{\underline{c} N_0}\rt\}.
  \end{align*}

    
    Moreover, under Condition \ref{cond:easy-spec}, $\|F_\textsc{l}\|_\infty = O(Q^{-1})$ and $N=O(|\cQ_\textsc{s}|)$, we have
    \begin{align*}
    B_N \le \frac{2vc H^{1/2}}{(c'\overline{n}^{-1}\underline{S})^{1/2}N^{1/2}}
    \end{align*}
and the BEB \eqref{eqn:BN-special} holds.

\end{enumerate}

\end{lemma}

We have a thorough understanding of the distribution of $T_0$. By eigenvalue decomposition, 
$$
T_0 \sim   \sum_{h=1}^H  \varrho_h \tilde{\xi}_{0,h}^2 \lesssim \varrho_1\chi^2_H, 
  $$
 where $\varrho_1\ge\cdots\ge\varrho_H$ are eigenvalues of $ W$ and ${\tilde{\xi}}_{0,1},\ldots,{\tilde{\xi}}_{0,H}$ are i.i.d. $\mathcal{N}(0,1)$. Therefore,
$T_0$ is stochastically dominated by $\chi^2_H$. 
When $H$ is fixed, the asymptotic distribution of $T$ follows immediately. 
When $H$ diverges, we need to further use the asymptotic distribution for the sum of independent random variables based on the Lindeberg--L\'{e}vy CLT and classical BEBs. Corollary \ref{cor:quad-clt-v2} below summarizes the results. 

\begin{corollary}[Limiting distribution of the quadratic form]\label{cor:quad-clt-v2} 
Let $N\rightarrow \infty$. 
Assume the upper bound in \eqref{eqn:quad-clt} vanishes:
\begin{align*}
    CH^{13/4}NB_N(B_N^2 + N^{-1}) + C H^{3/4}B_N + CH^{13/8}N^{1/4}B_N^{3/2}
    + CH^{11/8}N^{1/2}B_N^2 \to 0.
\end{align*}
\begin{enumerate}
    \item If $H$ is fixed, then
$
        T \rightsquigarrow T_0 .
        $   
 
    \item If $H$ diverges, then
    \begin{align*}
        \frac{T - \operatorname{Tr}( W)}{\sqrt{2\operatorname{Tr}( W^2)}} \rightsquigarrow \cN(0,1).
    \end{align*}
\end{enumerate}
\end{corollary}




 

\subsection{High order moments of  $\hY$}
\label{sec:high-moments}
In this subsection, we present some delicate characterizations of the high order moments of the sample average $\hY$, which are crucial for the proof of our main results and might be of independent interest for other problems. 

\begin{lemma}[High order moments of $\hY$]\label{lem:high-moment}
Assume complete randomization and Condition \ref{cond:moments}. 
\begin{enumerate}[label=(\roman*)]
    \item $\bbE\{(\hY_q - \overline{Y}(q))^2\} \le \frac{C\Delta^2}{N_q} $; 
    \item $\bbE\{(\hY_q - \overline{Y}(q))^4\} \le \frac{C\Delta^4}{N_q^2} $;
    \item $\Cov{(\hY_q - \overline{Y}(q))^2}{(\hY_{q'} - \overline{Y}(q'))^2} \le \frac{C(N_q + N_{q'})\Delta^4}{N_qN_{q'}(N-1)} $.
\end{enumerate}
\end{lemma}

\begin{lemma}[High order moments under unreplicated designs]\label{lem:moments-U}
Assume the potential outcomes are centered: $\overline{Y}(q) = 0$ for all $q\in[Q]$. Assume complete randomization and Condition \ref{cond:moments}. For the unreplicated design in Definition \ref{def:non-uniform-design}, there exists a universal constant $C>0$, such that
\begin{enumerate}[label = (\roman*)]
    \item for $q_1\in[Q]$, $\Var{Y_{q_1}^2} = (1-N^{-1})S_{Y^2}(q_1,q_1) \le C\Delta^4$;
    
    \item for $q_1\neq q_2$, $\Cov{Y_{q_1}^2}{Y_{q_2}^2} = -N^{-1}S_{Y^2}(q_1,q_2)$ and $|\Cov{Y_{q_1}^2}{Y_{q_2}^2} = -N^{-1}S_{Y^2}(q_1,q_2)| \le \frac{C\Delta^4}{N} $;
    
    \item for $q_1\neq q_2$, $ |\Cov{Y_{q_1}^2}{Y_{q_1}Y_{q_2}}| \le \frac{C\Delta^4}{N}$;
    
    \item for $q_1\neq q_2\neq q_3$, $ |\Cov{Y_{q_1}^2}{Y_{q_2}Y_{q_3}}| \le \frac{C\Delta^4}{N}$;
    
    \item for $q_1\neq q_2\neq q_3 \neq q_4$, $ |\Cov{Y_{q_1}Y_{q_2}}{Y_{q_3}Y_{q_4}}| \le \frac{C\Delta^4}{N^2}$.
\end{enumerate}
\end{lemma}

We assume the potential outcomes are centered in Lemma \ref{lem:moments-U} to simplify the formulas. Without this assumption, all results in  Lemma \ref{lem:moments-U}  hold if we subtract the means of the potential outcomes from the corresponding observed outcomes.

\subsection{Tail probability of variance estimation for nearly uniform design}
\label{sec:tail-uniform}

For an arbitrary set of indices $\cQ \subset [Q]$,  define 
\begin{align*}
    \hv = \sum_{q\in\cQ}w_q  N_q^{-1} \hat{S}(q,q)
\end{align*}
if $N_q \geq 2$ for all $q \in \cQ.$
Lemma \ref{lem:tail} below gives the tail probability of $\hv$. 

\begin{lemma}[Tail probability of variance estimation]\label{lem:tail}
Consider the nearly uniform design satisfying Definition \ref{def:uniform-design}. Assume Condition \ref{cond:moments}  and $\min_{q\in[Q]}~N_q \ge 2 $. Assume $(w_q)_{q\in[Q]}$ is a sequence of bounded real numbers: 
\begin{align*}
    \max_{q\in[Q]}|w_q| \le \overline{w}.
\end{align*}
Then there exists a universal constant $C>0$, such that 
\begin{align*}
    \bbP\lt\{\lt|\hv - \E{\hv}\rt|\ge t\rt\} \le \frac{C\overline{c}\underline{c}^{-4} \overline{w}^2 |\cQ| N_0^{-3}\Delta^4}{t^2}.
\end{align*}
\end{lemma}

\subsection{Tail probability of variance estimation for unreplicated arms}
\label{sec:tail-non-uniform}

Recall the notation in Section \ref{sec:var-unreplicate}. Define 
\begin{align}
     \hat{v}  =  \sum_{{\langle g \rangle}\in{\cG}}\sum_{q\in{\langle g \rangle}}{w_q}   \lt (Y_q - \hY_{{\langle g \rangle}}\rt)^2 .
    \label{eqn:var-est-3-U}
\end{align}

Lemma \ref{lem:hv-U} below gives the tail probability of $     \hat{v}  $.

\begin{lemma}[Analysis of $\hat{v}$ under unreplicated arms]\label{lem:hv-U} 
Assume Conditions \ref{cond:moments} and \ref{cond:bounded-bgv}. Assume $(w_q)_{q\in\cQ_{\textsc{u}}}$ is a sequence of bounded real numbers: 
\begin{align*}
    \max_{q\in\cQ_{\textsc{u}}}|w_q| \le \overline{w}.
\end{align*}
Then there exists a universal constant $C>0$, such that 
\begin{align*}
    \Prob{|\hat{v} - \E{\hv}|\ge t}
    \le  \frac{C\overline{w}^2(\Delta^4+\Delta^2\zeta^2) N_{\textsc{u}}}{t^2}.
\end{align*}
\end{lemma}

\subsection{Extension to vector potential outcomes}
\label{sec:vec-outcome}
In some settings, we are interested in vector potential outcomes. \cite{li2017general} proved some CLTs for vector outcomes. For treatment-control experiments, \cite{wang2022rerandomization} proved some BEBs based on the coupling method. However, the general theory for BEB is still incomplete.

Let $\{\bY_i(q)\in\bbR^p:i\in[N],q\in[Q]\}$ be a collection of potential outcomes. Let $\bF_1,\ldots,\bF_Q$ be $Q$ coefficient matrices in $\bbR^{H\times p}$. Define
$\gamma = \sum_{q=1}^Q \bF_q \overline{\bY}(q)$, and the moment estimator is  $\hgamma =  \sum_{q=1}^Q \bF_q \hat{\bY}_q$.
\cite{li2017general} calculated the mean and covariance of $\hgamma$:
\begin{gather*}
    \bbE\{\hgamma\} = \sum_{q=1}^Q \bF_q\overline{\bY}(q),\quad
    \text{Var}(\hgamma) = \sum_{q=1}^Q N_q^{-1}\bF_q\bS(q,q)\bF_q^\top - N^{-1}\bS_\bF := \bV_\hgamma,
\end{gather*}
where
\begin{gather*}
    \bS(q,q') = (N-1)^{-1}\sum_{i=1}^N (\bY_i(q) - \overline{\bY}(q))(\bY_i(q') - \overline{\bY}(q'))^\top,~ q,q'\in[Q],\\
    \bS_\bF = (N-1)^{-1} \sum_{i=1}^N (\gamma_i - \overline{\gamma})(\gamma_i - \overline{\gamma})^\top, \quad
    \gamma_i =  \sum_{q=1}^Q\bF_q\bY_i(q), \quad
    \overline{\gamma} = N^{-1}\sum_{i=1}^N \gamma_i. 
\end{gather*}

Define 
\begin{align*}
    \breve{\bY}_i(q) = \bY_i(q)-\overline{\bY}(q) \text{ for all } q\in[Q].
\end{align*}
Theorem \ref{thm:be-proj-standard-vec} below gives BEBs for projections of the standardized $\hgamma$.
\begin{theorem}[BEB for projections of the standardized $\hgamma$]\label{thm:be-proj-standard-vec}
Let 
\begin{align*}
    \tilde{\gamma} = \{\bV_\hgamma\}^{-1/2}(\hgamma - \bbE\{\hgamma\}).
\end{align*}
Assume complete randomization. (i) There exists a universal constant $C>0$, such that for any $b\in\bbR^H$ with $\|b\|_2 = 1$, we have
\begin{align*}
    \lt|\bbP\{b^\top\tilde{\gamma} \le t\} - \Phi(t)\rt| \le C\max_{i\in[N],q\in[Q]}\lt|b^\top  \bV_\hgamma^{-1/2}N_{q}^{-1}\bF_{q}\breve{\bY}_i({q}) \rt|.
\end{align*}
(ii) If there exists $\sigma_F \ge 1$, such that
\begin{align}\label{eqn:well-conditioned-vec}
     \sum_{q=1}^Q N_q^{-1}\bF_q  \bS(q,q) \bF_q^\top \preceq \sigma^2_F \bV_\hgamma,
\end{align}
then
\begin{align}\label{eqn:uniform-be-vec}
    &\sup_{b\in\bbR^H,\|b\|_2 =1}\sup_{t\in\bbR}\lt|\bbP\{b^\top\tilde{\gamma} \le t\} - \Phi(t)\rt| \notag\\
    \le& C \max_{i\in[N],q\in[Q]} \min\lt\{2 {\sigma_F} \sqrt{N_q^{-1}\breve{\bY}_i(q)^\top \bS(q,q)^{-1} \breve{\bY}_i(q)},~ \frac{  \|\bF_q\|_{2,1}\cdot  N_q^{-1}\|\breve{\bY}_i(q)\|_\infty}{\sqrt{\varrho_{\min}\{ \bV_\hgamma\}}}\rt\}.
\end{align}

\end{theorem}

Theorem \ref{thm:be-proj-standard-vec} extends Theorem \ref{thm:be-proj-standard} to vector potential outcomes. If $p=1$, Theorem \ref{thm:be-proj-standard-vec} recovers Theorem \ref{thm:be-proj-standard}. The novel part of the extension is to decide the appropriate vector and matrix norms in the upper bound for the vector potential outcomes $\breve{\bY}_i(q)$'s and the coefficient matrices $\bF_q$'s. The proof provides more insights into the choices. Moreover, we can derive many corollaries from Theorem \ref{thm:be-proj-standard-vec} as in the main paper. To avoid repetitions, we omit the details.





\section{Proofs of the results in the main paper and Appendix \ref{sec:additional}}\label{sec:main-proof}

\subsection{Proof of Theorem \ref{thm:be-proj-standard}}\label{sec:pf-be-proj-standard}

The proof of Theorem \ref{thm:be-proj-standard} is based on Theorem \ref{thm:linear-projection}. There are two key steps: (i) formulate $\tilde{\gamma}$ as a linear permutation statistic that satisfies the conditions of Theorem \ref{thm:linear-projection}; (ii) find explicit bounds for the BEB in Theorem \ref{thm:linear-projection}.

\begin{proof}[Proof of Theorem \ref{thm:be-proj-standard}]
Recall
\begin{align}\label{eqn:standard-est}
    \tilde{\gamma} = (V_\hgamma)^{-1/2}(\hgamma - \gamma) =  \Var{F^\top\hY}^{-1/2}(F^\top\hY - \bbE\{F^\top\hY\}).
\end{align}
\textbf{Step 1: Reformulate $\tilde{\gamma}$ as a multivariate linear permutation statistic.} 

We show that, there exist population matrices $M''_1,\ldots, M''_H$ that satisfy Condition \ref{cond:str-Mk}, such that $\tilde{\gamma} = \lt(\trace{M''_h P}\rt)_{h=1}^H$. 

\textit{1. Construction of $M''_h$'s.} Define
\begin{align}\label{eqn:center-PO}
    \breve{Y}_i(q) = Y_i(q) - \overline{Y}(q), ~\breve{\tau}_{hi} = N^{-1}\sum_{q'=1}^Q f_{q'h} \breve{Y}_i(q') .
\end{align}
For each $i,j$, define
\begin{align*}
    M_h'(i,j) = N_q^{-1} f_{qh}\breve{Y}_i(q) -  \breve{\tau}_{hi}, ~ \sum_{q'=0}^{q-1} N_q + 1\le j \le \sum_{q'=0}^{q} N_q
\end{align*}
such that $M'_h$ is the centered version of
\begin{align}\label{eqn:potentials}
M_{h} = 
    \bordermatrix{
             &    &  Z = 1                    &  \cdots        &  Z = q                                     & { \cdots }         &  Z = Q                                                         \cr 
1        &  &  f_{1h}N_{1}^{-1}Y_1(1)\cdot 1^\top_{N_{1}}  &  \cdots    &  f_{qh}N_{q}^{-1}Y_1(q)\cdot 1^\top_{N_{q}}   &  \cdots   &  f_{Qh}N_{Q}^{-1}Y_1(Q)\cdot 1^\top_{N_{Q}}   \cr 
2        &  &  f_{1h}N_{1}^{-1}Y_2(1)\cdot 1^\top_{N_{1}}  &  \cdots    &  f_{qh}N_{q}^{-1}Y_2(q)\cdot 1^\top_{N_{q}}  &   \cdots   &  f_{Qh}N_{Q}^{-1}Y_2(Q)\cdot 1^\top_{N_{Q}}   \cr 
 \cdots  &  &  \cdots                &  \cdots                                          &   \cdots     &  \cdots  \cr
N        &  &  f_{1h}N_{1}^{-1}Y_N(1)\cdot 1^\top_{N_{1}}  &  \cdots    &  f_{qh}N_{q}^{-1}Y_N(q)\cdot 1^\top_{N_{q}}  &   \cdots   &  f_{Qh}N_{Q}^{-1}Y_N(Q)\cdot 1^\top_{N_{Q}}   \cr
    }.
\end{align}

Observe that 
\begin{align}\label{eqn:hgamma-vecM}
   \hgamma - \gamma = \lt(\trace{M_1'P}, \dots,\trace{M_H'P}\rt)^\top =
   \begin{pmatrix}
    \{\myvec{M'_1}\}^\top\\
    \vdots\\
    \{\myvec{M'_H}\}^\top
    \end{pmatrix}
    \myvec{P}.
\end{align}

Construct $M''_h$'s as follows:
\begin{align}\label{eqn:vecM}
    \begin{pmatrix}
    \{\myvec{M''_1}\}^\top\\
    \vdots\\
    \{\myvec{M''_H}\}^\top
    \end{pmatrix}
    =
    V_{\hgamma}^{-1/2}
    \begin{pmatrix}
    \{\myvec{M'_1}\}^\top\\
    \vdots\\
    \{\myvec{M'_H}\}^\top
    \end{pmatrix}.
\end{align}
Combining \eqref{eqn:vecM} and \eqref{eqn:hgamma-vecM}, we can show $\tilde{\gamma} = \lt(\trace{M_1''P}, \dots, \trace{M_H''P}\rt)^\top$. The next step is to show $M''_h$'s satisfy Condition \ref{cond:str-Mk}.

\textit{2. Verify Condition \ref{cond:str-Mk}.} To verify that $M_h''$'s have zero row and column sums, we notice that summation of $j$-th column (or row) corresponds to a linear mapping from $\bbR^{N\times N}$ to $\bbR$ that can be defined by the trace inner product:
\begin{align*}
    \sum_{i=1}^N M''_h(i,j) = \trace{{M_h''}^{\top} T_j}  \text{ with }  T_j = (0_N,\dots,\underbrace{1_N}_{\text{column $j$}},\dots,0_N).
\end{align*}
Given that $M'_h$'s are row and column centered, we can use \eqref{eqn:vecM} to show that 
\begin{align*}
    \begin{pmatrix}
    \{\myvec{M''_1}\}^\top\\
    \vdots\\
    \{\myvec{M''_H}\}^\top
    \end{pmatrix}\myvec{T_j}
    =
    V_{\hgamma}^{-1/2}
    \begin{pmatrix}
    \{\myvec{M'_1}\}^\top\\
    \vdots\\
    \{\myvec{M'_H}\}^\top
    \end{pmatrix}\myvec{T_j} = 0.
\end{align*}

To show $M''_h$'s are standardized and mutually orthogonal, we notice
\begin{align}\label{eqn:V-tgamma-1}
    \COV{\tilde{\gamma}} = I_H.
\end{align}
Now using Lemma \ref{lem:mean-var}(ii), we have 
\begin{align}\label{eqn:V-tgamma-2}
    \COV{\tilde{\gamma}} = \lt(\frac{1}{N-1}\tr{M''_h}{M''_l}\rt)_{h,l\in[H]}.
\end{align}
Comparing \eqref{eqn:V-tgamma-1} and \eqref{eqn:V-tgamma-2}, we obtain the desired conclusion.

\vskip 2mm
\noindent\textbf{Step 2: Apply Theorem \ref{thm:linear-projection} to prove Part (i) of Theorem \ref{thm:be-proj-standard}.} 
\vskip 2mm

Apply  Theorem \ref{thm:linear-projection} to obtain that: for any $b\in\bbR^H$ with $\|b\|_2 = 1$, we have 
\begin{align}\label{eqn:BE-M}
    \sup_{t\in\bbR}|\bbP\{b^\top\tilde{\gamma} \le t\} - \Phi(t)| \le C{\max_{i,j\in[N]} \lt|\sum_{h=1}^Hb_hM''_h(i,j)\rt|}.
\end{align}

Each column of the matrix \eqref{eqn:potentials} corresponds to a treatment group $q$. For ease of presentation, it is convenient to highlight this connection with notation $q_j$, meaning the $j$-th column is constructed based on potential outcomes from treatment level $q_j$. 

Based on \eqref{eqn:vecM}, we have
\begin{align}\label{eqn:bM}
    \lt|\sum_{h=1}^Hb_hM''_h(i,j)\rt| &= 
    \lt|b^\top V_{\hgamma}^{-1/2}
    \begin{pmatrix}
    N_{q_j}^{-1} f_{q_j1}\breve{Y}_i({q_j}) -  \breve{\tau}_{1i} \\
    \vdots\\
    N_{q_j}^{-1} f_{{q_j}H}\breve{Y}_i({q_j}) -  \breve{\tau}_{Hi}
    \end{pmatrix}\rt|\\
    & = 
    \lt|\underbrace{b^\top V_{\hgamma}^{-1/2}
    \begin{pmatrix}
    N_{q_j}^{-1} f_{q_j1}\breve{Y}_i({q_j}) \\
    \vdots\\
    N_{q_j}^{-1} f_{{q_j}H}\breve{Y}_i({q_j})
    \end{pmatrix}}_{\text{term I}}
    -
    \underbrace{b^\top V_{\hgamma}^{-1/2} 
    \begin{pmatrix}
    \breve{\tau}_{1i} \\
    \vdots \\
    \breve{\tau}_{Hi}
    \end{pmatrix}}_{\text{term II}}\rt|. \label{eqn:design-scaled-po}
\end{align}

From the definition \eqref{eqn:center-PO}, term II is the average of term I over $j\in[N]$. Therefore, if we can bound term I for all $i,j$, then we can also bound term II  by the triangle inequality. For term I, we have 
\begin{align*}
    |\text{term I}| &= 
    \lt|
    b^\top V_{\hgamma}^{-1/2}
    \begin{pmatrix}
    f_{q_j1} \\
    \vdots\\
    f_{{q_j}H}
    \end{pmatrix}
    \rt| \cdot |N_{q_j}^{-1} \breve{Y}_i({q_j})| 
    \le 
    \lt\|b^\top V_\hgamma^{-1/2} F^\top\rt\|_\infty \cdot 
\max_{q\in [Q], i\in[N]} N_q^{-1} |\breve{Y}_i(q)|,
\end{align*}
which leads to the conclusion. 

\vskip 2mm
\noindent\textbf{Step 3: Apply Theorem \ref{thm:linear-projection} to prove Part (ii) of Theorem \ref{thm:be-proj-standard}.} 
\vskip 2mm

Part (i) of Theorem \ref{thm:be-proj-standard} depends on choice of $b$. To derive a bound that is uniform over $b$, we use two different ways to bound term I. 

\textbf{First bound for term I}:  we have
\begin{align}
    \lt|b^\top V_{\hgamma}^{-1/2}
    \begin{pmatrix}
    N_{q_j}^{-1} f_{{q_j}1}\breve{Y}_i({q_j}) \notag\\
    \vdots\\
    N_{q_j}^{-1} f_{{q_j}H}\breve{Y}_i({q_j})
    \end{pmatrix}\rt|
    &=
    \lt|b^\top V_{\hgamma}^{-1/2}
    \begin{pmatrix}
     f_{{q_j}1}  \\
    \vdots\\
     f_{{q_j}H} 
    \end{pmatrix} \cdot \sqrt{N_{q_j}^{-1}S({q_j},{q_j})} \cdot \frac{N_{q_j}^{-1}\breve{Y}_i({q_j})}{\sqrt{N_{q_j}^{-1}S({q_j},{q_j})}}\rt|\notag\\
    &\le  
    \lt|b^\top V_{\hgamma}^{-1/2}
    \begin{pmatrix}
     f_{{q_j}1}  \\
    \vdots\\
     f_{{q_j}H} 
    \end{pmatrix} \cdot \sqrt{N_{q_j}^{-1}S({q_j},{q_j})}\rt| \cdot  \lt|\frac{N_{q_j}^{-1}\breve{Y}_i({q_j})}{\sqrt{N_{q_j}^{-1}S({q_j},{q_j})}}\rt|\notag\\
    &\le 
    \lt\|b^\top V_{\hgamma}^{-1/2}
    F^\top \diag{N_q^{-1}S(q,q)}^{1/2}\rt\|_\infty \cdot  \lt|\frac{N_{q_j}^{-1}\breve{Y}_i({q_j})}{\sqrt{N_{q_j}^{-1}S({q_j},{q_j})}}\rt|. \label{eqn:part2-bd1}
\end{align}
The infinity norm is upper bounded by the $\ell_2$ norm:
\begin{align}
    &\lt\|b^\top V_{\hgamma}^{-1/2}
    F^\top  \diag{N_q^{-1}S(q,q)}^{1/2}\rt\|_\infty \notag\\
    \le &\lt\|b^\top V_{\hgamma}^{-1/2}
    F^\top \diag{N_q^{-1}S(q,q)}^{1/2}\rt\|_2\notag\\
    = & \sqrt{b^\top V_{\hgamma}^{-1/2}
    F^\top  \diag{N_q^{-1}S(q,q)}F V_{\hgamma}^{-1/2}b}\notag\\
    \le &\sqrt{b^\top V_{\hgamma}^{-1/2}
    (\sigma_F^2V_{\hgamma}) V_{\hgamma}^{-1/2}b}
    \see{by \eqref{eqn:well-conditioned}} \notag\\
    \le & \sigma_F. \label{eqn:part2-bd2}
\end{align}
Combining \eqref{eqn:part2-bd1} and \eqref{eqn:part2-bd2}, we have
\begin{align}\label{eqn:second-bd}
\lt|\sum_{h=1}^Hb_hM''_h(i,j)\rt|\le 2 {\sigma_F} \lt|\frac{ \breve{Y}_i({q_j})}{\sqrt{N_q S({q_j},{q_j})}}\rt|.
\end{align}

\textbf{Second bound for term I}: for $b\in\bbR^H$ with $\|b\|_2 = 1$, construct $b_0 = V_{\hgamma}^{ - 1/2} b   /  \| V_{\hgamma}^{ - 1/2} b  \|_2 \in\bbR^H $ with $\|b_0\|_2 = 1$. We can verify that
\begin{align*}
    b = \frac{V_{\hgamma}^{1/2}b_0}{\sqrt{b_0^\top V_{\hgamma} b_0}}.
\end{align*}

We have
\begin{align*}
    \lt|b^\top V_{\hgamma}^{-1/2}
    \begin{pmatrix}
    N_{q_j}^{-1} f_{{q_j}1}\breve{Y}_i({q_j}) \\
    \vdots\\
    N_{q_j}^{-1} f_{{q_j}r}\breve{Y}_i({q_j})
    \end{pmatrix}\rt|
    &=
    \lt|b^\top V_{\hgamma}^{-1/2}
    \begin{pmatrix}
     f_{{q_j}1}  \\
    \vdots\\
     f_{{q_j}H} 
    \end{pmatrix}  \cdot {N_{q_j}^{-1}\breve{Y}_i({q_j})}\rt|\\
    &= 
    \lt|b_0^\top 
    \begin{pmatrix}
     f_{{q_j}1}  \\
    \vdots\\
     f_{{q_j}H} 
    \end{pmatrix} \rt| \cdot  \lt|\frac{N_{q_j}^{-1}\breve{Y}_i({q_j})}{\sqrt{b_0^\top V_{\hgamma} b_0}}\rt|\\
    &= 
    \lt|F({q_j}, \cdot)b_0\rt| \cdot  \lt|\frac{N_{q_j}^{-1}\breve{Y}_i({q_j})}{\sqrt{b_0^\top V_{\hgamma} b_0}}\rt|  \\
    & \leq  \frac{ \|F({q_j}, \cdot)\|_2\cdot  N_{q_j}^{-1}|Y_i({q_j})-\overline{Y}({q_j})|}{\sqrt{\varrho_{\min}\{V_{\hgamma}\}}}
\end{align*}
where the last uniform bound follows from 
\begin{align*}
    \lt|F({q_j}, \cdot)b_0\rt| \le \|F({q_j}, \cdot)\|_2 , \quad
     {b_0^\top V_{\hgamma} b_0} \ge  {\varrho_{\min}\{V_{\hgamma}\}}.
\end{align*}
Then we have
\begin{align}\label{eqn:first-bd}
\lt|\sum_{h=1}^Hb_hM''_h(i,j)\rt|\le \frac{ 2\|F({q_j}, \cdot)\|_2\cdot  N_{q_j}^{-1}|Y_i({q_j})-\overline{Y}({q_j})|}{\sqrt{\varrho_{\min}\{V_{\hgamma}\}}}.
\end{align}


Combining \eqref{eqn:first-bd} and \eqref{eqn:second-bd}, we have
\begin{align}\label{eqn:combined-bd}
&\lt|\sum_{h=1}^Hb_hM''_h(i,j)\rt|\\
\le &2\min\lt\{ {\sigma_F} \lt|\frac{ Y_i({q_j})-\overline{Y}({q_j})}{\sqrt{N_{q_j} S({q_j},{q_j})}}\rt|, 
\frac{ \|F({q_j}, \cdot)\|_2\cdot N_{q_j}^{-1}|Y_i({q_j})-\overline{Y}({q_j})|}{\sqrt{\varrho_{\min}\{V_{\hgamma}\}}}\rt\}.
\end{align}

Now we can take maximum over $i,j\in[N]$ in \eqref{eqn:combined-bd} and use \eqref{eqn:BE-M} to conclude the proof.
\end{proof}

\subsection{Proof of Theorem \ref{thm:quad-be-nearly-uniform}}
The proof is an application of Lemma \ref{lem:BN} (i).

\subsection{Proof of Lemma \ref{lem:suff-conds}}

\begin{proof}[Proof of Lemma \ref{lem:suff-conds}]

(i) Suppose the individual causal effects are constant. Then
\begin{align*}
    F^\top  \diag{N_q^{-1}S(q,q)}F = V_\hgamma.
\end{align*}

(ii) Suppose the condition number of the correlation matrix corresponding to $V_\hY$ is upper bounded by $\sigma^2$. Then
\begin{align*}
    Q = \sum_{q=1}^Q \varrho_{q}(V_{\hY}^\star) \le Q\cdot \varrho_{\max}(V^\star_{\hY}) \le \sigma^2 Q\cdot \rho_{\min}(V_{\hY}^\star) 
\end{align*}
which implies $\rho_{\min}(V_{\hY}^\star) \ge \sigma^{-2}.$
Let $D = \diag{(N_q^{-1} - N^{-1})S(q,q)}$. Then
\begin{align*}
    V_\hgamma &= F^\top V_{\hY} F \\
    & = F^\top D^{1/2} V_{\hY}^\star D^{1/2} F\\
    & \succeq F^\top D^{1/2} (\sigma^{-2} I_Q) D^{1/2} F \\
    & \succeq \sigma^{-2}F^\top \diag{(N_q^{-1} - N^{-1})S(q,q)} F \\
    & \succeq c\sigma^{-2}F^\top \diag{N_q^{-1} S(q,q)} F,\\
\end{align*}
where the last line follows from  $N_q \le (1-c)N$. 
\end{proof}

\commenting{
\subsection{Proof of Corollary \ref{cor:non-uniform-bound}}
\begin{proof}[Proof of Corollary \ref{cor:non-uniform-bound}]
The insight for this proof is to apply the bound \eqref{eqn:uniform-be} according to different subset of arms. Under the conditions of Theorem \ref{thm:be-proj-standard},
\begin{align*}
    & \sup_{b\in\bbR^H,\|b\|_2 =1}\sup_{t\in\bbR}\lt|\bbP\{b^\top \tilde{\gamma}  \le t\} - \Phi(t)\rt|\\
    \le & C \max_{i\in[N],q\in[Q]} \min\lt\{{\sigma_F} \lt|\frac{ Y_i(q)-\overline{Y}(q)}{\sqrt{N_q S(q,q)}}\rt|, \frac{  \|F(q, \cdot)\|_2 \cdot N_q^{-1}|Y_i(q)-\overline{Y}(q)|}{\sqrt{\varrho_{\min}\{F^\top V_\hY F\}}}\rt\},
\end{align*}
which is bounded by the maximum of following two parts:
\begin{itemize}
    \item For those arms in $\cQ_1$, we keep the first term in the upper bound \eqref{eqn:uniform-be}:
    \begin{align*}
       \max_{i\in[N],q\in\cQ_1} \sigma_F \lt|\frac{ Y_i(q)-\overline{Y}(q)}{\sqrt{N_q S(q,q)}}\rt|;
    \end{align*}
    \item For those arms in $\cQ_2$, we apply the second term in the upper bound \eqref{eqn:uniform-be}:
    \begin{align*}
       \max_{i\in[N],q\in\cQ_2} \frac{  \|F(q, \cdot)\|_2 \cdot N_q^{-1}|Y_i(q)-\overline{Y}(q)|}{\sqrt{\varrho_{\min}\{F^\top V_\hY F\}}}.
    \end{align*}
\end{itemize}
Hence we conclude the proof.

\end{proof}
}

\subsection{Proof of Corollary \ref{cor:uniform-design-be}}\label{sec:uniform-design-be}
\begin{proof}[Proof of Corollary \ref{cor:uniform-design-be}]
It suffices to further control term II of the upper bound in \eqref{eqn:uniform-be}:
\begin{align*}
    \frac{\|F(q, \cdot)\|_2\cdot N_q^{-1}|Y_i(q)-\overline{Y}(q)|}{\sqrt{\varrho_{\min}\lt(\text{Cov} \{F^\top\hY\}\rt)}}.
\end{align*}
By definition of 2-norm, $\|F(q, \cdot)\|_2 \le \sqrt{H}\|F\|_\infty$. By Definition \ref{def:uniform-design}, $N_q^{-1} \le \underline{c}^{-1}N_0^{-1}$.  
Under \eqref{eqn:well-conditioned},
\begin{align}
    \varrho_{\min}\left( \text{Cov} \{F^\top\hY\} \right ) 
    \ge& \sigma_F^{-2} \varrho_{\min}\{F^\top  \diag{N_q^{-1}S(q,q)}F\}\label{eqn:recheck} \\
    \ge & \sigma_F^{-2} \varrho_{\min}\{F^\top  F\} \min_{q\in[Q]}\lt\{N_q^{-1}S(q,q)\rt\} \\
    \ge & \overline{c}^{-1}N_0^{-1}\sigma_F^{-2} \varrho_{\min}\{F^\top  F\} \min_{q\in[Q]} S(q,q) .
\end{align}
Now \eqref{eqn:uniform-design-be} is obtained by using \eqref{eq::contrast-matrix} and plugging in these results.

\end{proof}

\begin{remark}\label{rmk:relax-minimal-S}
    In the comments following Corollary \ref{cor:uniform-design-be}, we mentioned that the positive minimal variance condition $\min_{q\in[Q]} S(q,q)$ can be relaxed to allow where some $S(q,q)$’s are close or equal to zero. The idea is to recheck Step \eqref{eqn:recheck} of the proof. Instead of taking the minimal $S(q,q)$, we can take a thresholding value $\tilde{S}$, collect those $q$'s that have $S(q,q) > \tilde{S}$ into a set ${\bbS}$,  then get a lower bound as follows:
    \begin{align*}
        \varrho_{\min}\{F^\top  \diag{N_q^{-1}S(q,q)}F\} \ge \overline{c}^{-1}N_0^{-1} \tilde{S}\varrho_{\min}\{F(\bbS,\cdot)^\top F(\bbS,\cdot)\}. 
    \end{align*}
\end{remark}

\subsection{Proof of Corollary \ref{cor:non-uniform-design-be}}\label{sec:pf-non-uniform-design-be}

The key idea is to find explicit bounds on the BEB given by Theorem \ref{thm:be-proj-standard} for general designs. We partition the arms into $\cQ_{\textsc{s}} \cup \cQ_{\textsc{l}}$, and apply different parts in the general BEB in \eqref{eqn::terms1and2-be-proj-standard} to these two groups respectively.

\begin{proof}[Proof of Corollary \ref{cor:non-uniform-design-be}]
Recall the bound \eqref{eqn::terms1and2-be-proj-standard} in Theorem \ref{thm:be-proj-standard}.
The two parts in the upper bound shall be applied to different categories of arms from Definition \ref{def:non-uniform-design}. Because each $N_q$ is large for $q\in\cQ_{\textsc{l}} $, we keep the first part of \eqref{eqn:uniform-be} for $\cQ_{\textsc{l}} $:
\begin{align}\label{eqn:bd-QL}
    \max_{i\in[N], q\in \cQ_{\textsc{l}}} {\sigma_F} \lt|\frac{ Y_i(q)-\overline{Y}(q)}{\sqrt{N_q S(q,q)}}\rt|.
\end{align}

For the small groups in $ \cQ_{\textsc{s}} $, we apply the second part of \eqref{eqn:uniform-be}. First, we have $N_q^{-1} \le 1$. Besides, under \eqref{eqn:well-conditioned},
\begin{align}\label{eqn:bd-QS}
    \rho_{\min}\left( \COV{F^\top\hY} \right) &\ge \sigma_F^{-2} \rho_{\min}\{F^\top \diag{N_q^{-1}S(q,q)}_{q\in[Q]} F\}\notag\\
    &\ge \sigma_F^{-2} \rho_{\min}\lt\{F_{\textsc{s}}^\top \diag{N_q^{-1}S(q,q)}_{q\in \cQ_{\textsc{s}}}  
     F_{\textsc{s}}
     \rt\}\notag\\
    & \ge \sigma_F^{-2} \min_{q\in \cQ_{\textsc{s}}} \{N_q^{-1}S(q,q)\}\varrho_{\min}\{ F_{\textsc{s}}^\top F_{\textsc{s}}\} \notag\\
    & \ge \sigma_F^{-2}\overline{n}^{-1} \min_{q\in \cQ_{\textsc{s}}} \{S(q,q)\}\varrho_{\min}\{ F_{\textsc{s}}^\top F_{\textsc{s}}\}.
\end{align}
Hence for $q\in\cQ_{\textsc{s}}$, we keep the second part of \eqref{eqn:uniform-be} and use \eqref{eqn:bd-QS} to obtain upper bound:
\begin{align*}
    \frac{\underline{c}^{-1}\overline{n}^{-1} \sigma_F\|F(q, \cdot)\|_2 \cdot |Y_i(q)-\overline{Y}(q)|}{ (\overline{n}^{-1}\min_{q\in \cQ_{\textsc{s}}} \{S(q,q)\}\varrho_{\min}\{ F_{\textsc{s}}^\top F_{\textsc{s}}\})^{1/2}}. 
\end{align*}

Under Condition \ref{condition::proper-non-uniform}, we  have 
$$
\|F(q,\cdot)\|_2  \le c|\cQ_\textsc{s}|^{-1}\sqrt{H}
$$
    and
\begin{align*}
    \varrho_{\min}\{ F_{\textsc{s}}^\top F_{\textsc{s}}\}
    \ge  c'|\cQ_\textsc{s}|^{-1}.
\end{align*}
Hence,
\begin{align}\label{eqn:F-2-over-sqrt-rho}
    \frac{\|F(q,\cdot)\|_2}{(\varrho_{\min}\{ F_{\textsc{s}}^\top F_{\textsc{s}}\})^{1/2}} \le \sqrt{\frac{c^2H}{c'|\cQ_\textsc{s}|}}.
\end{align}


Putting \eqref{eqn:bd-QL}, \eqref{eqn:bd-QS} and \eqref{eqn:F-2-over-sqrt-rho} into \eqref{eqn:uniform-be} concludes the proof.

\end{proof}

\subsection{Proof of Theorem \ref{thm:uniform-var}}\label{sec:pf-uniform-var}
\begin{proof}[Proof of Theorem \ref{thm:uniform-var}]

\begin{enumerate}[label=(\roman*)]
    \item It  is well known. 
    \item 
    For the stochastic order in $L_\infty$ norm, we shall apply Lemma \ref{lem:tail} with $\cQ = [Q]$.  We have 
\begin{align*}
    \hV_{\hgamma}(h,h') &=  \sum_{q \in \cQ} F(h,q)F(h',q) N_q^{-1} \hat{S}(q,q)\\
    &=  \sum_{q \in \cQ} w_q N_q^{-1}\hat{S}(q,q),
\end{align*}
where
\begin{align*}
    w_q = F(h,q)F(h',q),\quad  |w_q| \le \|F\|_\infty^2. 
\end{align*}
Applying  Lemma \ref{lem:tail}, we have
\begin{align*}
    \bbP\lt\{\lt|\hv - \E{\hv}\rt|\ge t\rt\} \le \frac{C\overline{c}\underline{c}^{-4} \|F\|_\infty^4 Q N_0^{-3}\Delta^4}{t^2} := \frac{\circledast_1}{t^2},
\end{align*}
which implies
\begin{align*}
    \forall h,h'\in[H],~\bbP\lt\{|\hV_{\hgamma}(h,h') - \bbE\{\hV_{\hgamma}(h,h')\}|> t\rt\} 
    \le \frac{\circledast_1}{t^2}.
\end{align*}
Taking union bound over $h,h'\in[H]$, we have
\begin{align*}
    \bbP\lt\{\max_{h,h'\in[H]}|\hV_{\hgamma}(h,h') - \bbE\{\hV_{\hgamma}(h,h')\}|> t\rt\} 
    \le \frac{\circledast_1 \cdot H^2}{t^2}.
\end{align*}
Therefore, 
\begin{align*}
    \|\hV_{\hgamma}-\bbE\{\hV_{\hgamma}\}\|^2_{\infty} = O_{\bbP}\lt(\circledast_1 \cdot H^2\rt).
\end{align*}
    \item
 It follows from    \eqref{eqn:op-inf}. 
\end{enumerate}

\end{proof}

\subsection{Proof of Theorem \ref{thm:wald-uniform}}\label{sec:pf-wald-uniform}

We present and prove a more general result below, from which Theorem \ref{thm:wald-uniform} can be deduced as a corollary.

\begin{theorem}[Wald-type inference under replicated nearly uniform design]\label{thm:wald-uniform-general} 
Consider the nearly uniform design given by Definition \ref{def:uniform-design} that satisfies $\min_{q\in[Q]}N_q\ge 2$ and Condition \ref{condition::proper}. Also assume \eqref{eqn:well-conditioned} and Conditions \ref{cond:moments} and \ref{cond:easy-spec}. 
Define 
$W_N = V_\hgamma^{1/2}\bbE\{\hV_\hgamma\}^{-1}V_\hgamma^{1/2} \in \bbR^{H\times H}
$, which is a deterministic quantity.
Let $N\rightarrow \infty$. 
\begin{enumerate}[label = (\roman*)]
    \item For a fixed $H$, the Wald-type confidence set is asymptotically valid. Moreover,  assume there exists a {$W_\infty\in\bbR^{H\times H}$} such that
    $ \lim_{N\to\infty}W_N = W_\infty $.
    Use $\cL$ to denote the distribution of
    $
    \xi_H^\top W_\infty \xi_H , \text{ where } \xi_H \sim \cN(0,I_H).
    $
    We have 
    \begin{align*}
        (\hgamma - \gamma )^\top \hV_{\hgamma}^{-1} (\hgamma - \gamma) \rightsquigarrow \cL   
    \end{align*}
    and $\cL \lesssim \chi^2_H$.

    \item For a diverging $H$ with $H\rightarrow \infty$ and $H^{19/4}N^{-1/2} \to 0$,  the Wald-type confidence set is asymptotically valid. Moreover, we have  
    \begin{align*}
        \frac{(\hgamma - \gamma)^\top  \hV_{\hgamma}^{-1} (\hgamma - \gamma) - \trace{W_N}}{\sqrt{2\trace{W_N^2}}} \rightsquigarrow \cN(0,1) .
    \end{align*}
\end{enumerate}
\end{theorem}

\begin{proof}[Proof of Theorem \ref{thm:wald-uniform-general}]

We prove the ``fixed $H$'' and ``diverging $H$'' scenarios separately. In each scenario, we apply BEBs to obtain CLTs with the true variances, and then apply the variance estimation results to justify the statistical properties   after plugging in the variance estimators. 

\begin{enumerate}[label = (\roman*)]
    \item Consider fixed $H$. By Corollary \ref{cor:uniform-design-be}, under \eqref{eqn:well-conditioned}, Conditions   \ref{condition::proper} and \ref{cond:easy-spec}, the property of joint asymptotic normality holds:
    \begin{align*}
        V_{\hgamma}^{-1/2}(\hgamma-\gamma) \rightsquigarrow \cN(0, I_H).
    \end{align*}
    The continuous mapping theorem implies
    \begin{gather*}
        (\hgamma - \gamma)^\top V_{\hgamma}^{-1} (\hgamma - \gamma) \rightsquigarrow \chi^2_H.
    \end{gather*}
    Because $\bbE\{\hV_{\hgamma}\} \succeq V_\hgamma$, we have the stochastic dominance
    \begin{align*}
        (\hgamma - \gamma)^\top V_{\hgamma}^{-1} (\hgamma - \gamma) \gtrsim (\hgamma - \gamma)^\top \bbE\{\hV_{\hgamma}\}^{-1} (\hgamma - \gamma). 
    \end{align*}
    
    For variance estimation, under \eqref{eqn:well-conditioned}, Conditions  \ref{condition::proper}, \ref{cond:moments} and \ref{cond:easy-spec}, the stochastic order in Theorem \ref{thm:uniform-var} implies
    $N\hV_{\hgamma} - N\bbE\{\hV_{\hgamma}\} = o_\bbP(1)$. Hence $N^{-1}\hV_{\hgamma}^{-1} - N^{-1}\bbE\{\hV_{\hgamma}\}^{-1} = o_\bbP(1)$.
    Moreover,
    \begin{align*}
        \|\hgamma - \gamma\|_2^2 = O_\bbP(\|V_\hgamma\|_{\operatorname{op}}) = O_\bbP(\|F^\top\diag{N_q^{-1}S(q,q)} F\|_{\operatorname{op}}) = O_\bbP(\Delta^2N_0^{-1}\|F^\top F\|_{\operatorname{op}}).
    \end{align*}
    Using
    \begin{align*}
        \|F^\top F\|_{\operatorname{op}} \le \trace{F^\top F} \le c^2Q^{-2} \cdot (QH) \le c^2Q^{-1}H,
    \end{align*}
    we can  derive
    \begin{align*}
        {N}\|\hgamma - \gamma\|_2^2 = O_\bbP(H).
    \end{align*}
    
    Therefore,
    \begin{align}
        (\hgamma - \gamma)^\top \hV_{\hgamma}^{-1} (\hgamma - \gamma) &= (\hgamma - \gamma)^\top \E{\hV_{\hgamma}}^{-1} (\hgamma - \gamma) \\
        &+ (\hgamma - \gamma)^\top \lt( \hV_{\hgamma}^{-1} - \E{\hV_{\hgamma}}^{-1}\rt) (\hgamma - \gamma), \label{eqn:infer-uniform-1}
    \end{align}
    where \eqref{eqn:infer-uniform-1} has the order
    \begin{align*}
    (\hgamma - \gamma)^\top \lt( \hV_{\hgamma}^{-1} - \E{\hV_{\hgamma}}^{-1}\rt) (\hgamma - \gamma) = O_\bbP(H) \cdot o_\bbP(1) = o_\bbP(1). 
    \end{align*}
    Now using Slutsky's theorem, we conclude the validity of the Wald-type confidence interval. 

    Moreover, under the assumption that $W_N \to W_\infty$, we can use the continuous mapping theorem to derive
    \begin{align*}
        (\hgamma - \gamma)^\top \bbE\{\hV_{\hgamma}\}^{-1} (\hgamma - \gamma) = (\hgamma - \gamma)^\top V_{\hgamma}^{-1/2} W_N V_{\hgamma}^{-1/2}(\hgamma - \gamma) \rightsquigarrow \cL.
    \end{align*}
    Because $\bbE\{\hV_{\hgamma}\} \succeq V_\hgamma$, we have $I_H \succeq W_\infty$ and $ \cL \lesssim \chi^2_H$.

    \item Consider diverging $H$. We use a quadratic form CLT stated in Corollary \ref{cor:quad-clt-v2}. By Corollary \ref{cor:quad-clt-v2}, under  \eqref{eqn:well-conditioned}, Conditions \ref{condition::proper} and \ref{cond:easy-spec}, when $H^{19/4}N^{-1/2} \to 0$, with $W_N = V_{\hgamma}^{1/2} \E{\hV_{\hgamma}}^{-1} V_{\hgamma}^{1/2} $, we have
    \begin{align} \label{eqn::qclt-true-variance}
        \frac{(\hgamma - \gamma)^\top \E{\hV_{\hgamma}}^{-1} (\hgamma - \gamma) - \trace{W_N^2}}{\sqrt{2\trace{W_N}}} \rightsquigarrow \cN(0,1).
    \end{align}
    Because $\bbE\{\hV_{\hgamma}\} \succeq V_\hgamma$, we have
    \begin{align}\label{eqn:trQ-Q2}
        \trace{W_N} \le H,\quad \trace{W_N^2} \le H. 
    \end{align}
    Now we consider the difference induced by plugging in the variance estimator:
    \begin{align*}
        &|(\hgamma - \gamma)^\top \lt(\hV_{\hgamma}^{-1} - \bbE\{\hV_{\hgamma}\}^{-1}\rt) (\hgamma - \gamma)|\\
        \le & |(\hgamma - \gamma)^\top V_{\hgamma}^{-1}(\hgamma - \gamma)|\cdot \|V_{\hgamma}\|_{\text{op}} \cdot \|\hV_{\hgamma}^{-1} - \bbE\{\hV_{\hgamma}\}^{-1}\|_{\operatorname{op}}.
    \end{align*}
Use the matrix identity 
    \begin{align*}
        \hV_{\hgamma}^{-1} - \bbE\{\hV_{\hgamma}\}^{-1} = -\hV_{\hgamma}^{-1} (\hV_{\hgamma} - \bbE\{\hV_{\hgamma}\}) \bbE\{\hV_{\hgamma}\}^{-1} , 
    \end{align*}
we can verify that
    \begin{align*}
        \|\hV_{\hgamma}^{-1} - \bbE\{\hV_{\hgamma}\}^{-1}\|_{\operatorname{op}} &= \|\hV_{\hgamma}^{-1} (\hV_{\hgamma} - \bbE\{\hV_{\hgamma}\}) \bbE\{\hV_{\hgamma}\}^{-1}\|_{\operatorname{op}}\\
        &\le \|\hV_{\hgamma}^{-1}\|_{\operatorname{op}} \|\hV_{\hgamma} - \bbE\{\hV_{\hgamma}\}\|_{\operatorname{op}}\|\bbE\{\hV_{\hgamma}\}^{-1}\|_{\operatorname{op}}.
    \end{align*}
     Theorem \ref{thm:uniform-var} ensures
    \begin{gather*}
        N\|\hV_{\hgamma} - \bbE\{\hV_{\hgamma}\}\|_{\operatorname{op}} = O_{\bbP}({H^{2}N^{-1/2}}), \\ \|\bbE\{\hV_{\hgamma}\}^{-1}\|_{\operatorname{op}} = O(N), \quad \|\hV_{\hgamma}^{-1}\|_{\operatorname{op}} = O_{\bbP}(N), \quad \|V_{\hgamma}\|_{\operatorname{op}} = O_{\bbP}(N^{-1}).  
    \end{gather*}
    Corollary \ref{cor:quad-clt-v2} ensures
    \begin{align*}
        (\hgamma - \gamma)^\top V_{\hgamma}^{-1}(\hgamma - \gamma) = O_{\bbP}(H).
    \end{align*}
    Under \eqref{eqn:well-conditioned}, we have
    \begin{align*}
        \trace{W_N^2} &= \trace{V_\hgamma^{1/2}\E{\hV_\hgamma}^{-1}V_\hgamma\E{\hV_\hgamma}^{-1}V_\hgamma^{1/2}} \\
        &\ge \sigma_F^{-2}\trace{V_\hgamma^{1/2}\E{\hV_\hgamma}^{-1} V_\hgamma^{1/2}} \\  
        & = \sigma_F^{-2}\trace{\E{\hV_\hgamma}^{-1} V_\hgamma} \\  
        &\ge \sigma_F^{-4} \trace{I_H} = \sigma_F^{-4} H. 
    \end{align*}
    Using these results, we obtain
    \begin{align}\label{eqn:small-dev}
        \frac{|(\hgamma - \gamma)^\top \lt(\hV_{\hgamma}^{-1} - \bbE\{\hV_{\hgamma}\}^{-1}\rt) (\hgamma - \gamma)|}{\sqrt{2\trace{ W_N^2}}} = O_{\bbP}(H^{5/2}N^{-1/2}),
    \end{align}
    which converges to $0$ if $H^{19/4}N^{-1/2} \to 0$. Combine \eqref{eqn::qclt-true-variance} and \eqref{eqn:small-dev} to establish the desired CLT.
    
    To prove the validity of the confidence set, we notice that 
    \begin{align}
        &\Prob{(\hgamma - \gamma)^\top  \hV_{\hgamma}^{-1} (\hgamma - \gamma) \ge q_{H, \alpha}} \notag\\
        \le &\bbP\Big\{(\hgamma - \gamma)^\top  \bbE\{\hV_{\hgamma}\}^{-1} (\hgamma - \gamma) \\
        \phantom{\le} & + |(\hgamma - \gamma)^\top \lt(\hV_{\hgamma}^{-1} - \bbE\{\hV_{\hgamma}\}^{-1}\rt) (\hgamma - \gamma)| \ge q_{H, \alpha}\Big\} \notag\\
        \le &\Prob{(\hgamma - \gamma)^\top  \bbE\{\hV_{\hgamma}\}^{-1} (\hgamma - \gamma)\ge q_{H, \alpha} - cH^4N^{-1/2}} \label{eqn:main-tail}\\
        + &\Prob{|(\hgamma - \gamma)^\top \lt(\hV_{\hgamma}^{-1} - \bbE\{\hV_{\hgamma}\}^{-1}\rt) (\hgamma - \gamma)| \ge cH^4N^{-1/2} }. \label{eqn:small-tail}
    \end{align}
    Using \eqref{eqn:trQ-Q2} and \eqref{eqn:small-dev}, \eqref{eqn:small-tail} converges to zero if $ H^{19/4}N^{-1/2}\to 0 $. 
    
    For \eqref{eqn:main-tail}, using Lemma \ref{lem:BN} (i), 
    \begin{align}\label{eqn:main-tail-1}
        \lt|\Prob{(\hgamma - \gamma)^\top  \bbE\{\hV_{\hgamma}\}^{-1} (\hgamma - \gamma)\ge q_{H, \alpha} - cH^4N^{-1/2}} -  \Prob{\xi_H^\top W_N \xi_H\ge q_{H, \alpha} - cH^4N^{-1/2}}\rt| = o(1).
    \end{align}
    
    Now because $W_N \preceq I_H$, 
    \begin{align}\label{eqn:main-tail-2}
        \Prob{\xi_H^\top W_N \xi_H\ge q_{H, \alpha} - cH^4N^{-1/2}} \le \Prob{\xi_H^\top \xi_H\ge q_{H, \alpha} - cH^4N^{-1/2}}.
    \end{align}
    Moreover,  
    \begin{align}\label{eqn:main-tail-3}
       \sup_{t\in\bbR}\lt|\Prob{\frac{\xi_H^\top  \xi_H - H}{\sqrt{2H}}\le t} - \Phi(t)\rt| = o(1) \text{ as } H \to \infty.
    \end{align}
    Hence, 
    \begin{align}\label{eqn:main-tail-4}
        &\lt|\Prob{\frac{\xi_H^\top  \xi_H - H}{\sqrt{2H}}\ge \frac{q_{H, \alpha} - H - cH^4N^{-1/2} }{\sqrt{2H}}} -  \Prob{\frac{\xi_H^\top  \xi_H - H}{\sqrt{2H}}\ge \frac{q_{H, \alpha} - H}{\sqrt{2H}}}\rt| \notag\\
        = & \lt| \Phi\lt(\frac{q_{H, \alpha} - H - cH^4N^{-1/2} }{\sqrt{2H}}\rt) - \Phi\lt(\frac{q_{H, \alpha} - H}{\sqrt{2H}}\rt)\rt| + o(1) 
        \le  \frac{1}{\sqrt{2\pi}}H^{7/2}N^{-1/2} + o(1) = o(1).
    \end{align}
    Combining \eqref{eqn:main-tail-1}--\eqref{eqn:main-tail-4}, we conclude that, if $ H^{19/4}N^{-1/2}\to 0$, then
    \begin{align}\label{eqn:main-tail-5}
        \lim_{N\to\infty}\Prob{(\hgamma - \gamma)^\top  \bbE\{\hV_{\hgamma}\}^{-1} (\hgamma - \gamma)\ge q_{H, \alpha} - cH^4N^{-1/2}} \le \alpha.
    \end{align}
    From \eqref{eqn:main-tail} and \eqref{eqn:small-tail}, we conclude the asymptotic validity of the Wald-type inference.
\end{enumerate}

\end{proof}

\subsection{Proof of Theorem \ref{thm:quad-be-unreplicated}}

The proof is an application of Lemma \ref{lem:BN} (i).

\subsection{Proof of Lemma \ref{lemma::mean-variance-grouping} and Theorem \ref{thm:unreplicated-var}}\label{pf:lem-mean-var-grouping-thm-unrep-var}

Lemma \ref{lemma::mean-variance-grouping} is a special case of Theorem \ref{thm:unreplicated-var}\ref{thm:unreplicated-var-1}. We first give a proof for Theorem \ref{thm:unreplicated-var}, then add some discussions on improving variance estimation under more assumptions. 

\begin{proof}[Proof of  Theorem \ref{thm:unreplicated-var}]

\begin{enumerate}[label = (\roman*)]
    \item We first compute the expectation of \eqref{eqn:hV-QU}:
\begin{align}
    &\bbE\{  (Y_q - \hY_{\langle g \rangle})^2\} \notag\\
    =& \bbE\{  (Y_q - \overline{Y}(q) + \overline{Y}(q) - \overline{Y}_{\langle g \rangle} + \overline{Y}_{\langle g \rangle} - \hY_{\langle g \rangle})^2\} \notag\\
    =& \bbE\{  (Y_q - \overline{Y}(q) +  \overline{Y}_{\langle g \rangle} - \hY_{\langle g \rangle})^2\} + \bbE\{ (\overline{Y}(q) - \overline{Y}_{\langle g \rangle})^2\}\notag\\
    = &   (1-|{\langle g \rangle}|^{-1})^2 \bbE\{(Y_q - \overline{Y}(q))^2\}\notag\\
    + &   |{\langle g \rangle}|^{-2}  \bbE\lt[\lt\{\sum_{q'\in{\langle g \rangle},q'\neq q}(Y_{q'} - \overline{Y}(q'))\rt\}^2\rt] \notag\\
    - & 2  (1-|{\langle g \rangle}|^{-1})|{\langle g \rangle}|^{-1} \sum_{q'\in{\langle g \rangle},q'\neq q} \bbE\{(Y_q - \overline{Y}(q))(Y_{q'} - \overline{Y}(q'))\}\notag\\
    + &   \bbE\{(\overline{Y}(q) - \overline{Y}_{\langle g \rangle})^2\} \notag\\
    = &   (1-|{\langle g \rangle}|^{-1})^2 (1-N^{-1})S(q,q)\label{eqn:var-q}\\
    + &   |{\langle g \rangle}|^{-2}  \bbE\lt[\lt\{\sum_{q'\in{\langle g \rangle},q'\neq q}(Y_{q'} - \overline{Y}(q'))\rt\}^2\rt] \label{eqn:var-qprime}\\
    - & 2  (1-|{\langle g \rangle}|^{-1})|{\langle g \rangle}|^{-1}N^{-1} \sum_{q'\in{\langle g \rangle},q'\neq q} S(q,q')\label{eqn:cov-q}\\
    + &    (\overline{Y}(q) - \overline{Y}_{\langle g \rangle})^2 .\label{eqn:var-bg} 
\end{align}

\eqref{eqn:var-q} reflects the within group variation for arm $q$. \eqref{eqn:var-qprime} reflects the pooled variation for the arms except $q$ in group ${\langle g \rangle}$. \eqref{eqn:cov-q} captures the correlation between arm $q$ and the rest in ${\langle g \rangle}$. \eqref{eqn:var-bg} represents the between-group variation.

For \eqref{eqn:var-qprime}, we have 
\begin{align}
    &  |{\langle g \rangle}|^{-2}  \bbE\lt[\lt\{\sum_{q'\in{\langle g \rangle},q'\neq q}(Y_{q'} - \overline{Y}(q'))\rt\}^2\rt] \notag\\
    = &   |{\langle g \rangle}|^{-2} 1^\top_{|\langle g \rangle|}[\diag{S(q',q')}_{q'\in{\langle g \rangle},q'\neq q} - N^{-1}(S(q',q''))_{q',q''\in\langle g\rangle\backslash \{q\}}] 1_{|\langle g \rangle|} \notag\\
    = &   |{\langle g \rangle}|^{-2} (1-N^{-1}\varrho_{\langle g \rangle})\sum_{q'\in{\langle g \rangle},q'\neq q} S(q',q') + N^{-1}\mu_{\langle g \rangle}^{-1} \Theta_1(q,q),
\end{align}
where
\begin{gather}
     \Theta_1(q,q) = \mu_{\langle g \rangle} |\langle g \rangle|^{-2} 1^{\top}_{|\langle g \rangle|}\{\varrho_{\langle g \rangle}\diag{S(q',q')}_{q'\in{\langle g \rangle},q'\neq q} - (S(q',q''))_{q',q''\in\langle g\rangle\backslash \{q\}}\} 1_{|\langle g \rangle|} \ge 0. \label{eqn:Theta-1}
\end{gather}

We can upper bound $\Theta_1(q,q)$ as follows:
\begin{align*}
    \Theta_1(q,q) &\le \mu_{\langle g \rangle} |\langle g \rangle|^{-2}1_{|\langle g \rangle|}^{\top}\{\varrho_{\langle g \rangle}\diag{S(q',q')}_{q'\in{\langle g \rangle},q'\neq q}\}1_{|\langle g \rangle|} \\
    & = \mu_{\langle g \rangle} \cdot \frac{\varrho_{\langle g \rangle}}{|\langle g \rangle|} \cdot |\langle g \rangle|^{-1} \sum_{q'\in{\langle g \rangle},q'\neq q} S(q',q') \\
    & \le \mu_{\langle g \rangle} |\langle g \rangle|^{-1} \sum_{q'\in{\langle g \rangle}} S(q',q') \le \mu_{\langle g \rangle}  \max_{q'\in{\langle g \rangle}} S(q',q').
\end{align*}

For \eqref{eqn:cov-q}, we have
\begin{align}
    &2  (1-|{\langle g \rangle}|^{-1})|{\langle g \rangle}|^{-1}N^{-1} \sum_{q'\in{\langle g \rangle},q'\neq q} S(q,q') \notag\\
    \le & 2  (1-|{\langle g \rangle}|^{-1})|{\langle g \rangle}|^{-1}N^{-1} \sum_{q'\in{\langle g \rangle},q'\neq q} \sqrt{S(q,q)S(q',q')} \see{by the Cauchy-Schwarz inequality} \notag\\
    \le & 2  (1-|{\langle g \rangle}|^{-1})|{\langle g \rangle}|^{-1}N^{-1} \sum_{q'\in{\langle g \rangle},q'\neq q} \lt\{\frac{S(q,q) + S(q',q')}{2}\rt\} \notag\\
    \le & 
      (1-|{\langle g \rangle}|^{-1})|{\langle g \rangle}|^{-1}N^{-1} (|{\langle g \rangle}|-1) S(q,q)
    +
      (1-|{\langle g \rangle}|^{-1})|{\langle g \rangle}|^{-1}N^{-1} \sum_{q'\in{\langle g \rangle},q'\neq q} S(q',q') \notag\\
    = & (1-|{\langle g \rangle}|^{-1})^2 N^{-1} S(q,q)
    +
      (1-|{\langle g \rangle}|^{-1})|{\langle g \rangle}|^{-1}N^{-1} \sum_{q'\in{\langle g \rangle},q'\neq q} S(q',q'). \label{eqn:cov-q-2}
\end{align}

Define 
\begin{align}\label{eqn:Theta-2}
    \Theta_2(q,q) = \mu_{\langle g \rangle} \frac{(1-|{\langle g \rangle}|^{-1})}{|{\langle g \rangle}|}\sum_{q'\in{\langle g \rangle},q'\neq q} \lt\{{S(q,q) + S(q',q') - 2S(q,q')}\rt\}.
\end{align}

We have $\Theta_2(q,q) \ge 0$. We can upper bound $\Theta_2(q,q)$ by
\begin{align*}
    \Theta_2(q,q) \le 4\mu_{\langle g \rangle} \max_{q'\in \langle g \rangle} S(q',q').
\end{align*}

Now using \eqref{eqn:rho-max-S} and \eqref{eqn:cov-q-2}, we have
\begin{align*}
    \bbE\{\hV_\hY(q,q)\} 
    = &   \mu_{\langle g \rangle} (1-|{\langle g \rangle}|^{-1})^2 (1- 2N^{-1})S(q,q) + \mu_{\langle g \rangle} \bbE\{(\overline{Y}(q) - \overline{Y}_{\langle g \rangle})^2\} \\
    + & \mu_{\langle g \rangle} |{\langle g \rangle}|^{-2}N^{-1} \{N- \varrho_{\langle g \rangle} -  (|{\langle g \rangle}|-1)\}\sum_{q'\in{\langle g \rangle}, q'\neq q} S(q',q')\\
    + & N^{-1}\Theta(q,q),
\end{align*}
where
\begin{align}\label{eqn:Theta-qq}
    \Theta(q,q) = \Theta_1(q,q) + \Theta_2(q,q), \quad 0\le \Theta(q,q) \le 5\mu_{\langle g \rangle} \max_{q'\in \langle g \rangle} S(q',q').
\end{align}

Using $\mu_{\langle g \rangle} = (1-|{\langle g \rangle}|^{-1})^{-2} (1- 2N^{-1})^{-1}$ and Condition \ref{cond:cond-N}, we obtain that 
\begin{align*}
    \bbE\{\hV_\hY(q,q)\}  \ge S(q,q) + \mu_{\langle g \rangle}  (\overline{Y}(q) - \overline{Y}_{\langle g \rangle})^2  \ge S(q,q).  
\end{align*}

  \item 
We can show that
\begin{align*}
     \hV_{\hgamma}(h,h')  =&   \sum_{q \in \cQ} F(h,q)F(h',q) \hV_\hY(q,q) \\
     =&
     \sum_{{\langle g \rangle}\in{\langle g \rangle}}\sum_{q\in{\langle g \rangle}}w_q \lt (Y_q - \hY_{{\langle g \rangle}}\rt)^2 
\end{align*}
where 
\begin{align*}
    w_q = \mu_{\langle g \rangle} F(h,q)F(h',q), \text{ if $q\in{\langle g \rangle}$}
\end{align*}
satisfies 
\begin{align*}
    |w_q| \le (\max_{g\in[G]}\mu_{\langle g \rangle}) \|F\|_\infty^2 := \overline{w}. 
\end{align*}

Applying Lemma \ref{lem:hv-U}, we have 
\begin{align*}
    &\Prob{ |\hV_{\hgamma}(h,h') - \E{\hV_{\hgamma}(h,h')}|\ge t}\\
    \le &\frac{C\overline{w}^2\Delta^2
    (\Delta^2+\zeta^2)N}{t^2} =
    \frac{C(\max_{g\in[G]}\mu_{\langle g \rangle})^2 \|F\|_\infty^4\Delta^2
    (\Delta^2+\zeta^2)N}{t^2} := \circledast_2 .
\end{align*}

Taking union bound over $h,h'\in[H]$, we obtain
\begin{align*}
     \Prob{\|\hV_{\hgamma} - \E{\hV_{\hgamma}} \|_\infty \ge t}
    \le\frac{\circledast_2\cdot H^2}{t^2}.
\end{align*}

\item It follows from \eqref{eqn:op-inf}.
\end{enumerate}

\end{proof}

\textbf{More discussions on the conservativeness of $\hV_\hY$}. Theorem \ref{thm:unreplicated-var}\ref{thm:unreplicated-var-1} shows
\begin{align*}
    \bbE\{\hV_\hY\} 
    =  V_\hY  +  \Omega  +  \diag{\mu_{\langle g \rangle}(\overline{Y}(q) - \overline{Y}_{\langle g \rangle})^2}_{q\in\cQ_{\textsc{u}}} + N^{-1} (\Theta + S).  
\end{align*}

Following Lemma \ref{lemma::mean-variance-grouping}, we commented that the conservativeness can be reduced under different assumptions:
\begin{itemize}
    \item If we assume homogeneity in means within subgroups, i.e.,
\begin{align}\label{eqn:weak-means-supp}
    \overline{Y}(q) = \overline{Y}_{\langle g \rangle},~ \text{ for all } q\in\langle g \rangle,
\end{align}
then the term 
$$
\diag{\mu_{\langle g \rangle}(\overline{Y}(q) - \overline{Y}_{\langle g \rangle})^2}_{q\in\cQ_{\textsc{u}}}
$$ vanishes.

   \item If we assume homoskedasticity across treatment arms within the same subgroup, i.e.,
\begin{align}\label{eqn:weak-vars-supp}
    S(q,q) = S(q',q'),~ \text{ for all } q,q'\in \langle g \rangle,
\end{align}
then $ \Omega $ has diagonals: 
\begin{align*}
    {\Omega}(q,q) = \mu_{\langle g \rangle}(|g|-1)|g|^{-2}\lt(1-\frac{\varrho_{\langle g \rangle}}{N} - \frac{|g|-1}{N}\rt)S(q,q),
\end{align*}
which can also contribute to $S(q,q)$ and suggest that we can use a smaller correction factor $\mu'_{\langle g \rangle}$ to reduce the conservativeness:
\begin{align*}
    \mu'_{\langle g \rangle} = (1-|g|^{-1})^{-1}\{(1-|g|^{-1})(1-2N^{-1}) + |g|^{-1}(1 - (2|g| - 1)/N) \}^{-1} \le \mu_{\langle g \rangle}.
\end{align*}
When $|g|$ is large (say of the same order as $N$), $\mu'_{\langle g \rangle}$ is close to $ \mu_{\langle g \rangle}$ because $|g|^{-1}(1 - (2|g| - 1)/N)$ is small. When $|g|$ is small, say for pairing, $|g| = 2$, 
$$
\mu'_{\langle g \rangle} \le 2(1-3N^{-1})^{-1}, \quad \mu_{\langle g \rangle} = 4(1-2N^{-1})^{-1}.
$$
Hence $ \mu'_{\langle g \rangle} $ induces much less conservativeness than $ \mu_{\langle g \rangle} $ under stronger assumptions.

  \item If we assume the strong null hypothesis within subgroups, i.e.,
  \begin{align}\label{eqn:strong-po-supp}
      Y_i(q) = Y_i(q'), \text{ for all }i\in[N] \text{ and } q,q'\in {\langle g \rangle},
  \end{align}
  then both \eqref{eqn:weak-means-supp} and \eqref{eqn:weak-vars-supp} are satisfied. Then 
  \begin{gather*}
      \diag{\mu_{\langle g \rangle}(\overline{Y}(q) - \overline{Y}_{\langle g \rangle})^2}_{q\in\cQ_{\textsc{u}}} = 0,\\
      \Theta = 0 \see{ by the definitions of $\Theta_1$ in \eqref{eqn:Theta-1} and $\Theta_2$ in \eqref{eqn:Theta-2}}.
  \end{gather*}
  
  Applying the correction factor $\mu'_{\langle g \rangle}$, we can show 
  \begin{align*}
      \bbE\{\hV_\hY(q,q)\} = S(q,q).
  \end{align*}
\end{itemize}

\subsection{Proof of Theorem \ref{thm:wald-unreplicated}}\label{sec:pf-wald-unreplicated}

Based on Corollary \ref{cor:uniform-design-be} and Theorem \ref{thm:unreplicated-var}, the proof can be done similarly as Theorem \ref{thm:wald-uniform}. We omit the details here.

\subsection{Proof of Theorem \ref{thm:quad-be-non-uniform}}\label{sec:quad-be-non-unifor}
The proof is an application of Lemma \ref{lem:BN} (ii).

\begin{remark}\label{rmk:relax-non-uniform-BEB}
In the paragraph following Theorem \ref{thm:quad-be-non-uniform}, we commented that the condition $\|F_\textsc{l}\|_\infty = O(Q^{-1})$ and $N=O(|\cQ_\textsc{s}|)$, can be relaxed. The idea is that $\|F_\textsc{l}\|_\infty = O(Q^{-1})$ and $N=O(|\cQ_\textsc{s}|)$, are useful for establishing the bound for $B_N$ in Lemma \ref{lem:BN}(ii). Therefore, we can easily posit a sufficient condition by letting the upper bound for $B_N$ converge to zero. 
\end{remark}

\subsection{Proof of Theorem \ref{thm:non-uniform-var}}\label{sec:pf-non-uniform-var}
\begin{proof}[Proof of Theorem \ref{thm:non-uniform-var}]

\begin{enumerate}[label = (\roman*)]
    \item Combining the decomposition \eqref{eqn:composite-var-2} and the results from Theorems \ref{thm:uniform-var} and \ref{thm:unreplicated-var}, we have 
\begin{align*}
    \E{\hV_\hgamma} &= \E{F_{\textsc{u}}^\top \hV_{\hY,\textsc{u}} F_{\textsc{u}} + F_{\textsc{r}}^\top \hV_{\hY,\textsc{r}} F_{\textsc{r}} + F_{\textsc{l}}^\top \hV_{\hY,\textsc{l}} F_{\textsc{l}}} \\
    &\succeq F_{\textsc{u}}^\top \diag{S(q,q)}_{q\in\cQ_{\textsc{u}}} F_{\textsc{u}} + F_{\textsc{u}}^\top\Omega F_{\textsc{u}} + F_{\textsc{u}}^\top \diag{\mu_{\langle g \rangle}(\overline{Y}(q)-\overline{Y}_{\langle g \rangle})^2}_{q\in\cQ_{\textsc{u}}} F_{\textsc{u}} \\
    &+ F_{\textsc{r}}^\top \diag{N_q^{-1}S(q,q)}_{q\in\cQ_{\textsc{r}}} F_{\textsc{r}} + F_{\textsc{l}}^\top \diag{N_q^{-1}S(q,q)}_{q\in\cQ_{\textsc{l}}} F_{\textsc{l}} .
\end{align*}
Therefore, $\bbE\{\hV_\hgamma\} \succeq F^\top V_\hY F \succeq V_\hgamma$.

   \item
Decompose $\hV_{\hgamma}(h,h')$ into three terms: 
\begin{align*}
     \hV_{\hgamma}(h,h')  =&   \sum_{q \in \cQ} F(k,q)F(k',q) \hV_\hY(q,q) \\
     =&
     \underbrace{  \sum_{{\langle g \rangle}\in{\langle g \rangle}}\sum_{q\in{\langle g \rangle}}F_{\textsc{u}}(k,q)F_{\textsc{u}}(k',q) \mu_{\langle g \rangle}\lt (Y_q - \hY_{{\langle g \rangle}}\rt)^2}_{\hv_\text{I}}\\
    +& \underbrace{  \sum_{q\in\cQ_{S}}F_{\textsc{r}}(k,q)F_{\textsc{r}}(k',q)N_q^{-1}\hat{S}(q,q)}_{\hv_{\text{II}}}\\ 
    +& \underbrace{  \sum_{q\in\cQ_{L}} F_{\textsc{l}}(k,q)F_{\textsc{l}}(k',q) N_q^{-1}\hat{S}(q,q)}_{\hv_{\text{III}}},
\end{align*}

Applying Lemma \ref{lem:hv-U}, we have 
\begin{align*}
    \Prob{ |\hv_\text{I} - \E{\hv_\text{I}}|\ge t}
    \le 
    \frac{C(\max_{g\in[G]}\mu_{\langle g \rangle})^2 \|F_{\textsc{u}}\|_\infty^4\Delta^2
    (\Delta^2+\zeta^2)|\cQ_{\textsc{u}}|}{t^2} := \circledast_4 .
\end{align*}

Applying Lemma \ref{lem:tail} with $\cQ=\cQ_{\textsc{r}}$, $\overline{c} = \overline{n}$, $\underline{c} = 1$, $N_0 = 1$,  we have 
\begin{align*}
    \Prob{ |\hv_\text{II} - \E{\hv_\text{II}}|\ge t}
    \le 
    \frac{C\overline{n} \|F_{\textsc{r}}\|_\infty^4 |\cQ_{\textsc{r}}| \Delta^4}{t^2} := \circledast_5 .
\end{align*}

Applying Lemma \ref{lem:tail} with $\cQ=\cQ_{\textsc{l}}$,  we have 
\begin{align*}
    \Prob{ |\hv_\text{III} - \E{\hv_\text{III}}|\ge t}
    \le 
    \frac{C\overline{c}\underline{c}^{-4} \|F_{\textsc{l}}\|_\infty^4 |\cQ_{\textsc{l}}| N_0^{-3} \Delta^4}{t^2} := \circledast_6 .
\end{align*}

Therefore, 
\begin{align*}
    &\Prob{ |\hV_\hgamma(h,h') - \E{\hV_\hgamma(h,h')}|\ge t} \\
    \le & \Prob{\{|\hv_{\text{I}} - \E{\hv_{\text{I}}}|\ge t/3\}\cup\{|\hv_{\text{II}} - \E{\hv_{\text{II}}}|\ge t/3\}\cup\{|\hv_{\text{III}} - \E{\hv_{\text{III}}}|\ge t/3\}} \\
    \le & 9 (\circledast_4  + \circledast_5  + \circledast_6).
\end{align*}

Taking union bound over $h,h'\in[H]$, we have
\begin{align*}
    \Prob{ \|\hV_\hgamma - \E{\hV_\hgamma}\|_\infty\ge t} 
    \le 9H^2 (\circledast_4  + \circledast_5  + \circledast_6).
\end{align*}

\item It follows from \eqref{eqn:op-inf}.
\end{enumerate}

\end{proof}

\subsection{Proof of Theorem \ref{thm:wald-non-uniform}}\label{sec:pf-wald-non-uniform}

Based on Corollary \ref{cor:non-uniform-design-be} and Theorem  \ref{thm:non-uniform-var}, the proof is similar to Theorem \ref{thm:wald-uniform}. We omit the details here.

\subsection{Proof of Theorem \ref{thm:quad-clt}}

\begin{proof}[Proof of Theorem \ref{thm:quad-clt}]
For a given matrix $W$, let $\bbB_t(x; W) = \{y\in\bbR^H : (y-x)^\top  W (y-x) \le t\}$, which is convex. By  Theorem \ref{thm:be-bounded},
\begin{align*}
    \sup_{t\in\bbR} |\bbP(T\le t)-\bbP(T_0\le t)| &= \sup_{t\in\bbR} |\bbP\{\tilde{\gamma} \in \bbB_t(0; W)\}-\bbP\{\xi_H \in \bbB_t(0; W)\}|\\
    &\le \sup_{A\in\cA} |\bbP\{\tilde{\gamma} \in A\}-\bbP\{\xi_H \in A\}|\\
    & \le CH^{13/4}NB_N(B_N^2 + N^{-1}) + C H^{3/4}B_N + CH^{13/8}N^{1/4}B_N^{3/2} \\
    &+ CH^{11/8}N^{1/2}B_N^2 + CH^{7/8}N^{1/4}B_N^{3/2},
\end{align*}
where $B_N = \max_{h\in[H]}\max_{i,j\in[N]}|M''_h(i,j)|$. Here $M''_h(i,j)$ is the standardized population matrix given by Lemma \ref{lem:reformulate}. Now applying \eqref{eqn:standardM-bd} in Lemma \ref{lem:reformulate}, we can further upper bound $B_N$:
\begin{align}\label{eqn:bd-BN}
    B_N \le \varrho_{\min}(V_{\hgamma})^{-1/2}\sqrt{H} \max_{h\in[H]}\max_{i,q\in[N]}|f_{qh}N_{q}^{-1}(Y_i(q)- \overline{Y}(q))|.
\end{align}

\end{proof}

\subsection{Proof of Lemma \ref{lem:BN}}
\begin{proof}[Proof of Lemma \ref{lem:BN}]
We derive upper bounds on $B_N$ by bounding the quantities $ \varrho_{\min}(V_{\hgamma})$ and $ \max_{i,q\in[N]}|f_{qh}N_{q}^{-1}(Y_i(q)- \overline{Y}(q))|$. When bounds on $B_N$ are obtained, the BEB for $W$ is a direct application of Theorem \ref{thm:quad-clt}.
\begin{enumerate}[label = (\roman*)]
\item Under \eqref{eqn:well-conditioned} and Conditions \ref{condition::proper} and \ref{cond:easy-spec}, we have
\begin{gather*}
    \varrho_{\min}(V_{\hgamma}) \ge \varrho_{\min}(F^\top F)\cdot  \min_{q\in[Q]} N_q^{-1} S(q,q), \\
    \max_{i,q\in[N]}|f_{qh}N_{q}^{-1}(Y_i(q)- \overline{Y}(q))| \le 2\|F\|_{\infty}\cdot\underline{c}^{-1}N_0^{-1}\max_{i\in[N],q\in[Q]}|Y_i(q) - \overline{Y}(q)|.
\end{gather*}
Now use Condition \ref{condition::proper} and the upper bound for $B_N$ \eqref{eqn:bd-BN} to obtain
\begin{align*}
    B_N \le \frac{2c^{1/2}\underline{c}^{-1}\max_{i\in[N],q\in[Q]}|Y_i(q) - \overline{Y}(q)|}{(\overline{c}^{-1}\min_{q\in[Q]}S(q,q))^{1/2}} \cdot \lt(\frac{H}{QN_0}\rt)^{1/2}.
\end{align*}

Then we can apply Theorem \ref{thm:quad-clt} to derive the BEB. 

If we further assume Condition \ref{cond:easy-spec}, then $B_N = O(H^{1/2}N^{-1/2})$. Then \eqref{eqn:BN-special} in Theorem \ref{thm:quad-clt} holds.

\item In general designs, we first give a lower bound on $\varrho_{\min}(V_\hgamma)$:
\begin{align}\label{eqn:non-unif-BN-0}
    \varrho_{\min}(V_\hgamma) \ge \varrho_{\min}(F^\top_{\textsc{s}}  F_{\textsc{s}}) \cdot (\overline{n}^{-1}\min_{q\in\cQ_{\textsc{s}}}S(q,q)).
\end{align}
Use Condition \ref{condition::proper-non-uniform} to obtain
\begin{align*}
    \varrho_{\min}\{ F_{\textsc{s}}^\top F_{\textsc{s}}\}\ge  c'|\cQ_\textsc{s}|^{-1}.
\end{align*}

Then we bound the maximum part of $B_N$ in \eqref{eqn:BN} by considering arms in $\cQ_{\textsc{s}}$ and $\cQ_{\textsc{l}}$ separately. For $q\in\cQ_{\textsc{s}}$, because $N_q\ge 1$, under Condition \ref{condition::proper-non-uniform} we have
\begin{align}\label{eqn:non-unif-BN-1}
    \max_{h\in[H]}\max_{i\in[N],q\in\cQ_{\textsc{s}}}|f_{qh}N_{q}^{-1}(Y_i(q)- \overline{Y}(q))| \le 2 c|\cQ_\textsc{s}|^{-1}\max_{i\in[N],q\in\cQ_{\textsc{s}}}|Y_i(q) - \overline{Y}(q)|.
\end{align}
For $q\in\cQ_{\textsc{l}}$,  we have
\begin{align}\label{eqn:non-unif-BN-3}
\max_{h\in[H]}\max_{i\in[N],q\in\cQ_{\textsc{l}}}|f_{qh}N_{q}^{-1}(Y_i(q)- \overline{Y}(q))| \le 2 c\underline{c}^{-1} \|F_\textsc{l}\|_\infty N_0^{-1} \max_{i\in[N],q\in\cQ_{\textsc{l}}}|Y_i(q) - \overline{Y}(q)|.
\end{align}
Now plugging \eqref{eqn:non-unif-BN-0}--\eqref{eqn:non-unif-BN-3} into \eqref{eqn:BN}, we have
\begin{align*}
    B_N \le& \frac{2c H^{1/2}\max_{i\in[N],q\in[Q]}|Y_i(q) - \overline{Y}(q)|}{(c'\overline{n}^{-1}|\cQ_\textsc{s}|^{-1}\min_{q\in\cQ_{\textsc{s}}}S(q,q))^{1/2}}\cdot \max\lt\{\frac{1}{|\cQ_\textsc{s}|}, \frac{\|F_\textsc{l}\|_\infty}{\underline{c} N_0}\rt\} \\
    \le& \frac{2c\max_{i\in[N],q\in[Q]}|Y_i(q) - \overline{Y}(q)|}{(c'\overline{n}^{-1}\min_{q\in\cQ_{\textsc{s}}}S(q,q))^{1/2}}\cdot \max\lt\{\frac{1}{|\cQ_\textsc{s}|^{1/2}}, \frac{\|F_\textsc{l}\|_\infty |\cQ_\textsc{s}|^{1/2}}{\underline{c} N_0}\rt\}.
\end{align*}
\end{enumerate}

Now we can apply Theorem \ref{thm:quad-clt} to derive the BEB. 

If we further assume Condition \ref{cond:easy-spec}, $\|F_\textsc{l}\|_\infty = O(Q^{-1})$ and $N=O(|\cQ_\textsc{s}|)$, then $B_N = O(H^{1/2}N^{-1/2})$. Then \eqref{eqn:BN-special} in Theorem \ref{thm:quad-clt} holds.

\end{proof}

\subsection{Proof of Corollary \ref{cor:quad-clt-v2}}
\begin{proof}[Proof of Corollary \ref{cor:quad-clt-v2}]
No matter $H$ is increasing or not, by the conditions and  Theorem \ref{thm:quad-clt}, we know that as $N\to\infty$,
\begin{align*}
    \sup_{t\in\bbR} |\bbP(T\le t)-\bbP(T_0\le t)| = o(1).
\end{align*}
\begin{enumerate}[label = (\roman*)]
    \item When $H$ is fixed, the proof is done. 

    \item When $H$ is increasing to infinity, by the classical Lindeberg CLT, we have for a standard Normal variable $Z$,
\begin{align*}
    \sup_{t\in\bbR} |\bbP\{T_0\le t\}-\bbP\lt\{\sqrt{\Var{T_0}}Z + \bbE(T_0)\le  t\rt\}| = o(1).
\end{align*}
Using the expectation and variance calculation of $T_0$ in \eqref{eqn::qclt-true-variance}, 
we conclude the second part.
\end{enumerate}

\end{proof}

\subsection{Proof of Lemma \ref{lem:high-moment}}

\begin{proof}[Proof of Lemma \ref{lem:high-moment}]

Without loss of generality, we assume the potential outcomes are centered:  $\overline{Y}(q) = 0$ for all $q\in[Q]$.

(i) The first part follows from the variance formula of $\hY_q$.

(ii) Now we bound  the fourth moment of $\hat{Y}_q$:
\begin{align*}
    \bbE\lt\{\hat{Y}_q^4\rt\}= & \underbrace{\frac{1}{N_q^4}\bbE\lt\{\sum_{i=1}^N Y_i(q)^4\ind{Z_i = q}\rt\}}_{\text{II.2-1}}\\
    + & \underbrace{\frac{4}{N_q^4}\bbE\lt\{\sum_{i\neq j}^N Y_i(q)^3 Y_j(q)\ind{Z_i = Z_j = q}\rt\}}_{\text{II.2-2}}\\
    + & \underbrace{\frac{3}{N_q^4}\bbE\lt\{\sum_{i \neq j}^N Y_i(q)^2 Y_j(q)^2\ind{Z_i = Z_j = q}\rt\}}_{\text{II.2-3}}\\
    + & \underbrace{\frac{3}{N_q^4}\bbE\lt\{\sum_{i\neq j \neq k}^N Y_i(q) Y_j(q) Y_k(q)^2\ind{Z_i = Z_j = Z_k = q}\rt\}}_{\text{II.2-4}}\\
    + & \underbrace{\frac{1}{N_q^4}\bbE\lt\{\sum_{i \neq j \neq k \neq l}^N Y_i(q) Y_j(q) Y_k(q) Y_l(q)\ind{Z_i = Z_j = Z_k = Z_l = q}\rt\}}_{\text{II.2-5}}.
\end{align*}

Compute
\begingroup
\allowdisplaybreaks
\begin{align*}
    \text{II.2-1} &= \frac{1}{N_q^3N}\sum_{i=1}^N Y_i(q)^4,\\
    \text{II.2-2} &= \frac{4}{N_q^4}\bbE\lt\{\sum_{i\neq j}^N Y_i(q)^3 Y_j(q)\ind{Z_i = Z_j = q}\rt\} \\
    &= \frac{4(N_q-1)}{N_q^3N(N-1)} \sum_{i\neq j}^N Y_i(q)^3 Y_j(q)  \\
    &= -\frac{4(N_q-1)}{N_q^3N(N-1)} \sum_{i\neq j}^N Y_i(q)^4,\\ 
    \text{II.2-3} &= \frac{3(N_q-1)}{N_q^3N(N-1)} \sum_{i \neq j}^N Y_i(q)^2 Y_j(q)^2,\\
    \text{II.2-4} & = \frac{3(N_q-1)(N_q-2)}{N_q^3N(N-1)(N-2)} \sum_{i\neq j \neq k}^N Y_i(q) Y_j(q) Y_k(q)^2 \\
    & = \frac{3(N_q-1)(N_q-2)}{N_q^3N(N-1)(N-2)} \sum_{ j \neq k}^N -(Y_j(q) + Y_k(q)) Y_j(q) Y_k(q)^2 \\
    & = -\frac{3(N_q-1)(N_q-2)}{N_q^3N(N-1)(N-2)} \sum_{ j \neq k}^N Y_j(q)^2 Y_k(q)^2  
    + \frac{3(N_q-1)(N_q-2)}{N_q^3N(N-1)(N-2)} \sum_{ k}^N   Y_k(q)^4, \\
    \text{II.2-5} &= \frac{N_q(N_q-1)(N_q-2)(N_q-3)}{N_q^4 N(N-1)(N-2)(N-3)} \sum_{i \neq j \neq k \neq l}^N Y_i(q) Y_j(q) Y_k(q) Y_l(q) \\
    &= -\frac{3N_q(N_q-1)(N_q-2)(N_q-3)}{N_q^4 N(N-1)(N-2)(N-3)} \sum_{i \neq j
    \neq k}^N Y_i(q) Y_j(q) Y_k(q)^2\\
    &= \frac{3N_q(N_q-1)(N_q-2)(N_q-3)}{N_q^4 N(N-1)(N-2)(N-3)}\sum_{ j \neq k}^N Y_j(q)^2 Y_k(q)^2 \\
    &- \frac{3N_q(N_q-1)(N_q-2)(N_q-3)}{N_q^4 N(N-1)(N-2)(N-3)} \sum_{ k}^N   Y_k(q)^4.
\end{align*}
\endgroup
Now bound these terms:
\begin{gather*}
    |\text{II.2-1}| \le \frac{\Delta^4}{N_q^3}, \quad |\text{II.2-2}| \le \frac{4\Delta^4}{N_q^2N}, \\
    |\text{II.2-3}| \le  \frac{6\Delta^4 }{N_q^2 } ~\left(\text{using $\sum_{i\neq j} Y_i(q)^2Y_j(q)^2 \le \sum_iY_i(q)^2\sum_{j}Y_j(q)^2$}\right),\\
    |\text{II.2-4}| \le \frac{6\Delta^4}{N_q(N-2)} + \frac{3\Delta^4}{N_q(N-1)(N-2)}, \\
    |\text{II.2-5}| \le \frac{6\Delta^4}{(N-2)(N-3)} + \frac{3\Delta^4}{(N-1)(N-2)(N-3)}.
\end{gather*}
Choose $N$ large enough to obtain
\begin{align*}
      \bbE\lt\{\hat{Y}_q^4\rt\}  \le \frac{C\Delta^4}{N_q^2}.
\end{align*}

(iii) Then we compute the covariance terms:
\begin{align*}
    &\bbE\lt\{\hat{Y}_q^2 \hat{Y}_{q'}^2\rt\} - \bbE\lt\{\hat{Y}_q^2\rt\} \bbE\lt\{\hat{Y}_{q'}^2\rt\}\\
    =&  \Bigg\{\underbrace{\frac{1}{N_q^2N_{q'}^2}\sum_{i\neq k}Y_i(q)^2Y_k(q')^2 \frac{N_qN_{q'}}{N(N-1)}}_{\text{II.2-1}}\\
    &+\underbrace{\frac{1}{N_q^2N_{q'}^2}\sum_{i\neq j \neq k}Y_i(q)Y_j(q)Y_k(q')^2 \frac{N_q(N_q-1)N_{q'}}{N(N-1)(N-2)}}_{\text{II.2-2}}\\
    &+\underbrace{\frac{1}{N_q^2N_{q'}^2}\sum_{i\neq k \neq l}Y_i(q)^2Y_k(q')Y_l(q') \frac{N_qN_{q'}(N_{q'}-1)}{N(N-1)(N-2)}}_{\text{II.2-3}}\\
    &+\underbrace{\frac{1}{N_q^2N_{q'}^2}\sum_{i \neq j \neq k \neq l}Y_i(q)Y_j(q)Y_k(q')Y_l(q') \frac{N_q(N_q-1)N_{q'}(N_{q'}-1)}{N(N-1)(N-2)(N-3)}}_{\text{II.2-4}}\Bigg\}\\
    &- \underbrace{\lt\{\frac{1}{N_q} - \frac{1}{N}\rt\}S(q,q)\cdot \lt\{\frac{1}{N_{q'}} - \frac{1}{N}\rt\}S(q',q')}_{\text{II.2-5}}.
\end{align*}
For II.2-1 and II.2-5:
\begin{align*}
   &\lt|\frac{1}{N_q^2N_{q'}^2}\sum_{i\neq k}Y_i(q)^2Y_k(q')^2 \frac{N_qN_{q'}}{N(N-1)} - \frac{1}{(N-1)^2}\lt(\frac{N-N_q}{N_qN}\rt) \lt(\frac{N-N_{q'}}{N_{q'}N}\rt)\lt\{\sum_{i=1}^NY_i(q)^2\rt\}\lt\{\sum_{k=1}^NY_i(q')^2\rt\}\rt| \\
   =& \Bigg|\lt\{\frac{1}{N_qN_{q'}N(N-1)} - \frac{1}{(N-1)^2}\lt(\frac{N-N_q}{N_qN}\rt) \lt(\frac{N-N_{q'}}{N_{q'}N}\rt)\rt\}\lt\{\sum_{i=1}^NY_i(q)^2\rt\}\lt\{\sum_{i=1}^NY_i(q')^2\rt\}\\
   -& \frac{1}{N_qN_{q'}N(N-1)}\sum_{i=1}^NY_i(q)^2Y_i(q')^2\Bigg| \\
   \le& \frac{(N_q+N_{q'}+1)N-N_qN_{q'}}{N^2(N-1)^2N_qN_{q'}} N^2\Delta^4 + \frac{\Delta^4}{N_qN_{q'}(N-1)} \le \frac{7(N_q + N_{q'})\Delta^4}{N_qN_{q'}(N-1)}.
\end{align*}
 For II.2-2:
\begin{align*}
    &\lt|\frac{1}{N_q^2N_{q'}^2}\sum_{i\neq j \neq k}Y_i(q)Y_j(q)Y_k(q')^2 \frac{N_q(N_q-1)N_{q'}}{N(N-1)(N-2)}\rt|\\
    =& \lt|-\frac{1}{N_q^2N_{q'}^2}\sum_{i \neq k}Y_i(q)\{Y_i(q)+Y_k(q)\}Y_k(q')^2 \frac{N_q(N_q-1)N_{q'}}{N(N-1)(N-2)}\rt|\\
    \le& \frac{N_q-1}{N_qN_{q'}N(N-1)(N-2)} \sum_{i\neq k} \lt\{\frac{1}{2}Y_i(q)^4 +\frac{1}{2}Y_k(q')^4 + \frac{1}{4}Y_i(q)^4 + \frac{1}{4}Y_k(q)^4 + \frac{1}{2}Y_k(q')^4 \rt\}\\
    \le& \frac{N_q-1}{N_qN_{q'}N(N-1)(N-2)} \{N(N-1)\cdot 2\Delta^4\} \le \frac{2(N_q + N_{q'})\Delta^4}{N_qN_{q'}(N-2)}.
\end{align*}
II.2-3 is similar to II.2-2:
\begin{align*}
    &\lt|\frac{1}{N_q^2N_{q'}^2}\sum_{i\neq k \neq l}Y_i(q)^2Y_k(q')Y_l(q') \frac{N_qN_{q'}(N_{q'}-1)}{N(N-1)(N-2)}\rt|\le\frac{2(N_q + N_{q'})\Delta^4}{N_qN_{q'}(N-2)}.
\end{align*}
For II.2-4:
\begin{align*}
    &\lt|\frac{1}{N_q^2N_{q'}^2}\sum_{i \neq j \neq k \neq l}Y_i(q)Y_j(q)Y_k(q')Y_l(q') \frac{N_q(N_q-1)N_{q'}(N_{q'}-1)}{N(N-1)(N-2)(N-3)}\rt|\\
    = &\lt|-\frac{1}{N_q^2N_{q'}^2}\sum_{i \neq j \neq k}Y_i(q)Y_j(q)Y_k(q')\{Y_i(q')+Y_j(q')+Y_k(q')\} \frac{N_q(N_q-1)N_{q'}(N_{q'}-1)}{N(N-1)(N-2)(N-3)}\rt|\\
    \le & \frac{(N_q-1)(N_{q'}-1)}{N_qN_{q'}N(N-1)(N-2)(N-3)}\cdot{6N(N-1) \Delta^4} \see{reduce terms like II.2-2}\\
    \le &\frac{3(N_q-1)(N_{q'}-1)\Delta^4}{N_qN_{q'}(N-2)(N-3)}.
\end{align*}
Summarizing II.2-1 to II.2-5,
\begin{align}\label{eqn:cov-bound}
    \lt|\text{Cov}\lt\{\hat{Y}_q^2, \hat{Y}_{q'}^2\rt\}\rt| \le \frac{C(N_q + N_{q'})\Delta^4}{N_qN_{q'}N}.
\end{align}

\end{proof}

\subsection{Proof of Lemma \ref{lem:moments-U}}
\begin{proof}[Proof of Lemma \ref{lem:moments-U}]
Part (i) and Part (ii) can be shown by constructing new potential outcomes $\{Y_i(q)^2\}$ and applying the variance formula for the sample average. Thus we omit the proof.

For Part (iii), we have
\begin{align*}
    |\Cov{Y_{q_1}^2}{Y_{q_1}Y_{q_2}}| &= |\E{Y_{q_1}^3Y_{q_2}} - \E{Y_{q_1}^2}\E{Y_{q_1}Y_{q_2}}|\\
    &= |\frac{1}{(N)_2}\sum_{i\neq j} Y_i(q_1)^3Y_j(q_2) + (1-N^{-1})S_{Y}(q_1,q_1) \cdot N^{-1}S(q_1,q_2)| \le \frac{C\Delta^4}{N}. 
\end{align*}

For Part (iv), we have
\begin{align*}
    |\Cov{Y_{q_1}^2}{Y_{q_2}Y_{q_3}}| &= |\E{(Y_{q_1}^2 - \E{Y_{q_1}^2})Y_{q_2}^2Y_{q_3}^2}|\\
    &= \lt|\E{\sum_{i\neq j\neq k} \{Y_i(q_1)^2 - N^{-1}\sum_{i\in[N]}Y_i(q_1)^2\}Y_j(q_2)Y_k(q_3)\ind{Z_i = q_1,Z_j = q_2, Z_k = q_3}}\rt| \\
    &=  \lt|\frac{1}{(N)_3}\sum_{i\neq j\neq k} \{Y_i(q_1)^2 - N^{-1}\sum_{i\in[N]}Y_i(q_1)^2\}Y_j(q_2)Y_k(q_3)\rt| \\
    & = \lt|-\frac{1}{(N)_3}\sum_{j\neq k} \{Y_j(q_1)^2 + Y_k(q_1)^2 - 2 N^{-1}\sum_{i\in[N]}Y_i(q_1)^2\}Y_j(q_2)Y_k(q_3)\rt| \le \frac{C\Delta^4}{N}.
\end{align*}

For Part (v), we have
\begin{align*}
    &|\Cov{Y_{q_1}Y_{q_2}}{Y_{q_3}Y_{q_4}}| \\
    =& \lt|\frac{1}{(N)_4}\sum_{i\neq j \neq k \neq l} \lt\{Y_i(q_1)Y_j(q_2) - \frac{1}{(N)_2}\sum_{i\neq j} Y_i(q_1)Y_j(q_2)\rt\}Y_k(q_3)Y_l(q_4)\rt| \\
    =& \lt|-\frac{1}{(N)_4}\sum_{i\neq j \neq k} \lt\{Y_i(q_1)Y_j(q_2) - \frac{1}{N(N-1)}\sum_{i\neq j} Y_i(q_1)Y_j(q_2)\rt\}Y_k(q_3)(Y_i(q_4) + Y_j(q_4) + Y_k(q_4))\rt|.
\end{align*}
Further, we have
\begin{align*}
    &\frac{1}{(N)_4}\lt|\sum_{i\neq j\neq k} \lt\{Y_i(q_1)Y_j(q_2) - \frac{1}{(N)_2}\sum_{i\neq j} Y_i(q_1)Y_j(q_2)\rt\}Y_k(q_3)Y_i(q_4)\rt|\\
    = &\lt|-\frac{1}{(N)_4}\sum_{i\neq j} \lt\{Y_i(q_1)Y_j(q_2) - \frac{1}{(N)_2}\sum_{i\neq j} Y_i(q_1)Y_j(q_2)\rt\}(Y_i(q_3) + Y_j(q_3))Y_i(q_4)\rt| \\
    \le & \frac{C\Delta^4}{N^2}.
\end{align*}
Similar to the summation
\begin{align*}
    \lt|\frac{1}{(N)_4}\sum_{i\neq j\neq k} \lt\{Y_i(q_1)Y_j(q_2) - \frac{1}{(N)_2}\sum_{i\neq j} Y_i(q_1)Y_j(q_2)\rt\}Y_k(q_3)Y_j(q_4)\rt| \le \frac{C\Delta^4}{N^2}.
\end{align*}
Last, it remains to bound
\begin{align*}
    &\lt|\frac{1}{(N)_4}\sum_{i\neq j\neq k} \lt\{Y_i(q_1)Y_j(q_2) - \frac{1}{(N)_2}\sum_{i\neq j} Y_i(q_1)Y_j(q_2)\rt\}Y_k(q_3)^2\rt|\\
    = &\lt|\frac{1}{(N)_4}\sum_{j\neq k} \lt\{-(Y_j(q_1) + Y_k(q_1))Y_j(q_2) + \frac{N-2}{(N)_2}\sum_{i} Y_i(q_1)^2\rt\}Y_k(q_3)^2\rt|\\
    \le & \frac{C\Delta^4}{N^2}.
\end{align*}
Hence we conclude the proof by combining the above parts.
\end{proof}

\subsection{Proof of Lemma \ref{lem:tail}}
\begin{proof}[Proof of Lemma \ref{lem:tail}] The proof is based on Chebyshev's inequality and bounding the variance of 
\begin{align*}
    \sum_{q\in\cQ}w_q  N_q^{-1} \hat{S}(q,q) 
    =& \underbrace{\sum_{q\in\cQ}w_q N_q^{-1}(N_q-1)^{-1} \sum_{q_i=q}(Y_i-\overline{Y}(q))^2 }_{\text{II}.1} \\
    -& \underbrace{ \sum_{q\in\cQ}w_q  (N_q-1)^{-1}\lt(\hat{Y}(q)-\overline{Y}(q)\rt)^2}_{\text{II}.2} .
\end{align*}
The above decomposition ensures that we can assume $Y_i(q)$'s are centered without loss of generality.
For II.1, we have
\begin{align}\label{eqn:II.1}
     \Var{\text{II.1}} \le  \sum_{q\in\cQ} w_q^{2}(N_q-1)^{-1}N_q^{-2} S_{Y^2}(q,q) \le 4\underline{c}^{-3}\overline{w}^{2}|\cQ| N_0^{-3} \Delta^4.
\end{align}
For II.2, we have 
\begin{align}
    \text{Var}\{\text{II.2}\} &\le \sum_{q\in\cQ} w_q^2 (N_q-1)^{-2}\Var{\hat{Y}_q^2} \notag\\
    & + \sum_{q\neq q'\in\cQ} w_q w_{q'} (N_q-1)(N_{q'} - 1)\Cov{\hat{Y}_q^2}{\hat{Y}_{q'}^2}  \notag\\
    &\le \sum_{q\in\cQ} w_q^2 (N_q-1)^{-2}\bbE\lt\{ \hat{Y}_q^4 \rt\} \notag\\
    & + \sum_{q\neq q'\in\cQ} w_q w_{q'} (N_q-1)(N_{q'} - 1)\Cov{\hat{Y}_q^2}{\hat{Y}_{q'}^2}  \notag\\
    &\le  \sum_{q\in\cQ} w_q^2 (N_q-1)^{-2}(C\Delta^4 N_q^{-2}) \label{eqn:var-II.2}\\
    & + \sum_{q\neq q'\in\cQ} w_q w_{q'} (N_q-1)^{-1}(N_{q'} - 1)^{-1} \frac{C(N_q + N_{q'})\Delta^4}{N_qN_{q'}N}\label{eqn:cov-II.2}\\
    &\see{By Lemma \ref{lem:high-moment}}.\notag
\end{align}
For \eqref{eqn:var-II.2}, we have
\begin{align}\label{eqn:var-II.2-1}
    \sum_{q\in\cQ} w_q^2 (N_q-1)^{-2}(C\Delta^4 N_q^{-2}) \le C\underline{c}^{-4}\overline{w}^2|\cQ| N_0^{-4}\Delta^4.
\end{align}
For \eqref{eqn:cov-II.2}, we have
\begin{align}
     &\lt|\sum_{q\neq q'\in\cQ} w_q w_{q'} (N_q-1)^{-1}(N_{q'} - 1)^{-1} \frac{C(N_q + N_{q'})\Delta^4}{N_qN_{q'}N}\rt|\notag\\
     \le & \sum_{q\neq q'\in\cQ}\overline{w}^2 \cdot 4(\underline{c}N_0)^{-4} \cdot \frac{C\overline{c}N_0\Delta^4}{\underline{c}QN_0}\notag\\
     \le & \overline{w}^2 \cdot 4(\underline{c}N_0)^{-4} \cdot \frac{C\overline{c} |\cQ|^2 \Delta^4}{\underline{c}Q }\notag\\
     \le &
     C\underline{c}^{-4}\overline{w}^2 (\overline{c}/\underline{c}) |\cQ| N_0^{-4} \Delta^4\notag\\
     \le & C\overline{c}\underline{c}^{-4} \overline{w}^2 |\cQ| N_0^{-3}\Delta^4 ,\label{eqn:cov-II.2-1}
\end{align}
where in the last inequality \eqref{eqn:cov-II.2-1}, we use the fact that as the lower bound for the size of the arms, $\underline{c}N_0 $
is in general greater than some absolute constant (in many cases just use $1$). 

Combining \eqref{eqn:II.1}--\eqref{eqn:cov-II.2-1}, we have
\begin{align*}
    \Var{\sum_{q\in\cQ}w_q  N_q^{-1} \hat{S}(q,q)} \le  C\overline{c}\underline{c}^{-4} \overline{w}^2 |\cQ| N_0^{-3}\Delta^4.
\end{align*}

We apply Chebyshev's inequality to complete the proof.

\end{proof}

\subsection{Proof of Lemma \ref{lem:hv-U}}
\begin{proof}[Proof of Lemma \ref{lem:hv-U}]
The proof is based on Chebyshev's inequality and bounding the variance of $\hv$.

For any ${\langle g \rangle}$,  we have
\begin{align}
    \sum_{q\in{\langle g \rangle}}w_q  \lt (Y_q - \hY_{{\langle g \rangle}}\rt)^2 
        & =  \sum_{q\in{\langle g \rangle}}w_q   \lt (Y_q - \overline{Y}(q) + \overline{Y}(q) - \overline{Y}_{\langle g \rangle} + \overline{Y}_{\langle g \rangle} - \hY_{{\langle g \rangle}}\rt)^2 \notag\\
        & =  \underbrace{\sum_{q\in{\langle g \rangle}}w_q(Y_q - \overline{Y}(q))^2}_{\text{Term I}} 
        + 
        \underbrace{\sum_{q\in{\langle g \rangle}}w_q(\overline{Y}(q) - \overline{Y}_{\langle g \rangle})^2}_{\text{Term II}}
        +
        \underbrace{\sum_{q\in{\langle g \rangle}}w_q(\overline{Y}_{\langle g \rangle} - \hY_{{\langle g \rangle}})^2}_{\text{Term III}}   \label{eqn:decomp-var-U-1}\\
        & 
        +
        2 \underbrace{\sum_{q\in{\langle g \rangle}}w_q   \lt \{(Y_q - \overline{Y}(q)) (\overline{Y}(q) - \overline{Y}_{\langle g \rangle}) \rt\}}_{\text{Term IV}}  \label{eqn:decomp-var-U-2}\\
        & 
        + 2\underbrace{(\overline{Y}_{\langle g \rangle} - \hY_{{\langle g \rangle}})  \sum_{q\in{\langle g \rangle}}w_q   \lt \{(Y_q - \overline{Y}(q))  \rt\}}_{\text{Term V}} \label{eqn:decomp-var-U-3}\\
        & 
        + 
        2\underbrace{(\overline{Y}_{\langle g \rangle} - \hY_{{\langle g \rangle}})  \sum_{q\in{\langle g \rangle}}w_q   \lt \{(\overline{Y}(q) - \overline{Y}_{\langle g \rangle}) \rt\}}_{\text{Term VI}} . \label{eqn:decomp-var-U-4}
\end{align}

There are six terms in \eqref{eqn:decomp-var-U-1} to \eqref{eqn:decomp-var-U-4}. We deal with them separately. 

\textbf{Bound summations involving Term I, IV and VI.} We first show upper bounds for the variance of Term I, IV and VI (summed over $g\in[G]$):
\begin{gather}
    \Var{\sum_{g\in{G}}\sum_{q\in{\langle g \rangle}}w_q(Y_q - \overline{Y}(q))^2} \le C\sum_{q\in[Q]} w_q^2 \Delta^4,\label{eqn:V-term-I}\\
    \Var{\sum_{g\in[G]}\sum_{q\in{\langle g \rangle}}w_q   \lt \{(Y_q - \overline{Y}(q)) (\overline{Y}(q) - \overline{Y}_{\langle g \rangle}) \rt\}} \le C\sum_{q\in[Q]} w_q^2  \Delta^2 \zeta^2, \label{eqn:V-term-IV}\\
    \Var{\sum_{g\in[G]}(\overline{Y}_{\langle g \rangle} - \hY_{{\langle g \rangle}})  \sum_{q\in{\langle g \rangle}}w_q   \lt \{(\overline{Y}(q) - \overline{Y}_{\langle g \rangle}) \rt\}} \le C\sum_{q\in[Q]} w_q^2\Delta^2\zeta^2.\label{eqn:V-term-VI}
\end{gather}

The key idea for proving \eqref{eqn:V-term-I}--\eqref{eqn:V-term-VI} is to treat the summations as linear combinations of sample averages and apply Lemma \ref{lem:moments-U}. Take \eqref{eqn:V-term-I} for example. We can treat $Y'_i(q ) = (Y_i(q) - \overline{Y}(q))^2$ as pseudo potential outcomes and obtain:
\begin{align*}
    \Var{\sum_{g\in{G}}\sum_{q\in{\langle g \rangle}}w_q(Y_q - \overline{Y}(q))^2} \le \sum_{g\in\cG}\sum_{q\in\langle g \rangle} w_q^2 S_{Y'}(q,q) 
    \le 
    C\sum_{q\in[Q]} w_q^2 \Delta^4.
\end{align*}
Similar derivation holds for \eqref{eqn:V-term-IV} and \eqref{eqn:V-term-VI}. 

\textbf{Bound summations involving Term II.} Term II is a non-random quantity.
Therefore, it will not make any contribution  to the variance.

\textbf{Bound summations involving Terms III.}
Now we bound
\begin{gather}
    \Var{\sum_{g\in[G]}|{\langle g \rangle}| \overline{w}_{\langle g \rangle}(\overline{Y}_{\langle g \rangle} - \hY_{{\langle g \rangle}})^2}, \see{where $\overline{w}_{\langle g \rangle} = |{\langle g \rangle}|^{-1}\sum_{q\in{\langle g \rangle}}w_q$} \label{eqn:hard-I}.
\end{gather}
We calculate
\begin{align*}
    \eqref{eqn:hard-I} = &\underbrace{\sum_{g\in[G]} |{\langle g \rangle}|^2 \overline{w}_{\langle g \rangle}^2 \Var{(\hY_{\langle g \rangle} - \overline{Y}_{\langle g \rangle})^2} }_{\text{Term III.1}}\\
    + & \underbrace{\sum_{g\neq g'\in[G]} |{\langle g \rangle}||{\langle g \rangle}'| \overline{w}_{\langle g \rangle} \overline{w}_{{\langle g \rangle}'} \Cov{(\hY_{\langle g \rangle} - \overline{Y}_{\langle g \rangle})^2}{(\hY_{{\langle g \rangle}'} - \overline{Y}_{{\langle g \rangle}'})^2}}_{\text{Term III.2}}.
\end{align*}
For Term III.1, we can show
\begin{align}
    &\sum_{g\in[G]} |{\langle g \rangle}|^2 \overline{w}_{\langle g \rangle}^2 \Var{(\hY_{\langle g \rangle} - \overline{Y}_{\langle g \rangle})^2}\notag \\
    = & \sum_{g\in[G]} |{\langle g \rangle}|^2 \overline{w}_{\langle g \rangle}^2 \Cov{(\hY_{\langle g \rangle} - \overline{Y}_{\langle g \rangle})^2}{(\hY_{\langle g \rangle} - \overline{Y}_{\langle g \rangle})^2}\notag \\
    = & \sum_{g\in[G]}|{\langle g \rangle}|^2\overline{w}_{\langle g \rangle}^2 |{\langle g \rangle}|^{-4} \Bigg\{\sum_{q\in{\langle g \rangle}}\Var{(Y_q - \overline{Y}(q))^2} + \sum_{q_1 \neq q_2\in[{\langle g \rangle}]} \Cov{(Y_{q_1} - \overline{Y}(q_1))^2}{(Y_{q_2} - \overline{Y}(q_2))^2}\notag \\
    & \phantom{=\sum_{g\in[G]}|{\langle g \rangle}|^2\overline{w}_{\langle g \rangle}^2  } +\sum_{q_1\neq q_2\in{\langle g \rangle}} \Cov{(Y_{q_1} - \overline{Y}(q_1))^2}{(Y_{q_1} - \overline{Y}(q_1))(Y_{q_2} - \overline{Y}(q_2))}\Bigg\}\notag \\
     & \phantom{=\sum_{g\in[G]}|{\langle g \rangle}|^2\overline{w}_{\langle g \rangle}^2  } +\sum_{q_1\neq q_2\neq q_3\in{\langle g \rangle}} \Cov{(Y_{q_1} - \overline{Y}(q_1))^2}{(Y_{q_2} - \overline{Y}(q_2))(Y_{q_3} - \overline{Y}(q_3))}\Bigg\}\notag \\
     & \phantom{=\sum_{g\in[G]}|{\langle g \rangle}|^2\overline{w}_{\langle g \rangle}^2  } +\sum_{q_1\neq q_2\neq q_3 \neq q_4\in{\langle g \rangle}} \Cov{(Y_{q_1} - \overline{Y}(q_1))(Y_{q_2} - \overline{Y}(q_2))}{(Y_{q_3} - \overline{Y}(q_3))(Y_{q_4} - \overline{Y}(q_4))}\Bigg\}\notag \\
    \le &  \sum_{g\in[G]}|{\langle g \rangle}|^{-2}\overline{w}_{\langle g \rangle}^2   \lt\{C|{\langle g \rangle}|\Delta^4 +
    C|{\langle g \rangle}|^2\Delta^4/N +
    C|{\langle g \rangle}|^2\Delta^4/N + C|{\langle g \rangle}|^3\Delta^4/N + C|{\langle g \rangle}|^4\Delta^4/N^2\rt\}\notag \\
    \le & C\sum_{g\in[G]} \overline{w}_{\langle g \rangle}^2  \Delta^4 \le C\overline{w}^2\Delta^4G \le C\overline{w}^2\Delta^4N_{\textsc{u}}. \label{eqn:Term-III.1}
\end{align}
To bound Term III.2, we first obtain the following bound using Lemma \ref{lem:moments-U}:
\begin{gather}\label{eqn:Cov-g-gprime}
     \lt|\Cov{(\hY_{\langle g \rangle} - \overline{Y}_{\langle g \rangle})^2}{(\hY_{{\langle g \rangle}'} - \overline{Y}_{{\langle g \rangle}'})^2}\rt| \le \frac{C\Delta^4(|{\langle g \rangle}| + |{\langle g \rangle}'|)}{|{\langle g \rangle}||{\langle g \rangle}'|N}, \forall~ g\neq g'\in[G].
\end{gather}
The derivation is similar to what we did when handling Term III.1, thus we omit the details here. Using \eqref{eqn:Cov-g-gprime}, we have
\begin{align}\label{eqn:Term-III.2}
    \lt|\sum_{g\neq g'\in[G]} |{\langle g \rangle}||{\langle g \rangle}'| \overline{w}_{\langle g \rangle} \overline{w}_{{\langle g \rangle}'} \Cov{(\hY_{\langle g \rangle} - \overline{Y}_{\langle g \rangle})^2}{(\hY_{{\langle g \rangle}'} - \overline{Y}_{{\langle g \rangle}'})^2}\rt|
    \le  \frac{CN_{\textsc{u}}^2\overline{w}^2\Delta^4}{N} \le C\overline{w}^2\Delta^4 N_{\textsc{u}}.
\end{align}

Combine \eqref{eqn:Term-III.1} and \eqref{eqn:Term-III.2} to obtain
\begin{align}\label{eqn:hard-I-bd}
    \text{\eqref{eqn:hard-I}} \le C\overline{w}^2\Delta^4 N_U.
\end{align}

\textbf{Bound summations involving Term V.} Now we bound
\begin{gather}
    \Var{\sum_{g\in[G]}(\overline{Y}_{\langle g \rangle} - \hY_{{\langle g \rangle}})  \sum_{q\in{\langle g \rangle}}w_q   \lt \{(Y_q - \overline{Y}(q))  \rt\}}. \label{eqn:hard-II}
\end{gather}

We can show 
\begin{align*}
    \eqref{eqn:hard-II} &= \sum_{g\in[G]} \Var{(\overline{Y}_{\langle g \rangle} - \hY_{{\langle g \rangle}})  \sum_{q\in{\langle g \rangle}}w_q (Y_q - \overline{Y}(q)) }\\
    & + \sum_{g\neq g\in[G]} \Cov{(\overline{Y}_{\langle g \rangle} - \hY_{{\langle g \rangle}})  \sum_{q\in{\langle g \rangle}}w_q (Y_q - \overline{Y}(q))}{(\overline{Y}_{{\langle g \rangle}'} - \hY_{{\langle g \rangle}'})  \sum_{q\in{\langle g \rangle}'}w_q (Y_q - \overline{Y}(q))}.
\end{align*}
The analysis is very similar to \eqref{eqn:hard-I}. We omit the proof and directly state the conclusion:
\begin{gather}
    \sum_{g\in[G]} \Var{(\overline{Y}_{\langle g \rangle} - \hY_{{\langle g \rangle}})  \sum_{q\in{\langle g \rangle}}w_q (Y_q - \overline{Y}(q)) } \le C\overline{w}^2\Delta^4 G \le C\overline{w}^2\Delta^4 N_{\textsc{u}},\notag\\
    \sum_{g\neq g\in[G]} \Cov{(\overline{Y}_{\langle g \rangle} - \hY_{{\langle g \rangle}})  \sum_{q\in{\langle g \rangle}}w_q (Y_q - \overline{Y}(q))}{(\overline{Y}_{{\langle g \rangle}'} - \hY_{{\langle g \rangle}'})  \sum_{q\in{\langle g \rangle}'}w_q (Y_q - \overline{Y}(q))} \le C\overline{w}^2\Delta^4N_{\textsc{u}},\notag\\
    \text{\eqref{eqn:hard-II}} \le C\overline{w}^2\Delta^4 N_U. \label{eqn:hard-II-bd}
\end{gather}

\textbf{Summarize results.} Combining \eqref{eqn:V-term-I},  \eqref{eqn:V-term-IV},  \eqref{eqn:V-term-VI}, \eqref{eqn:hard-I-bd} and \eqref{eqn:hard-II-bd}, for the unreplicated design, we have 
\begin{align*}
    \Var{\hv} \le C\overline{w}^2(\Delta^4+\Delta^2\zeta^2) N_{\textsc{u}}.
\end{align*}
Now the tail bound can be obtained by Chebyshev's inequality.
\end{proof}

\subsection{Proof of Theorem \ref{thm:be-proj-standard-vec}}

\begin{proof}[Proof of Theorem  \ref{thm:be-proj-standard-vec}]
The proof extends that of Theorem \ref{thm:be-proj-standard}. 

\noindent\textit{Part (i) of Theorem \ref{thm:be-proj-standard-vec}.}  The main difference is that we need to carefully choose the norms and get more delicate bounds.
By Lemma \ref{lem:reformulate},  there are population matrices $M''_1,\ldots, M''_H$ that satisfy Condition \ref{cond:str-Mk} and $\tilde{\gamma} = \lt(\trace{M''_h P}\rt)_{h=1}^H$. We apply  Theorem \ref{thm:linear-projection} to obtain that for any $b\in\bbR^H$ with $\|b\|_2 = 1$, 
\begin{align}\label{eqn:BEB-M-vec}
    \sup_{t\in\bbR}|\bbP\{b^\top\tilde{\gamma} \le t\} - \Phi(t)| \le C{\max_{i,j\in[N]} \lt|\sum_{h=1}^Hb_hM''_h(i,j)\rt|}.
\end{align}
Here following the proof of  Lemma \ref{lem:reformulate}, $M''_h$ is obtained through the following definition of $M'_h$. Define
\begin{align}\label{eqn:vec-center-PO}
    \breve{\bY}_i(q) = \bY_i(q) - \overline{\bY}(q), ~\breve{\gamma}_{i} = N^{-1}\sum_{q'=1}^Q \bF_{q'} \breve{\bY}_i(q') .
\end{align}
For each $i,j\in[N]$, define
\begin{align}\label{eqn:vec-M}
    M_h'(i,j) = N_q^{-1} \bF_q(h,\cdot)\breve{\bY}_i(q) -  \breve{\gamma}_{hi}, ~ \sum_{q'=0}^{q-1} N_q + 1\le j \le \sum_{q'=0}^{q} N_q.
\end{align}
\eqref{eqn:vec-M} indicates a natural mapping from column $j$ to a particular treatment level ${q_j}$. Then
\begin{align*}
    b^\top 
    \begin{pmatrix}
    \{\myvec{M''_1}\}^\top\\
    \vdots\\
    \{\myvec{M''_H}\}^\top
    \end{pmatrix}
    =
    b^\top \bV_\hgamma^{-1/2}
    \begin{pmatrix}
    \{\myvec{M'_1}\}^\top\\
    \vdots\\
    \{\myvec{M'_H}\}^\top
    \end{pmatrix}.
\end{align*}
Hence
\begin{align}
    \lt|\sum_{h=1}^Hb_hM''_h(i,j)\rt| &= 
    \lt|b^\top  \bV_\hgamma^{-1/2}
    \begin{pmatrix}
    N_{q_j}^{-1} \bF_{q_j}(1,\cdot)\breve{\bY}_i({q_j}) -  \breve{\gamma}_{1i} \\
    \vdots\\
    N_{q_j}^{-1} \bF_{q_j}(H,\cdot)\breve{\bY}_i({q_j}) -  \breve{\gamma}_{Hi}
    \end{pmatrix}\rt|\\
    & = 
    \lt| b^\top  \bV_\hgamma^{-1/2}
    \begin{pmatrix}
    N_{q_j}^{-1} \bF_{q_j}(1,\cdot)\breve{\bY}_i({q_j}) \\
    \vdots\\
    N_{q_j}^{-1} \bF_{q_j}(H,\cdot)\breve{\bY}_i({q_j})
    \end{pmatrix}
    -
    b^\top  \bV_\hgamma^{-1/2} 
    \begin{pmatrix}
    \breve{\gamma}_{1i} \\
    \vdots \\
    \breve{\gamma}_{Hi}
    \end{pmatrix}\rt| \\
    & = |\underbrace{b^\top \bV_\hgamma^{-1/2}N_{q_j}^{-1}\bF_{q_j}\breve{\bY}_i({q_j}) }_{\text{term I}}- \underbrace{b^\top \bV_\hgamma^{-1/2}\breve{\gamma}_i }_{\text{term II}}| \label{eqn:termI-termII}
\end{align}

From \eqref{eqn:vec-center-PO}, term II is the average of term I over $j\in[N]$. Therefore, we can bound term I for all $i,q$ and use triangle inequality to obtain a bound for term II. Combining \eqref{eqn:BEB-M-vec}-\eqref{eqn:termI-termII}, we conclude Part (i) of Theorem \ref{thm:be-proj-standard-vec}.

\vskip 3mm
\noindent\textit{Part (ii) of Theorem \ref{thm:be-proj-standard-vec}.} To prove Part (ii) of Theorem \ref{thm:be-proj-standard-vec}, we use two ways to further obtain a bound for term I that is uniform over $b$.

\textbf{First bound for term I}. Revisit term I. We have
\begin{align}\label{eqn:vec-1}
    \lt|b^\top  \bV_\hgamma^{-1/2}N_{q_j}^{-1}\bF_{q_j}\breve{\bY}_i({q_j}) \rt|
    &=\lt|b^\top  \bV_\hgamma^{-1/2} \bF_{q_j}\{N_{q_j}^{-1}\bS({q_j},{q_j})\}^{1/2}\{N_{q_j}^{-1}\bS({q_j},{q_j})\}^{-1/2}\{N_{q_j}^{-1}\breve{\bY}_i({q_j})\} \rt|  \notag\\
    &\le 
    \lt\|b^\top  \bV_\hgamma^{-1/2}\bF_{q_j}\{N_{q_j}^{-1}\bS({q_j},{q_j})\}^{1/2}\rt\|_2 \cdot  \lt\|\{N_{q_j}^{-1}\bS({q_j},{q_j})\}^{-1/2}\{N_{q_j}^{-1}\breve{\bY}_i({q_j})\} \rt\|_2.
\end{align}
We further bound the first term in \eqref{eqn:vec-1} as follows:
\begin{align}\label{eqn:vec-2}
    &\lt\|b^\top  \bV_\hgamma^{-1/2} \bF_{q_j}\{N_{q_j}^{-1}\bS({q_j},{q_j})\}^{1/2}\rt\|_2^2 \\
    \le & {\sum_{q=1}^Q \lt\|b^\top  \bV_\hgamma^{-1/2} \bF_q\{N_q^{-1}\bS(q,q)\}^{1/2}\rt\|_2^2 } \notag\\
    \le & {\sum_{q=1}^Q  b^\top  \bV_\hgamma^{-1/2} \bF_q\{N_q^{-1}\bS(q,q)\}\bF_q \bV_\hgamma^{-1/2} b  } \notag\\
    \le &{b^\top  \bV_\hgamma^{-1/2}
    (\sigma_\bF^2 \bV_\hgamma)  \bV_\hgamma^{-1/2}b}
    \see{by Condition \eqref{eqn:well-conditioned-vec}}\notag\\
    \le & \sigma_F^2.\notag
\end{align}
Combining \eqref{eqn:vec-1} and \eqref{eqn:vec-2}, we have
\begin{align}\label{eqn:first-bd-vec}
\lt|\sum_{h=1}^Hb_hM''_h(i,j)\rt|^2\le 4 {\sigma_F^2} {N_{q_j}^{-1}\breve{\bY}_i({q_j})^\top \bS({q_j},{q_j})^{-1} \breve{\bY}_i({q_j})}.
\end{align}

Combining  \eqref{eqn:second-bd-vec} and \eqref{eqn:first-bd-vec}, we have
\begin{align*}
\lt|\sum_{h=1}^Hb_hM''_h(i,j)\rt|\le \min\lt\{2 {\sigma_F} \sqrt{N_{q_j}^{-1}\breve{\bY}_i({q_j})^\top \bS({q_j},{q_j})^{-1} \breve{\bY}_i({q_j})}, \frac{  \|\bF_{q_j}\|_{2,1}\cdot  N_{q_j}^{-1}\|\bY_i({q_j})-\overline{\bY}({q_j})\|_\infty}{\sqrt{\varrho_{\min}\{ \bV_\hgamma\}}}\rt\}.
\end{align*}

\textbf{Second bound for term I}. For $b\in\bbR^H$ with $\|b\|_2 = 1$, construct $b_0 = \bV_\hgamma^{-1/2} b / \|  \bV_\hgamma^{-1/2} b  \|_2 \in\bbR^H $ with $\|b_0\|_2 = 1$. We can verify that
\begin{align*}
    b = \frac{ \bV_\hgamma^{1/2}b_0}{\sqrt{b_0^\top  \bV_\hgamma b_0}}.
\end{align*}
Then 
\begin{align*}
    \lt|b^\top  \bV_\hgamma^{-1/2}N_{q_j}^{-1}\bF_{q_j}\breve{\bY}_i({q_j}) \rt|
    &= 
    \lt|b_0^\top  N_{q_j}^{-1}\bF_{q_j}\breve{\bY}_i({q_j}) \rt| \cdot  \lt|\frac{1}{\sqrt{b_0^\top  \bV_\hgamma b_0}}\rt|\\
    &\le 
    \lt\|b_0^\top \bF_{q_j} \rt\|_1 \cdot  \frac{N_{q_j}^{-1}\|\breve{Y}_i({q_j})\|_\infty}{\sqrt{b_0^\top  \bV_\hgamma b_0}}.
\end{align*}

This gives the bounds that depend on the choice of $b $.

To get a uniform bound, we need to bound $\lt\|b_0^\top \bF_{q_j}  \rt\|_1$ and  $b_0^\top  \bV_\hgamma b_0$. We can show
\begin{align*}
    \|b_0^\top \bF_{q_j} \|_1 = \sum_{k\in[p]} |b_0^\top \bF_{q_j}(\cdot,k)| \le \sum_{k\in[p]} \|\bF_{q_j}(\cdot,k)\|_2 =\|\bF_{q_j}\|_{2,1},~
    {b_0^\top  \bV_\hgamma b_0} \ge {\varrho_{\min}\{ \bV_\hgamma\}}.
\end{align*}

Hence
\begin{align}\label{eqn:second-bd-vec}
    \lt\|b_0^\top \bF_{q_j}\rt\|_1 \cdot  \frac{N_{q_j}^{-1}\|\breve{\bY}_i({q_j})\|_\infty}{\sqrt{b_0^\top  \bV_\hgamma b_0}} \le \frac{ \max_{q\in[Q]}\|\bF_{q_j}\|_{2,1}\cdot  N_{q_j}^{-1}\|\bY_i({q_j})-\overline{\bY}({q_j})\|_\infty}{\sqrt{\varrho_{\min}\{ \bV_\hgamma\}}}.
\end{align}

\end{proof}

\subsection{Proof of Theorem \ref{thm:hVo}}\label{sec:pf-hVo}
\begin{proof}[Proof of Theorem \ref{thm:hVo}]
    
(i) Taking expectation, we have
\begin{align*}
    &\E{\sum_{q\in[Q]} \lt(\frac{F(q,\cdot)^\top Y_q}{Q^{-1}} - \gamma\rt)  \lt(\frac{F(q,\cdot)Y_q}{Q^{-1}} - \gamma^\top\rt)}\\
    = & \bbE\bigg\{\sum_{q\in[Q]} \lt\{\frac{F(q,\cdot)^\top (Y_q - \overline{Y}(q))}{Q^{-1}} - \lt(\gamma - \frac{F(q,\cdot)^\top\overline{Y}(q)}{Q^{-1}}\rt)\rt\} \cdot\\
    & \phantom{\bbE\bigg\{\sum_{q\in[Q]}}\lt\{\frac{F(q,\cdot)(Y_q - \overline{Y}(q))}{Q^{-1}} - \lt(\gamma^\top - \frac{F(q,\cdot)\overline{Y}(q)}{Q^{-1}}\rt)\rt\}\bigg\}\\
    = & \E{ \sum_{q\in[Q]}\frac{F(q,\cdot)^\top F(q,\cdot) (Y_q - \overline{Y}(q))^2}{Q^{-2}} } + \sum_{q\in[Q]} \lt(\gamma - \frac{F(q,\cdot)^\top\overline{Y}(q)}{Q^{-1}}\rt) \lt(\gamma^\top - \frac{F(q,\cdot)\overline{Y}(q)}{Q^{-1}}\rt)\\
    =& Q^2(1-Q^{-1})F^\top \diag{S(q,q)}_{q\in[Q]} F + \sum_{q\in[Q]} \lt(\gamma - \frac{F(q,\cdot)^\top\overline{Y}(q)}{Q^{-1}}\rt) \lt(\gamma^\top - \frac{F(q,\cdot)\overline{Y}(q)}{Q^{-1}}\rt). 
\end{align*}
Therefore, we can show
\begin{align}
    \E{ \hV_{\hgamma} } = & \underbrace{\mu_Q Q^2(1-Q^{-1})F^\top \diag{S(q,q)}_{q\in[Q]} F  - \mu_Q Q \COV{\hgamma}}_{\ostar_1} \notag\\
    & + \underbrace{\mu_Q \sum_{q\in[Q]} \lt(\gamma - \frac{F(q,\cdot)^\top\overline{Y}(q)}{Q^{-1}}\rt) \lt(\gamma^\top - \frac{F(q,\cdot)\overline{Y}(q)}{Q^{-1}}\rt)}_{\ostar_2}. \label{eqn:E-hVo-2}
\end{align}
Recall the variance formula for $\hgamma$ is given by $\COV{\hgamma} = F^\top V_\hgamma F$. It follows that the two parts of \eqref{eqn:E-hVo-2} are lower bounded by 
\begin{align*}
    \ostar_1 \succeq \mu_Q Q(Q-2) \COV{\hgamma}, \quad \ostar_2 \succeq 0.
\end{align*} 
Therefore, this motivates us to choose $ \mu_Q= \{Q(Q-2)\}^{-1}$ to obtain a conservative estimator. 

(ii) The proof of (ii) can be done by applying Chebyshev's inequality and the moment inequalities given in Lemma \ref{lem:moments-U}. It is omitted here.

(iii) The proof of (iii) is based on part (ii) and \eqref{eqn:op-inf}.
\end{proof}

\subsection{Statement and proof of extension of Condition \ref{cond:easy-spec}}\label{sec:extend-conditions}
From a super population perspective, we consider potential outcomes that are generated from some probability distributions.
\begin{proposition}\label{prop:extend-conditions}
    Assume the potential outcomes are a sample from the super population where $Y_i(q)$'s are independent and each
    $Y_i(q)$ has mean $\mu_q$, variance $s_q^2$ and is sub-gaussian with parameter $\sigma^2$:
    \begin{align*}
        \Prob{|Y_i(q) - \mu_q|\ge t} \le C e^{-\frac{t^2}{2\sigma^2}}.
    \end{align*}
    Then there exists a universal constant $C > 0$, such that with probability great than $1 - C(NQ)^{-1}$, we have 
    \begin{gather*}
        \max_{q\in[Q],i\in[N]}|Y_i(q) - \overline{Y}(q)|\le 3\sigma \sqrt{\log(QN)},\\
        \min_{q\in[Q]} S(q,q) \ge \min_{q\in[Q]}s_q/2.
    \end{gather*}
\end{proposition}

Proposition \ref{prop:extend-conditions} suggests that $M_N(q)$'s are upper bounded by logarithms of $QN$ and $S(q,q)$'s are lower bounded by constants. These orders are smaller than $O(N)$ and thus allow the use of the general BEB results in Section \ref{sec:PCLT-projection}.

\begin{proof}[Proof of Proposition \ref{prop:extend-conditions}]
    Let $\nu = \max_{q\in[Q]}\mu_q$ and  $\underline{s} = \min_{q\in[Q]} s_q^2$.
    By Bonferroni correction, we have
    \begin{align*}
        \Prob{\max_{q\in[Q],i\in[N]}|Y_i(q) - \mu_q|\ge 2\sigma\sqrt{\log(NQ)}}
        \le C N\sum_{q\in[Q]}e^{-\frac{4\sigma^2\log(NQ)}{2\sigma^2}} \le C(NQ)^{-1}.
    \end{align*}
    That is, with probability higher than $1 - C(NQ)^{-1}$, we have
    \begin{align}\label{eqn:bound-Yiq}
        \max_{q\in[Q],i\in[N]}|Y_i(q) - \mu_q|\le 2\sigma\sqrt{\log(NQ)}.
    \end{align}
    By Hoeffding's inequality \citep{wainwright2019high}, 
    \begin{align}
        \Prob{|\overline{Y}(q) - \mu_q| \ge \nu} \le Ce^{-\frac{N\nu^2}{2\sigma^2}}.
    \end{align}
    By a Bonferroni union bound, we can show that with probability greater than $1 - C(QN)^{-1} $,
    \begin{align} \label{eqn:bound-Yq}
        \max_{q\in[Q]}|\overline{Y}(q) - \mu_q|\le 2\sigma\sqrt{\frac{\log(NQ)}{N}}.
    \end{align}
    Besides, the squared variables $(Y_i(q) - \mu_{q})^2$ are sub-exponential  \citep[Example 2.8 and Theorem 2.13]{wainwright2019high}. Therefore with probability greater than $1 - CQ\exp({-C'N}) $, we have
    \begin{align}\label{eqn:bound-Yiq2}
        \max_{q\in[Q]} \lt|\frac{1}{N-1}\sum_{i=1}^N (Y_i(q) - \mu_q )^2 - s_q^2\rt| \le \frac{\min_{q\in[Q]}s_q}{2}.
    \end{align}
    Summarizing the results \eqref{eqn:bound-Yiq}, \eqref{eqn:bound-Yq} and \eqref{eqn:bound-Yiq2}, we complete the proof.
\end{proof}

\subsection{Proof of the statements in Example \ref{exp:two-arm}}\label{sec:pf-two-arm}
    \begin{itemize}
      \item When the potential outcomes under treatment and control are non-negatively correlated, i.e. $S(1,0) \ge 0$,  it always holds that
    \begin{align*}
        V_\hgamma \ge \frac{p_0}{N_1}S(1,1) + \frac{p_1}{N_0} S(0,0) \ge \min\{p_0,p_1\} \left\{\frac{1}{N_1}S(1,1) + \frac{1}{N_0} S(0,0)\right\}.
    \end{align*}
    Hence \eqref{eqn:well-conditioned} holds with $\sigma_F^{-2} = \min\{p_0,p_1\}$.

      \item When the potential outcomes are negatively correlated, the   lower bound for $S(1,0)$ given by Cauchy-Schwarz inequality is:
    \begin{align}\label{eqn:lower-bd-cs}
        S(1,0) \ge - \sqrt{S(1,1)S(0,0)}.
    \end{align}
    If the correlation does not attain the worst-case bound, say
    \begin{align*}
        S(1,0) \ge - c\sqrt{S(1,1)S(0,0)}
    \end{align*}
    for some universal constant $c\in (0,1)$, then we have bound
    \begin{align*}
        V_\hgamma &\ge \frac{p_0}{N_1}S(1,1) + \frac{p_1}{N_0} S(0,0) - \frac{2c}{N}\sqrt{S(1,1)S(0,0)}\\
        &\ge \frac{p_0}{N_1}S(1,1) + \frac{p_1}{N_0} S(0,0) - \frac{c}{N}\lt\{\frac{N_0}{N_1}S(1,1) + \frac{N_1}{N_0}S(0,0)\rt\}\\
        &\ge (1-c)\min\{p_0,p_1\} \left\{\frac{1}{N_1}S(1,1) + \frac{1}{N_0} S(0,0)\right\}.
    \end{align*}
    Then \eqref{eqn:well-conditioned} holds with $\sigma_F^{-2} = (1-c)\min\{p_0,p_1\}$.

    When the bound in \eqref{eqn:lower-bd-cs} is attained, \eqref{eqn:well-conditioned} might be violated. We can verify that
    \begin{align*}
        V_\hgamma = \frac{1}{N}\lt(\sqrt{\frac{N_0}{N_1}S(1,1)} - \sqrt{\frac{N_1}{N_0}S(0,0)}\rt)^2.
    \end{align*}
    Hence if 
    \begin{align*}
        \sqrt{\frac{N_0}{N_1}S(1,1)} = \sqrt{\frac{N_1}{N_0}S(0,0)} \quad \text{ or equivalently } \quad \frac{S(1,1)}{S(0,0)} = \frac{N_1^2}{N_0^2},
    \end{align*}
    then $V_\hgamma = 0$ and $\hgamma$ has a degenerate covariance matrix. 
    \end{itemize}

\newpage

\end{appendix}

\end{document}